\documentclass[twocolumn,aps,superscriptaddress,showpacs,floatfix,prd,noshowpacs]{revtex4-2}
\usepackage{mathrsfs}
\usepackage{amssymb}
\usepackage{amsmath}
\usepackage{graphicx}
\usepackage[normalem]{ulem}
\usepackage[dvips]{color}
\usepackage{bm}
\usepackage{longtable}
\usepackage{slashed}

\usepackage{enumitem}

\usepackage{times}
\usepackage{fouriernc}

\usepackage[bb=ams, cal=cm, scr=boondox, frak=euler]{mathalpha}

\usepackage[breaklinks=true]{hyperref}
\hypersetup{
  colorlinks=true,
  citecolor=cyan,
  linkcolor=black,
  urlcolor=cyan,
}

\renewcommand\sout{\bgroup \color{red} \ULdepth=-.5ex \ULset}

\renewcommand{\rm}[1]{\textrm{#1}}
\renewcommand{\d}{\mathrm{d}}

\begin{document}
\title{Strong Gravity Extruding Peaks in Speed of Sound Profiles of Massive Neutron Stars}

\author{Bao-Jun Cai\footnote{bjcai87@gmail.com}}
\affiliation{Quantum Machine Learning Laboratory, Shadow Creator Inc., Shanghai 201208, China} 
\author{Bao-An Li\footnote{Bao-An.Li$@$tamuc.edu}}
\affiliation{Department of Physics and Astronomy, Texas A$\&$M
University-Commerce, Commerce, TX 75429-3011, USA}

\date{\today}

\begin{abstract}

The speed of sound squared (SSS) $s^2$ in massive neutron stars (NSs) characterizes not only the stiffness of supradense neutron-rich matter within but also equivalently properties of the curved geometry due to the strong-field gravity and matter-geometry coupling. A peaked density or radius profile of $s^2$ has been predicted for massive NSs using various NS Equation of State (EOS) models. However, the nature, cause, location and size of the peak in $s^2$ profiles are still very EOS model dependent. In this work, we investigate systematically $s^2$ profiles in massive NSs in a new approach that is independent of the nuclear EOS model and without any presumption about the NS structure and/or composition. In terms of the small quantities (reduced radius, the energy density and pressure scaled by their central values), we perform double-element perturbative expansions in solving perturbatively the scaled Tolman--Oppenheimer--Volkoff (TOV) equations and analyzing $s^2$ profiles from the Newtonian limit to the general relativistic (GR) case. The GR term in the TOV equations plays a twofold role: it compresses NS matter and modifies the pressure/energy density ratio from small values in Newtonian stars showing no $s^2$ peak to large ones for massive NSs possessing a peak in their $s^2$ profiles, and eventually takes away the peak in extremely compact/massive NSs approaching the causality limit. 
{In particular,  the peaked behavior in $s^2$ is expected to emerge near the center of massive NSs like PSR J0740+6620, while a sharp phase transition is unlikely to occur there.}
These features revealed from our analyses are universal as they are intrinsic properties of the GR stellar structure equations independent of the still very uncertain EOS of supradense neutron-rich matter in NSs.

\end{abstract}

\maketitle

\section{Introduction}

The speed of sound squared (SSS) defined as $s^2\equiv \d P/\d\varepsilon$
characterizes effectively the stiffness of the Equation of State (EOS: $P(\varepsilon)$) of dense nuclear matter\,\cite{Baym18}.
Here $P$ and $\varepsilon$ are the pressure and energy density of the system under consideration, respectively.
Great attention has been paid recently to the density dependence of $s^2$ in both neutron stars (NSs) and heavy-ion collisions. In particular, its possible peaked behavior which may reveal novel structures/phases of superdense matter has stimulated much interest and extensive investigations\,\cite{Tews18,Baym19,Cas19,McL19,Dua20,Fer20,Mot20,Mal20,Zhao20,Min21,Jok21,Sen21,Stone21,Kap21,Tan22,Tan22-a,Alt22,Dri22,Huang22,Kojo22,Ecker22,Ecker23,Fuji22,Fuji23,Han23,Mar23,Pro23,Som23,Liu23,Chen23}.
For example,  the approximate conformal symmetry of quark matter at extremely high densities and its possible realization in massive NS cores\,\cite{Ann18,Ann20N,Ann23,Gorda23,Gorda21} may induce a peak in $s^2$, indicating possibly the occurrence of a sharp phase transition or a continuous crossover.
However, it has also been shown recently that a purely nucleonic EOS model may also generate such a peak in dense neutron-rich matter accessible in massive NSs and/or relativistic heavy-ion collisions within the current uncertainty ranges of its high-density parameters\,\cite{ZLi23}. It is also interesting to note that the $s^2$ in self-bound quark stars made purely of absolutely stable deconfined quark matter, on the other hand,  may not show the peaked behavior\,\cite{Cao23}. Moreover, a very  recent Bayesian analysis\,\cite{Mro23} of X-ray measurements and gravitational wave (GW) observations of NSs\,\cite{Abbott2017,Abbott2018,Abbott2020} incorporating the perturbative QCD (pQCD) predictions\,\cite{Gorda21} shows that the peaked behavior in $s^2$ is consistent with but not required by these astrophysical data and pQCD predictions, see also Refs.\,\cite{Brandes23,Brandes23-a,Tak23,Pang23,Fan23} for more discussions on the related issues. What is the nature or cause of SSS's peaked behavior in massive NSs? Given the fact that essentially all previous studies on this question have been done within some selected NS EOS models leading to the various alternatives/possibilities mentioned above, here we try to answer this question in an approach that is independent of the nuclear EOS model and without any presumption about the NS structure and/or composition.

\renewcommand*\figurename{\small FIG.}
\begin{figure}[h!]
\centering
\includegraphics[width=7.5cm]{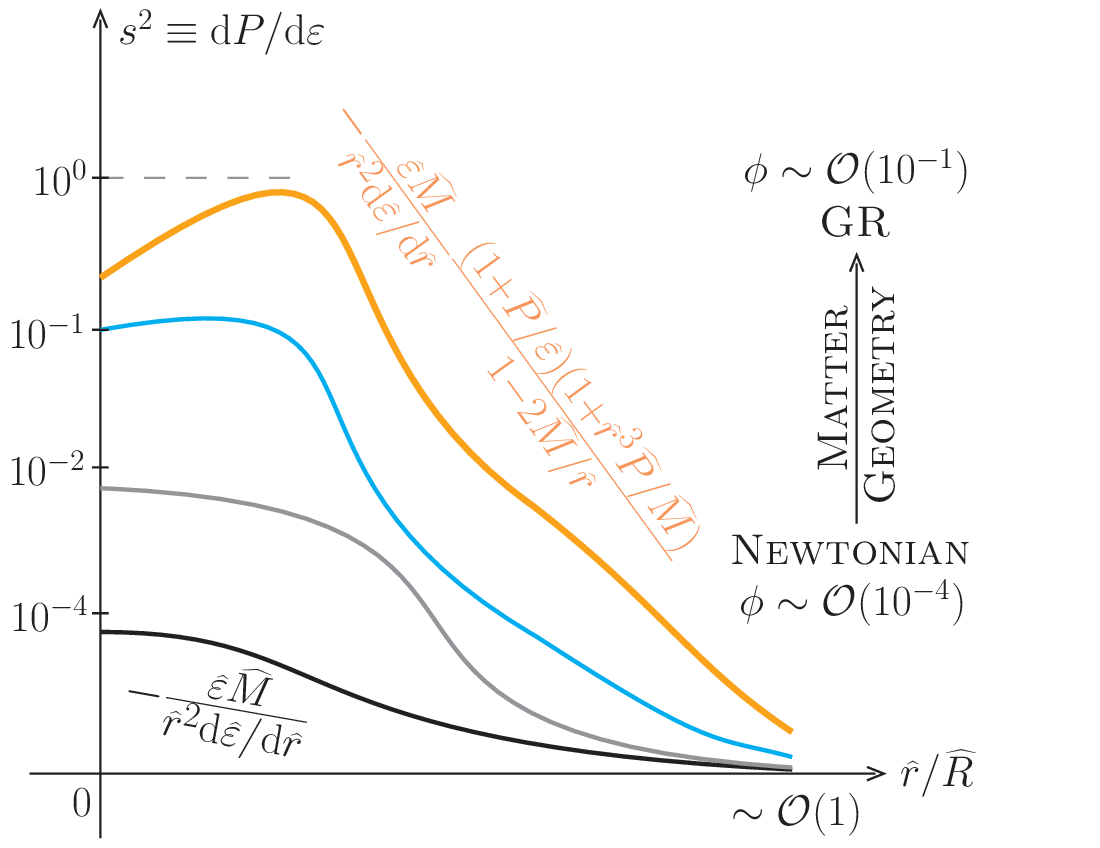}
\caption{(Color Online). A sketch of the evolution of $s^2$'s radial profile in NSs.
{At low densities characterized by small $\phi=P/\varepsilon$ (Newtonian limit), the $s^2$ monotonically decreases from the center to the surface (black line),  and as the $\phi$ increases approaching the GR case (orange line) a peak eventually emerges near the NS center.
The general expressions for $s^2$ are also captioned near the curves,  see Section \ref{SEC_EOS} and Section \ref{SEC_SS} for the notations and detail quantitative discussions.  We adopt $c=1$ in our discussion.}
}\label{fig_s2_GRNewt}
\end{figure}

Firstly, an order-of-magnitude analysis for some special cases can provide us useful clues and set the relevant scales for understanding the behavior of $s^2$.
Because the $s^2$ is dimensionless (adopting $c=1$) and $P$ and $\varepsilon$ have the same units,  we can write generally $s^2= \phi f(\phi)$ as $s^2\to0$ if no matter exists ($\phi\to0$), here $\phi\equiv P/\varepsilon$.
The function $f$ can be expanded around $\phi\approx0$ as $f\approx f_0+f_1\phi+f_2\phi^2+\cdots$ with $f_0>0$ (required by the stability condition $s^2\geq0$).
Considering stars as white dwarfs (WDs), we have $P\lesssim 10^{22\mbox{-}23}\,\rm{dynes}/\rm{cm}^2\approx10^{-(11\mbox{-}10)}\,\rm{MeV}/\rm{fm}^3$ and $\varepsilon\lesssim10^{8\mbox{-}9}\,\rm{kg}/\rm{m}^3\sim10^{-6}\,\rm{MeV}/\rm{fm}^3$, thus $\phi\lesssim10^{-(5\mbox{-}4)}$. 
Similarly, for NS matter around nuclear saturation density $\rho_{\rm{sat}}\approx0.16\,\rm{fm}^{-3}$, one estimates $P\lesssim3\,\rm{MeV}/\rm{fm}^3$ and $
\varepsilon\approx150\,\rm{MeV}/\rm{fm}^3$,  and therefore $\phi\lesssim0.02$.
For these systems one can safely neglect the terms $f_1\phi+f_2\phi^2+\cdots$ in $f$, therefore $s^2\approx f_0\phi$ or equivalently $P\sim \varepsilon^{f_0}$ is obtained\,\cite{Shapiro1983}, indicating $s^2$ is probably an increasing function of $\phi$ (or energy density $\varepsilon$). The absence of a peak in $s^2$ in NSs around $\rho_{\rm{sat}}$ was confirmed by the chiral effective field theory (CEFT)\,\cite{Tews18,Ess21}, e.g., Ref.\,\cite{Ess21} predicted that $s^2$ is monotonic around $\rho_{\rm{sat}}$ and $s^2(\rho_{\rm{sat}})\approx0.03\sim\phi$.
On the other hand,  $\phi$ could be sizable $\gtrsim\mathcal{O}(0.1)$ in massive NSs especially in their cores and possible peaks in $s^2$ may emerge.
In this sense, massive NSs like PSR J0740+6620 observed by Neutron Star Interior Composition Explorer (NICER)/XMM Newton\,\cite{Fon21,Riley21,Miller21,Salmi22}, PSR J1614-2230 via Shapiro delay\,\cite{Dem10} and PSR J0348+0432 using its spectroscopy\,\cite{Ant13} provide us excellent opportunities to study the possible peaked structure in $s^2$ profiles.

In this work,  we demonstrate that the super-strong gravity in massive NSs which makes $P/\varepsilon\gtrsim\mathcal{O}(0.1)$ unavoidably induces a peak in $s^2$ in their cores under some general conditions, and there would be no peak in $s^2$ for Newtonian stars with small $P/\varepsilon$ values.
Our results are obtained directly from solving perturbatively the scaled general-relativistic (GR) stellar structure equations without using any NS EOS model. We thus essentially removed the elusiveness of superdense matter EOS in understanding properties of $s^2$ profiles in NSs.
FIG.\,\ref{fig_s2_GRNewt} sketches the evolution of the radial $s^2$ profile in NSs (near their centers) from the Newtonian limit (with small $P/\varepsilon$) to the GR case (see the following two sections on the notations and detailed analyses).
Due to the perturbative nature of our method (Section \ref{SEC_EOS}), we primarily focus on the peaked structure near the NS center, although other structures (e.g., plateau, bump and spike, etc.) may emerge at various locations or densities in NSs\,\cite{Tan22}.

The rest of this paper is organized as follows: In Section \ref{SEC_EOS} our perturbative method is briefly reviewed and as an application we derive the NS core EOS $P$-$\varepsilon$ to positions around one-fifth of the NS radius from the center.
Section \ref{SEC_SS} discusses the SSS as a connection between the supradense matter and the compact space geometry, we then use the general boundary conditions and requirement for a peaked $s^2$ to argue why there is probably no peaked behavior in $s^2$ at low $P/\varepsilon$ values.
We show in Section \ref{SEC_NEWT} that  this is the case for Newtonian stars.
Section \ref{SEC_GR} is devoted to studying the emergence of a peak in $s^2$ near massive NS centers when the $P/\varepsilon$ is sizable $\gtrsim\mathcal{O}(0.1)$.
In Section \ref{SEC_COM}, we connect our predictions for the peaked structure in $s^2$ with some other predictions in the literature.
Section \ref{SEC_SUM} summarizes our work.
In Appendix \ref{App0}, we give a general proof on the absence of odd terms in expanding $\widehat{\varepsilon}$ and $\widehat{P}$ over $\widehat{r}$,  in Appendix \ref{App} we estimate the maximum size of the coefficient $a_4$ used in the radial-expansion of the reduced energy density to the fourth power of the reduced radial coordinate using a meta EOS model for NSs consisting of neutrons, protons and electrons (npe) at $\beta$-equilibrium by varying its parameters within their currently known uncertainty ranges.

\section{Method Review and EOS in NS Cores}\label{SEC_EOS}

\subsection{Perturbative Approach for Solving the Scaled TOV Equations without Using any Nuclear EOS Model}

The GR stellar equations for NSs,  i.e., the celebrated Tolman--Oppenheimer--Volkoff (TOV) equations\,\cite{TOV39-1,TOV39-2,Misner1973}, effectively coupling geometry and matter are the very basis for investigating properties of supradense matter in NSs\,\cite{Walecka1974,Chin1976,Baym1976,Freedman1977,Akmal1998,
LP01,Alford2008,LCK08,Wat16,Oertel2017,Isa18,Dri21,Lovato22,Soren2023}. 
By anatomizing the intrinsic structures of the TOV equations,  nearly model-independent EOS properties could be obtained.
Specifically, we write the TOV equations in the dimensionless form as\,\cite{CLZ23-a,CLZ23-b},
\begin{equation}\label{def-1}
\frac{\d\widehat{P}}{\d\widehat{r}}
=-\frac{\widehat{\varepsilon}\widehat{M}}{\widehat{r}^2}
\frac{(1+{\widehat{P}}/{\widehat{\varepsilon}})
(1+{\widehat{r}^3\widehat{P}}/{\widehat{M}})}{1-
{2\widehat{M}}/{\widehat{r}}},~~\frac{\d\widehat{M}}{\d\widehat{r}}=\widehat{r}^2\widehat{\varepsilon},
\end{equation}
here the (reduced) pressure $\widehat{P}=P/\varepsilon_{\rm{c}}$,  energy density $\widehat{\varepsilon}=\varepsilon/\varepsilon_{\rm{c}}$ and $\widehat{M}=M/W$ are all functions of the (reduced) radius $\widehat{r}=r/Q$ with $W\equiv G^{-1}(4\pi G\varepsilon_{\rm{c}})^{-1/2}$ and $Q\equiv (4\pi G \varepsilon_{\rm{c}})^{-1/2}$.

{Considering the fact that $\widehat{P}$ and $\widehat{\varepsilon}$ are even functions of $\widehat{r}$ while $\widehat{M}$ is an odd function of $\widehat{r}$\,\cite{CLZ23-a}},
one can expand them according to\,\cite{CLZ23-a}
\begin{align}
\widehat{P}\approx& \widehat{P}_{\rm{c}}+b_2\widehat{r}^2+b_4\widehat{r}^4+\cdots,\\
\widehat{\varepsilon}\approx& 1+a_2\widehat{r}^2+a_4\widehat{r}^4+\cdots,\\
\widehat{M}\approx&\widehat{r}^3/3+a_2\widehat{r}^5/5+a_4\widehat{r}^7/7+\cdots.\label{u-M}
\end{align}
{An alternative and more general mathematical proof on the evenness or oddness of these functions is given in Appendix \ref{App0},  besides the original proof given in\,\cite{CLZ23-a} where the relevant expansion coefficients are evaluated individually. In fact, the structure of $\widehat{M}$ as a function of $\widehat{r}$ could be seen immediately from the mass evolution part of TOV equations, e.g. the evenness of $\widehat{\varepsilon}(\widehat{r})$ implies that $\widehat{M}$ is an odd function of $\widehat{r}$ and 
consequently $\widehat{M}(0)=0$.
Now, by putting $\widehat{P}$, $\widehat{\varepsilon}$ and $\widehat{M}$ into the pressure evolution equation, one can find that} $b_2=-6^{-1}(1+3\widehat{P}_{\rm{c}}^2+4\widehat{P}_{\rm{c}})$, $a_2=b_2/s_{\rm{c}}^2$ and 
{\begin{equation}\label{def-b4}
b_4=-\frac{1}{2}b_2\left(\widehat{P}_{\rm{c}}
+\frac{4+9\widehat{P}_{\rm{c}}}{15s_{\rm{c}}^2}\right),
\end{equation}}etc., here $s_{\rm{c}}^2\equiv \d P_{\rm{c}}/\d\varepsilon_{\rm{c}}$ is the central SSS \cite{CLZ23-a}.
{By truncating to order $\widehat{r}^2$, the pressure is given as $\widehat{P}\approx \widehat{P}_{\rm{c}}+b_2\widehat{r}^2$,  from which and the definition of NS radius $\widehat{R}$ via $\widehat{P}(\widehat{R})=0$ we obtain\,\cite{CLZ23-a}:
\begin{equation}\label{skd-1}
R=\widehat{R}Q\sim\frac{1}{\sqrt{\varepsilon_{\rm{c}}}}\left(
\frac{\widehat{P}_{\rm{c}}}{1+3\widehat{P}_{\rm{c}}^2+4\widehat{P}_{\rm{c}}}
\right)^{1/2}\equiv \nu_{\rm{c}}.
\end{equation}
Then the mass could be found to scale as\,\cite{CLZ23-a}
\begin{equation}\label{skd-2}
M_{\rm{NS}}=\widehat{M}W\sim\frac{1}{\sqrt{\varepsilon_{\rm{c}}}}\left(
\frac{\widehat{P}_{\rm{c}}}{1+3\widehat{P}_{\rm{c}}^2+4\widehat{P}_{\rm{c}}}
\right)^{3/2}\equiv \Gamma_{\rm{c}}.
\end{equation}
Here the $\nu_{\rm{c}}$ and $\Gamma_{\rm{c}}$ factors are both in unit of $\rm{fm}^{3/2}/\rm{MeV}^{1/2}$, namely the unit of their front term $1/\sqrt{\varepsilon_{\rm{c}}}$
since the $\widehat{P}_{\rm{c}}$ is dimensionless.

As discussed in detail and demonstrated quantitatively in\,\cite{CLZ23-a}, the above scalings relates directly the two most general NS observables with the central energy and pressure. Knowing either the mass or the radius, one can obtain the NS central EOS $P_{\rm{c}}(\varepsilon_{\rm{c}})$, while knowing both simultaneously one can get the individual values of the central pressure $P_{\rm{c}}$ and energy density $\varepsilon_{\rm{c}}$. 

The maximum-mass configuration of NSs is uniquely useful for investigating the EOS of the densest visible matter\,\cite{Lat05,Of1,Of2} existing in our Universe.
The scalings $R$-$\nu_{\rm{c}}$ and $M_{\rm{NS}}$-$\Gamma_{\rm{c}}$ at the maximum-mass configuration $M_{\rm{NS}}^{\max}\equiv M_{\rm{TOV}}$ on the mass-radius (M-R) curve were verified by using 104 widely used microscopic as well as phenomenological EOSs in solving the TOV equations with the traditional integration approach. They can be written as\,\cite{CLZ23-a}
\begin{align}
M_{\rm{NS}}^{\max}/M_{\odot}\approx&1.73\times10^3\left(\frac{\Gamma_{\rm{c}}}{\rm{fm}^{3/2}/\rm{MeV}^{1/2}}\right)-0.106,\label{sdd-1}\\
R_{\max}/\rm{km}\approx&1.05\times10^3\left(\frac{\nu_{\rm{c}}}{\rm{fm}^{3/2}/\rm{MeV}^{1/2}}\right)+0.64,\label{sdd-2}
\end{align}
where $R_{\max}$ is the NS radius at $M_{\rm{NS}}^{\max}$.
As a numerical example, for a NS having $P_{\rm{c}}=200\,\rm{MeV}/\rm{fm}^3$ and $\varepsilon_{\rm{c}}=800\,\rm{MeV}/\rm{fm}^3$, we have $\nu_{\rm{c}}\approx0.012\,\rm{fm}^{3/2}/\rm{MeV}^{1/2}$ and $\Gamma_{\rm{c}}\approx0.0014\,\rm{fm}^{3/2}/\rm{MeV}^{1/2}$, respectively. The Eqs.\,(\ref{sdd-1}) and (\ref{sdd-2}) then give $M_{\rm{TOV}}\equiv M_{\rm{NS}}^{\max}\approx2.26M_{\odot}$ and $R_{\max}\approx11.9\,\rm{km}$, respectively. In fact, the $M_{\rm{NS}}^{\max}$ from this example is consistent with the $M_{\rm{TOV}}=2.25^{+0.08}_{-0.07}$ M$_{\odot}$ from a very recent Bayesian analysis of currently available multi-messenger astronomical data under knowing constraints from nuclear physics \cite{Fan23}. The $R$-$\nu_{\rm{c}}$ scaling for other NS masses was also verified\,\cite{CLZ23-b}.
It is interesting to note that very recently Lattimer verified independently the validity of the scalings in Eqs.\,(\ref{skd-1}) and (\ref{skd-2}) by using a large and diverse set of nuclear EOSs available in the literature\,\cite{Jim}. He found that at $M_{\rm{NS}}^{\max}$, the accuracies of Eqs.\,(\ref{skd-1}) and (\ref{skd-2}) are 7\% and 8\%; and at $1.4M_{\odot}$, they are 2\% and 6\%, respectively\,\cite{Lat24-talk}.

In addition, the $s_{\rm{c}}^2$ for the maximum-mass configuration $M_{\rm{NS}}^{\max}$ could be obtained as\,\cite{CLZ23-a},
\begin{equation}\label{def_sc2}
s_{\rm{c}}^2=\widehat{P}_{\rm{c}}\left(1+\frac{1}{3}\frac{1+3\widehat{P}_{\rm{c}}^2+4\widehat{P}_{\rm{c}}}{1-3\widehat{P}_{\rm{c}}^2}\right),
\end{equation}
{via the condition $\d M_{\rm{NS}}/\d\varepsilon_{\rm{c}}=0$ with $M_{\rm{NS}}\sim\widehat{R}^3/\sqrt{\varepsilon_{\rm{c}}}$\,\cite{CLZ23-a}.}
We focus on results of the maximum-mass configuration $M_{\rm{NS}}^{\max}$ in this work and give related results for NSs along the M-R curve only in Subsection \ref{sub4}.
It is also necessary to note that the causality condition $s_{\rm{c}}^2\leq1$ was found by us earlier to give an upper limit for the reduced central pressure as $\widehat{P}_{\rm{c}}\lesssim0.374$\,\cite{CLZ23-a}.

{To investigate the nature and source of the peaked $s^2$ profiles in massive NSs, it is useful to note here that} the terms on the right side of the TOV pressure evolution equation could be classified as follows\,\cite{Sil04,Smi12},
\begin{enumerate}[leftmargin=*,label=(\alph*)]
\item The front factor $-{\widehat{\varepsilon}\widehat{M}}/{\widehat{r}^2}$ is from the classical equation for NSs in hydrostatic equilibrium under Newtonian limit\,\cite{Chan10}. 
\item The two terms in the numerator represent special relativity (SR) corrections of order $(v/c)^2$ arising from Einstein's mass-energy relation and the fact that the pressure goes like $v^2$ at the non-relativistic limit. The ratio $\widehat{P}/\widehat{\varepsilon}=\phi$ is a matter effect (due to the absence of $\widehat{r}$), and it should be zero if $\widehat{P}=0$ is taken. The $\widehat{r}^3\widehat{P}/\widehat{M}$ is the coupling between matter (characterized by $\widehat{P}$) and geometry (by $\widehat{r}^3/\widehat{M}\approx3-9a_2\widehat{r}^2/5+\cdots$, see Eq.\,(\ref{u-M})). It also vanishes by taking $\widehat{P}=0$. 
\item The denominator $1-2\widehat{M}/\widehat{r}$ is a pure GR effect. It remains even if $\widehat{P}=0$ is taken.
Specifically, as we shall show with Eq.\,(\ref{u-M}) that $2\widehat{M}/\widehat{r}\approx 2\widehat{r}^2/3+2a_2\widehat{r}^4/5+\cdots$, which involves only $\widehat{r}$ and is small for Newtonian stars (e.g., $\widehat{r}^2\leq\widehat{R}^2\lesssim10^{-3}$, see Section \ref{SEC_NEWT}).
The factor $2\widehat{M}/\widehat{R}$ can be sizable for massive and compact NSs, making the GR factor $(1-2\widehat{M}/\widehat{r})^{-1}$ in the TOV equations large.
\end{enumerate}
The presence of strong-gravity in NSs is a necessary condition for making the $P/\varepsilon$ sizable.
For Newtonian stars with relatively weak gravity, both the SR and GR corrections are small, leading to the classic Newtonian evolution equation for pressure as $\d\widehat{P}/\d\widehat{r}=-\widehat{\varepsilon}\widehat{M}/\widehat{r}^2$.

{To ease our following discussions and appreciate better the benefits of analyzing perturbatively the dimensionless TOV equations using polynomials of reduced variables with respect to the traditional integration approach for solving the TOV equation given a model nuclear EOS, we notice here that:
\begin{enumerate}[leftmargin=*,label=(\alph*)]
\item The TOV equations can be obtained from varying the total action of NS matter and gravity according to the Hamiltonian variational principle. 
As such, there is a well-known competition/degeneracy between gravity/geometry and NS matter in describing global properties of NSs.
In the classical picture of Newtonian stars, the repulsive nuclear pressure has to balance the attractive gravity. Thus, given the mass and radius observed for a NS, one should be able to extract the same NS EOS from investigating either the NS gravity/geometry or matter inside it. The EOSs extracted independently from both directions should match/constrain each other.

\item The TOV equations are blind about the composition of NSs in the sense that regardless how/what the energy density is made of, as long as the same
EOS $P(\varepsilon)$ is given, a unique mass-radius sequence is determined. This feature is independent of the techniques (using integral or polynomials) people may use to solve the TOV equations. As well documented in the literature, combinations of different mechanisms including formations of various new particles, e.g., hyperons, baryon resonances and possible phase transitions to quarks and gluons with/without considering dark matter, can lead to the same NS EOS. The resulting mass-radius curve can not distinguish the composition of NSs with the same EOS unless one looks into observables from microscopic processes happening inside NSs. 

\item Solving the TOV equations requires an input nuclear EOS which is presently model dependent. On the other hand, our perturbative analysis starting from the NS center using polynomials reveals directly scalings of the NS mass and radius with combinations of the central pressure and energy density through the $\Gamma_{\rm{c}}$ and $\nu_{\rm{c}}$ factors without using any input nuclear EOS. Essentially, we expand the physical solution of the TOV equations close to the NS center. The most important outcome is that once the $\Gamma_{\rm{c}}$ or $\nu_{\rm{c}}$ factor is known via observations of $M_{\rm{NS}}^{\max}$ or $R_{\max}$, the central EOS $P_{\rm{c}}(\varepsilon_{\rm{c}})$ is determined, although theoretically different nuclear physics mechanisms may lead to the same $P_{\rm{c}}(\varepsilon_{\rm{c}})$. In this sense, the features of the central EOS we extracted, i.e., the peaked radial/density profile of SSS $s^2$, are determined mostly by the NS strong-field gravity or geometry instead of the nuclear EOS models as we shall demonstrate. One of our major findings is that a sharp phase transition (PT) near the NS center is basically ruled out, while a continuous crossover signaled by a reduction of $s^2$ in NS cores (or equivalently a peaked behavior at some places near the center) and PTs occurring far from the centers are not excluded. 
Similarly, our approach can not tell the microscopic mechanisms (e.g., formation of hyperons, a mixed phase or high-order transitions to purely quark matter) leading to the reduction of $s^2$. Consequently,  PTs could not alter the conclusions of our analyses using the perturbative approach here. 

\item Both the traditional approach for solving the TOV equations and our perturbative analysis of its central solutions have their own advantages and limitations. Nevertheless, main features of the central EOSs from the two approaches have to match. Thus, studies of some NS properties using both approaches may be beneficial. They may provide complementary information leading to a deeper understanding of superdense matter in strong gravitational fields.

\end{enumerate}

\subsection{TOV Prediction on the Core EOS in NSs without Using any Nuclear Physics Input}

The expansion coefficients of $\widehat{P}$ and $\widehat{\varepsilon}$ over $\widehat{r}$ are heuristic as they contain fundamentally the intrinsic information on the NS structure starting from the center.
For example,  the coefficients $a_2$, $b_2$ and $b_4$ are all independent of the denser matter EOS while $a_4$ has some ambiguities due to the uncertainties of the EOS {especially at high densities relevant to NSs}, but as we shall show later our conclusions are independent of the specific values of $a_4$ within its maximum uncertainty range. 
{In fact, since all quantities are made dimensionless and considering typical NSs with $R\approx10\mbox{-}15\,\rm{km}$ and $\varepsilon_{\rm{c}}\approx10^{2\mbox{-}3}\,\rm{MeV}/\rm{fm}^3$, the length scale $Q$ is on the same order as $R$ and so its reduced value is about $\widehat{R}\sim\mathcal{O}(1)$.
The form of the energy density $\widehat{\varepsilon}\approx 1+a_2\widehat{r}^2+a_4\widehat{r}^4+\cdots$ together with $0\leq\widehat{\varepsilon}\leq1$ and $0\leq\widehat{r}\leq\widehat{R}\sim\mathcal{O}(1)$ then essentially indicate the magnitude of $a_2$ and $a_4$, etc., is $\sim\mathcal{O}(1)$.
In particular,  combing the expressions for $b_2$ and $s_{\rm{c}}^2$ (of Eq.\,(\ref{def_sc2})) gives $-1.5\lesssim a_2\lesssim-0.5$ for $0.1\lesssim\widehat{P}_{\rm{c}}\lesssim0.374$, with the latter being a typical range for $\widehat{P}_{\rm{c}}$ in NSs.
Moreover,  we give an example in the appendix on how the coefficient $a_4$ is related to the dense matter EOS and its maximum size is found to be around 1 based on our current best knowledge about nuclear EOS. }
Therefore, at places with small $\widehat{r}$ (i.e., near the NS center) where higher-order coefficients have weak impact,  the resulted EOS are expected to be nearly model-independent.
Specifically, the contribution from the term $a_4\widehat{r}^4$ to $\widehat{\varepsilon}$, e.g.,  for $\widehat{r}\approx0.20$ (about one-fifth of the reduced radius $\widehat{R}$), is $\lesssim0.4\%$ (compared with 1) using $-2\lesssim a_4\lesssim2$ (here and for some other discussions we purposely use a rather large magnitude for $a_4$ compared to the best estimate of its value given in the appendix for making more conservative predictions). Nevertheless, we note that although $a_4$ has tiny impact on the $\widehat{\varepsilon}$, it may become relevant for forming the peaked profiles of $s^2$ in massive NSs. We shall thus study carefully its effects.
Similarly,  the $b_4$-term contributes $\lesssim0.5\%$ to $\widehat{P}/\widehat{P}_{\rm{c}}$ for $\widehat{r}\approx0.20$ and $\widehat{P}_{\rm{c}}\gtrsim0.1$.
See FIG.\,\ref{fig_s2_prep_r} for the radial dependence of $\widehat{\varepsilon}$ and $\widehat{P}/\widehat{P}_{\rm{c}}$ to $\widehat{r}\approx0.20$.
The $b_6$- and $a_6$-terms have even smaller contributions to $\widehat{P}/\widehat{P}_{\rm{c}}$ and $\widehat{\varepsilon}$, respectively.

\renewcommand*\figurename{\small FIG.}
\begin{figure}[h!]
\centering
\includegraphics[width=6.5cm]{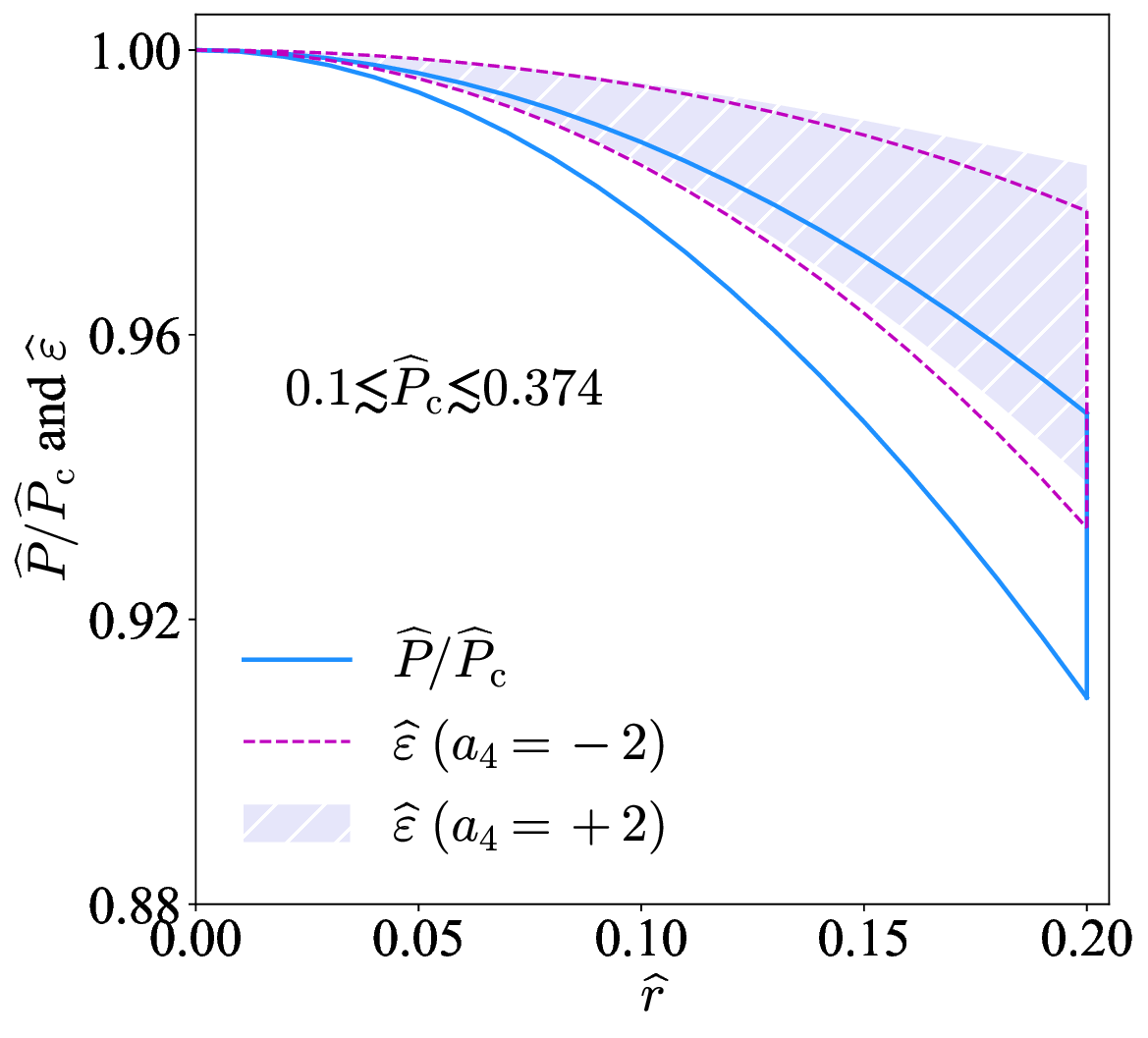}
\caption{(Color Online). Radial dependence of $\widehat{\varepsilon}$ and $\widehat{P}/\widehat{P}_{\rm{c}}$ to $\widehat{r}\approx0.20$, here $0.1\lesssim\widehat{P}_{\rm{c}}\lesssim0.374$ is adopted.
{The $\widehat{\varepsilon}$ with $a_4=\pm2$ are showed by the lavender and magenta bands, respectively.}
}\label{fig_s2_prep_r}
\end{figure}

In FIG.\,\ref{fig_s2_prep}, we show the core EOS $\widehat{P}$-$\widehat{\varepsilon}$ near $\widehat{r}=0$ by expanding $\widehat{P}$ and $\widehat{\varepsilon}$ over $\widehat{r}$ to order $\widehat{r}^4$ with $0.1\lesssim\widehat{P}_{\rm{c}}\lesssim0.374$ and $-2\lesssim a_4\lesssim2$,  an example with $\widehat{P}_{\rm{c}}\approx0.24$ and $a_4\approx-1$ is given with the dashed purple curve.
Notice that the plum band (with light-plum background) of FIG.\,\ref{fig_s2_prep} marked by ``TOV (this work)'' originates mainly from the band for $\widehat{P}_{\rm{c}}$, and the ``cone'' would be shrunk if $\widehat{P}_{\rm{c}}$ is further refined.
The ripples on the curve of $\widehat{P}$-$\widehat{\varepsilon}$ characterize the variation of $s^2$, as examplified by the light-blue instance (in which case there is an increasing of $s^2$). 
The dotted black line represents configurations with $s^2>1$ (which violate the causality principle), while the black dashed line (marked by $P=\widehat{P}_{\rm{c}}\varepsilon$) is the boundary given by $s^2/\widehat{P}_{\rm{c}}=1$,  which leads to the conformal limit ($\gamma_{\rm{c}}\equiv[\d\ln P/\d\ln\varepsilon]_{\rm{c}}= s_{\rm{c}}^2/\widehat{P}_{\rm{c}}=1$).
A few empirical EOSs (solid grey lines) mostly based on nuclear theory predictions under astrophysical and experimental nuclear physics constraints in the literature are also shown for comparisons. More specifically, they are the APR EOS\,\citep{Akmal1998}, the Dirac-Brueckner-Hartree-Fock EOS (MPA1\,\cite{MPA1} and ENG\,\cite{ENG}),  the Skryme\,\cite{Sky} energy density functional EOS SLy\,\cite{SLy}, the relativistic mean-field model\,\cite{Serot1986} with (GNH1\,\cite{Glen85}  and H4\,\cite{H4}) and without hyperons (NL3\,\cite{NL3} and FSU\,\cite{FSU}), the quark mean field model\,\cite{ALi20}, the crossover EOS of Ref.\,\cite{Kap21} and the hybrid ALF2 EOS\,\cite{ALF2} (the last two EOSs fall outside the tan band).
Here, the $(\varepsilon_{\rm{c}},P_{\rm{c}})$ for each EOS is taken as the one producing its $M_{\rm{NS}}^{\max}=M_{\rm{TOV}}$ configuration, and the curve is reduced with $(\varepsilon_{\rm{c}},P_{\rm{c}})$ using the correspondingly $\beta$-stable EOS.
A recent constraint on the NS EOS incorporating pQCD predictions\,\cite{Gorda23} is shown by the dash-dotted cyan band ($\varepsilon_{\rm{c}}\approx1000\,\rm{MeV}/\rm{fm}^3$ and $P_{\rm{c}}\approx250\mbox{-}400\,\rm{MeV}\rm{fm}^3$ are taken).
Because the pQCD theory predicts an approximate conformal symmetry at very high densities ($\gtrsim 40\rho_{\rm{sat}})$,  the upper boundary of its constraining band is close to the line $P=\widehat{P}_{\rm{c}}\varepsilon$ (conformal limit, black dashed).

\begin{figure}[h!]
\centering
\includegraphics[width=6.5cm]{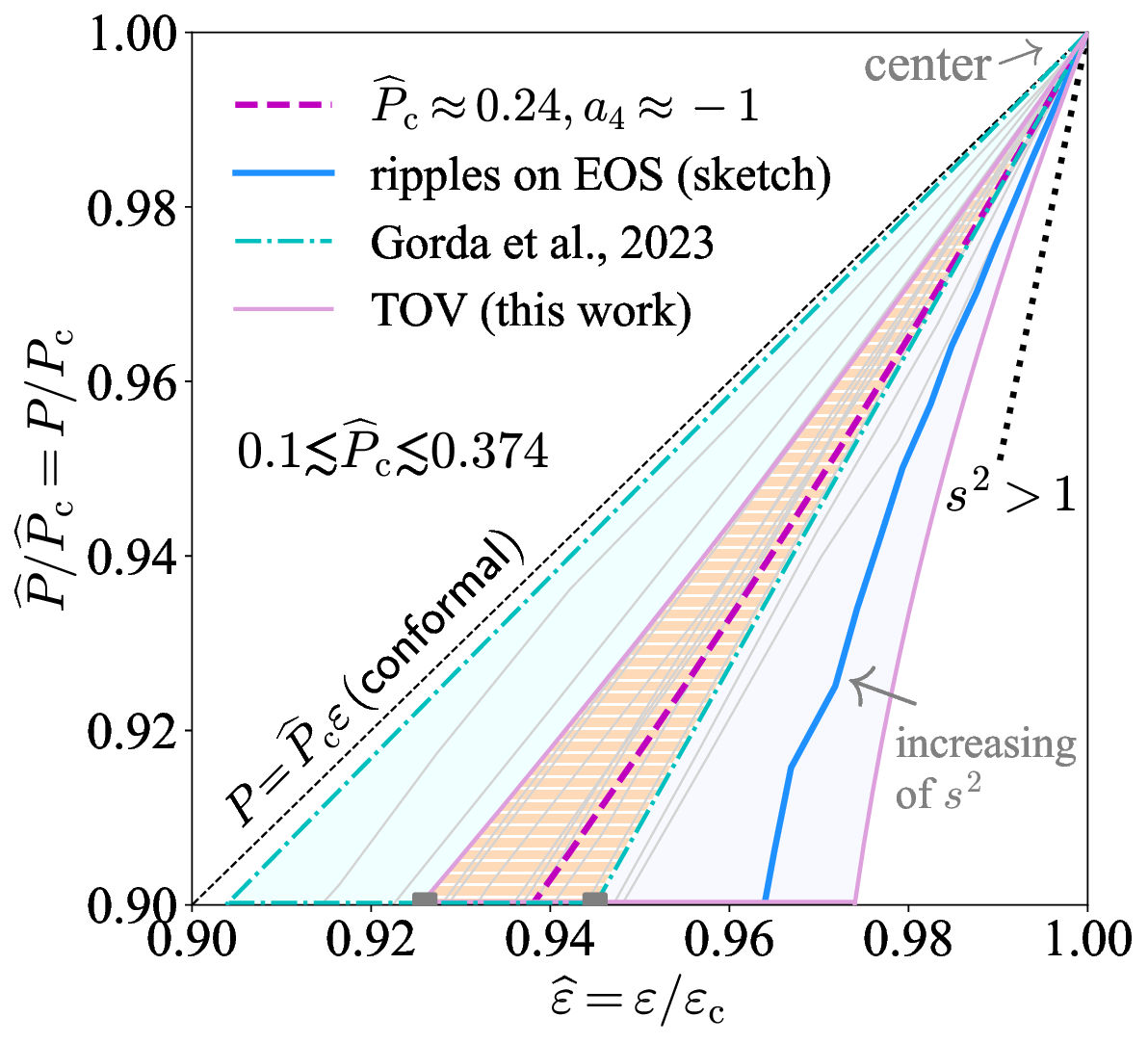}
\caption{(Color Online). Core EOS of NSs by expanding $\widehat{P}$ and $\widehat{\varepsilon}$ over $\widehat{r}$ to order $\widehat{r}^4$ (plum band) where $0.1\lesssim\widehat{P}_{\rm{c}}\lesssim0.374$ and $-2\lesssim a_4\lesssim2$ are adopted. An instance with $\widehat{P}_{\rm{c}}\approx0.24$ and $a_4\approx-1$  (dashed magenta) and a few empirical EOSs are shown (solid grey).
{A recent constraint on the dense matter EOS incorporating pQCD effects\,\cite{Gorda23} is also shown using the cyan dash-dotted band,
$P=\widehat{P}_{\rm{c}}\varepsilon$ marks the conformal boundary, the solid blue line shows an example of increasing of $s^2$ occurring there.
The overlapped region of Ref.\,\cite{Gorda23} and this work is indicated by the hatched tan band.
}
}\label{fig_s2_prep}
\end{figure}

{The core EOS shown by the plum band is directly inferred from the gravity (geometry of compact NSs) encapsulated in the TOV equations without any presumption about it, unlike in the traditional approach. Indeed, the EOSs from the two approaches match with each other as demonstrated by the hatched tan band in FIG.\,\ref{fig_s2_prep}.
Particularly, we have $0.925\lesssim\widehat{\varepsilon}\lesssim0.945$ (indicated by two grey solid rectangles) for $\widehat{P}/\widehat{P}_{\rm{c}}\approx0.9$ by combining these two EOSs.
Moreover,  we have $s_{\rm{c}}^2\neq0$ according to Eq.\,(\ref{def_sc2}), therefore
a sharp PT near the center or equivalently a plateau on the $P$-$\varepsilon$ curve occurring there is basically excluded.
However, a continuous crossover signaled by a smooth reduction of $s^2$ in NS cores (or equivalently a peaked behavior at some places near the center) and PTs occurring far from the centers are not excluded.
Analyzing circumstances under which the peaked structure in $s^2$ may emerge is the main task of the present work, and we will present more details on it in the following sections.
Actually, our prediction on the core EOS (plum band in FIG.\,\ref{fig_s2_prep}) is consistent with several empirical EOSs (grey lines) having a continuous crossover near the centers.
Moreover,  the index $\gamma_{\rm{c}}=s_{\rm{c}}^2/\widehat{P}_{\rm{c}}$ should not approach 1 due to the nonlinear feature of $s_{\rm{c}}^2$ shown in Eq.\,(\ref{def_sc2}) and in fact we have $\gamma_{\rm{c}}\geq4/3\approx1.33$. This implies the matter in NS cores can hardly be conformal, and this is very different from the predictions in Refs.\,\cite{Ann23,Gorda23}.
}

Perturbatively, we have the core EOS as,
\begin{align}\label{zf-1}
\widehat{P}/\widehat{P}_{\rm{c}}\approx&1+\frac{4}{3}\mu+\frac{16}{15}\mu^2
+\left(\frac{4}{3}-\frac{4}{5}\mu\right)\mu\widehat{P}_{\rm{c}}\notag\\
&+\left(2+\frac{28-256a_4}{3}\mu\right)\mu\widehat{P}_{\rm{c}}^2\notag\\
&+\left(4+\frac{6400a_4-262}{15}\mu\right)\mu\widehat{P}_{\rm{c}}^3+\mathcal{O}(\widehat{P}_{\rm{c}}^4,\mu^3),
\end{align}
where $\mu\equiv \widehat{\varepsilon}-1<0$, and the last three terms are the finite-$\widehat{P}_{\rm{c}}$ corrections (which disappear if $\widehat{P}_{\rm{c}}\to0$ is taken).
Clearly, the $a_6$-term has no effect on the core EOS at this order, and as more higher-order terms like $b_6\widehat{r}^6$ and $a_6\widehat{r}^6$ are included, the core EOS of the form like Eq.\,(\ref{zf-1}) could then be extrapolated to even larger radii $\widehat{r}$.

Since the plum cone of $P$-$\varepsilon$ in FIG.\,\ref{fig_s2_prep} is a straightforward consequence of the TOV equations without using any nuclear physics input, it is expected to hold universally. Owing to the smallness of $\mu$ (near the center) and the general smallness of $\widehat{P}_{\rm{c}}$, Eq.\,(\ref{zf-1}) provides a controllable expansion which predicts the core EOS near $\widehat{r}=0$.

\section{Speed of Sound Connecting Properties of Supradense Matter with Curved Geometry of Compact and Massive Objects}\label{SEC_SS}

Using the scaled TOV equations of (\ref{def-1}), the SSS $s^2$ can be written as
\begin{align}\label{st_s2}
\textsc{GR}:\;s^2
=\frac{\d\widehat{P}}{\d\widehat{\varepsilon}}
=-\frac{\widehat{\varepsilon}\widehat{M}}{\widehat{r}^2\d\widehat{\varepsilon}/\d\widehat{r}}\frac{(1+\widehat{P}/\widehat{\varepsilon})(1+\widehat{r}^3\widehat{P}/\widehat{M})}{1-2\widehat{M}/\widehat{r}}.
\end{align}
It clearly shows that not only dense matter {(characterized by $\widehat{P}$, $\widehat{\varepsilon}$ as well as their ratio $\widehat{P}/\widehat{\varepsilon}$)} but also the strong curved geometry {(characterized by the factor $1-2\widehat{M}/\widehat{r}$)} could affect $s^2$.
{In this sense, the speed of sound acts as a bridge connecting properties of supra-dense matter with curved geometry of compact/massive NSs.}
On the other hand, if one starts directly from the Newtonian evolution equation $\d\widehat{P}/\d\widehat{r}=-\widehat{\varepsilon}\widehat{M}/\widehat{r}^2$ without incorporating the SR and GR effects discussed earlier\,\cite{Chan10}, then,
\begin{equation}\label{s2-N}
\textsc{Newtonian}:\;s^2
=\frac{\d\widehat{P}}{\d\widehat{\varepsilon}}
=-\frac{\widehat{\varepsilon}\widehat{M}}{\widehat{r}^2\d\widehat{\varepsilon}/\d\widehat{r}}.
\end{equation}
Because there is no factor of $\widehat{P}$ on the right side, i.e., the pressure (matter effect) could not affect the $s^2$ explicitly (though $\widehat{\varepsilon}$ contains implicitly the effects of $\widehat{P}$).
Moreover, since $1+\widehat{P}/\widehat{\varepsilon}>1$, $1+\widehat{r}^3\widehat{P}/\widehat{M}>1$ and $0<1-2\widehat{M}/\widehat{r}<1$, we have $(1+\widehat{P}/\widehat{\varepsilon})(1+\widehat{r}^3\widehat{P}/\widehat{M})/(1-2\widehat{M}/\widehat{r})>1$, i.e., $s^2$ in NSs is enhanced compared with its Newtonian limit.
In addition, because $s^2\leq1$ due to the principle of causality, it could not always increase (even if matter/geometry effects are strong enough), implying under certain circumstances $s^2$ may decrease with decreasing $\widehat{r}$ (or increasing $\widehat{\varepsilon}$) when going into the core.

{After this general discussion on the evolutionary behavior of $s^2$, we now figure out under which circumstance a possible peak may emerge in $s^2$.}
By inserting the perturbative expansions of $\widehat{P}$, $\widehat{\varepsilon}$ and $\widehat{M}$ into Eq.\,(\ref{st_s2}), one can obtain an expression for $s^2$ similar to Eq.\,(\ref{zf-1}) for the core EOS.
In particular, we have to order $\widehat{r}^2$,
\begin{equation}
 s^2\approx s_{\rm{c}}^2+l_2\widehat{r}^2,
 \end{equation}
where
 \begin{equation}
 l_2=\frac{2s_{\rm{c}}^2}{b_2}\left(b_4-s_{\rm{c}}^2a_4\right).
\end{equation}
Because $s_{\rm{c}}^2>0$ (see Eq.\,(\ref{def_sc2})) and in order for $s^2$ to obtain a peak near $\widehat{r}=0$, it is necessary that $l_2>0$ and therefore $s^2>s_{\rm{c}}^2$.
{Since $b_2=-6^{-1}(1+3\widehat{P}_{\rm{c}}^2+4\widehat{P}_{\rm{c}})<0$,
the condition $l_2>0$ is equivalent to $b_4-s_{\rm{c}}^2a_4<0$,
or $a_4>b_4/s_{\rm{c}}^2$. By using the expression of $b_4$ given in Eq.\,(\ref{def-b4}) which involves the coefficient $b_2$, we then obtain an equivalent condition of $l_2>0$ as,
}
\begin{align}\label{def_a4ineq}
a_4>\frac{1}{12}\frac{1+3\widehat{P}_{\rm{c}}^2+4\widehat{P}_{\rm{c}}}{s_{\rm{c}}^2}\left(\widehat{P}_{\rm{c}}
+\frac{4+9\widehat{P}_{\rm{c}}}{15s_{\rm{c}}^2}\right).
\end{align}
Besides,  $a_4$ should fulfill some extra general constraints: 
\begin{enumerate}[leftmargin=*,label=(\alph*)]
\item Due to the decreasing feature of $\widehat{\varepsilon}\approx1+a_2\widehat{r}^2+a_4\widehat{r}^4$ with $\widehat{r}$ (i.e., $\d\widehat{\varepsilon}/\d\widehat{r}<0$ or $2a_2\widehat{r}+4a_4\widehat{r}^3<0$), it is necessary that $a_4<-a_2/2\widehat{R}^2$.
\item {On the other hand}, by considering the increase of $\widehat{M}$ with $\widehat{r}$ (i.e., $\d\widehat{M}/\d\widehat{r}=\widehat{r}^2\widehat{\varepsilon}>0$), one has the criterion $a_4\widehat{R}^4>-1-a_2\widehat{R}^2$.
\end{enumerate}
Here, $\widehat{R}\approx1$ is the reduced radius, and $a_2=b_2/s_{\rm{c}}^2$ is fixed at a certain $\widehat{P}_{\rm{c}}$.
{Meeting the above three criteria, namely inequality (\ref{def_a4ineq}), $a_4<-a_2/2\widehat{R}^2$ and $a_4\widehat{R}^4>-1-a_2\widehat{R}^2$ guarantees a peak in $s^2$ at some $\widehat{r}\neq0$ with certain ranges of $\widehat{P}_{\rm{c}}$.}
In this work, we primarily focus on the peaked behavior of $s^2(\widehat{r})$ which could be mapped onto, e.g., the peaked profile of $s^2(\widehat{\varepsilon})$ or $s^2(\rho)$.

\begin{figure}[h!]
\centering
\includegraphics[width=6.5cm]{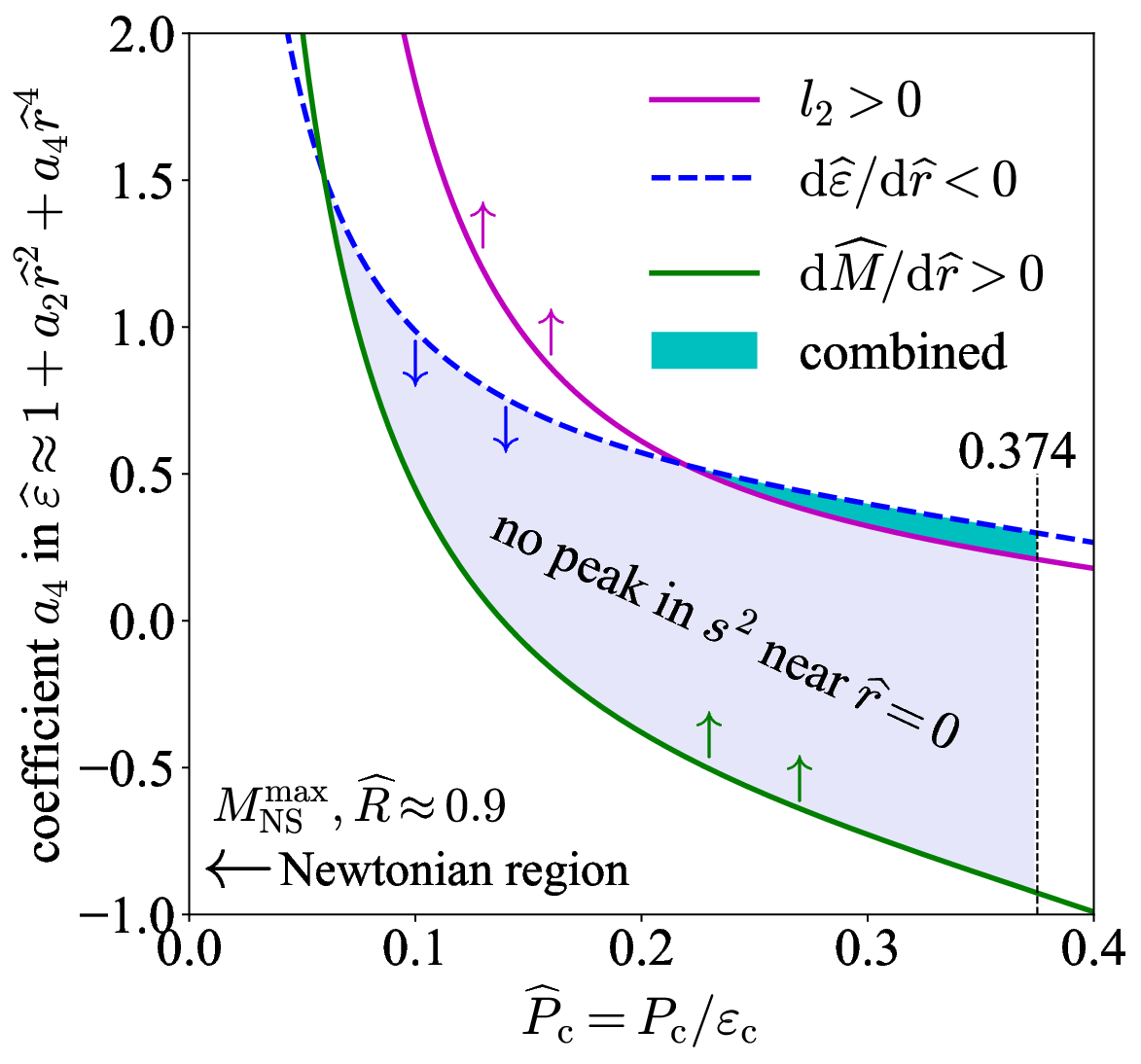}
\caption{(Color Online). Combined region for $a_4$ (cyan band) for the onset of a peaked profile of $s^2$ {under the inequality (\ref{def_a4ineq}) and the criteria (a) and (b)},  here $\widehat{R}\approx0.9$ is adopted for illustrations.
The vertical dashed line marks $\widehat{P}_{\rm{c}}\approx 0.374$ by setting $s_{\rm{c}}^2$ of Eq.\,(\ref{def_sc2}) $\leq1$, $l_2$ appears in $s^2\approx s_{\rm{c}}^2+l_2\widehat{r}^2+\cdots$.
}\label{fig_a4peak}
\end{figure}

The resulted regions for $a_4$ are shown in FIG.\,\ref{fig_a4peak}, here the colored arrows indicate whether the corresponding curve is a lower- or an upper-limit.
The criterion $\d\widehat{\varepsilon}/\d\widehat{r}<0$ {(criterion (a))} in the small-$\widehat{P}_{\rm{c}}$ limit becomes $a_4\lesssim\mathcal{O}(\widehat{P}_{\rm{c}}^{-1})$ while the criterion $l_2>0$ {(inequality (\ref{def_a4ineq}))} gives $a_4\gtrsim\mathcal{O}(\widehat{P}_{\rm{c}}^{-2})$, implying there would be no overlapped region for $a_4$ for such {small} $\widehat{P}_{\rm{c}}$.
In fact, criteria (a) and (b) above together lead to $a_4\lesssim\widehat{R}^{-4}\approx1$, which certainly becomes incompatible with the magenta bound for $\widehat{P}_{\rm{c}}\lesssim0.1$ as it requires $a_4\gtrsim2$ for $\widehat{P}_{\rm{c}}\lesssim0.1$.
{This discrepancy becomes even larger as $\widehat{P}_{\rm{c}}$ decreases even smaller.}
On the other hand, the combined region for $a_4$ for the onset of a peak in $s^2$ profile may eventually emerge (shown as the cyan band) with increasing $\widehat{P}_{\rm{c}}$.  
We would like point out that our analysis is consistent with many model calculations on the $s^2$ profile, in which the $s^2$ is a monotonic function (of energy density or radial distance), see, e.g., Ref.\,\cite{Pro23} for some examples.
The result shown in FIG.\,\ref{fig_a4peak} is also consistent with a recent study showing that the peaked $s^2$ is not necessary\,\cite{Mro23}, even after considering several recent observational as well as theoretical constraints.
In fact,  {although $a_2$ is definitely negative,} the coefficient $a_4$ could be either positive or negative,  depending on the dense matter EOS, e.g., see the appendix for an illustration on how the nuclear parameters could affect the coefficient $a_4$.

{As a short summary of this section, we find} that in order to have a peaked behavior in $s^2$, it is necessary that $a_4>0$. This is because a positive $b_4$ (of Eq.\,(\ref{def-b4})) slows down the decrease of $\widehat{P}$ (due to $b_2<0$) and for the $s^2$ to be larger than $s_{\rm{c}}^2$, a positive $a_4$ is necessary to slow down the decrease of $\widehat{\varepsilon}$ (due to $a_2<0$) too since approximately 
\begin{equation}\label{s2app}
s^2\approx{\Delta \widehat{P}}/{\Delta\widehat{\varepsilon}},
\end{equation} obtained by two nearby points on the EOS curve.
{In the next two sections, we carry out detailed calculations on the $s^2$ to verify the qualitative analysis given in this section.} 
In particular, we may find that $l_2<0$ holds (see Eq.\,(\ref{for_s2Newt}) and Eq.\,(\ref{for_s2Newt-1})) in the Newtonian limit even if $a_4$ is positive, {i.e., no peak would emerge in $s^2$ for Newtonian stars.} In this sense,  it is the GR effect that extrudes a peak in the $s^2$ profile (see Section \ref{SEC_GR} for details).

\section{Newtonian Stars: No Peak in $s^2$ near $\widehat{r}=0$}\label{SEC_NEWT}

In this section, we work out the detailed expression for the $s^2$ for Newtonian stars.
Going away from the center by expanding the right side of Eq.\,(\ref{s2-N}) we obtain the $s^2$ to order $\widehat{r}^4$ as
\begin{equation}
s^2\approx s_{\rm{c}}^2+l_2^{\rm{N}}\widehat{r}^2+l_4^{\rm{N}}\widehat{r}^4,
\end{equation}
where a superscript ``N'' is added to specify the Newtonian case. 
Explicitly, we have
\begin{align}\label{for_s2Newt}
s^2\approx &s_{\rm{c}}^2+\left(12a_4s_{\rm{c}}^4-\frac{4}{15}\right)\widehat{r}^2\notag\\
&+
\left(
144a_4^2s_{\rm{c}}^6+18a_6s_{\rm{c}}^4-\frac{62}{35}a_4s_{\rm{c}}^2+\frac{1}{60s_{\rm{c}}^2}\right)
\widehat{r}^4.
\end{align}
A main feature for the Newtonian $s^2$ radial profile is that it does not depend on $\widehat{P}_{\rm{c}}$ explicitly, although $s_{\rm{c}}^2$ may implicitly contain factors of $\widehat{P}_{\rm{c}}$.

For Newtonian stars, the reduced radius $\widehat{R}$ is generally small.
We estimate the order-of-magnitude of each terms in Eq.\,(\ref{for_s2Newt}).
{Taking WDs as a typical illustration and using $P\lesssim 10^{22\mbox{-}23}\,\rm{dynes}/\rm{cm}^2\approx10^{-(11\mbox{-}10)}\,\rm{MeV}/\rm{fm}^3$,  $\varepsilon\lesssim10^{8\mbox{-}9}\,\rm{kg}/\rm{m}^3\approx10^{-6}\,\rm{MeV}/\rm{fm}^3$ together with  $R\approx10^4\,\rm{km}$,  we obtain $\widehat{R}=R/Q\approx0.05$ and $\widehat{R}^2\approx3\times10^{-3}$ as well as $\phi=P/\varepsilon\approx10^{-4}\sim10^{-5}$. In fact, the ratio $\phi$ of pressure over energy density could be even smaller for main-sequence stars like the sun. 
Specifically, the pressure and energy density in solar core are about $10^{-16}\,\rm{MeV}/\rm{fm}^3$ and $10^{-10}\,\rm{MeV}/\rm{fm}^3$\,\cite{NASA}, respectively, and therefore $\phi\approx 10^{-6}$.
It is reasonable/conservative in this sense to use $\phi\lesssim\mathcal{O}(10^{-k})$ with $k\gtrsim4\mbox{-}5$ for Newtonian stars.}
Considering $s_{\rm{c}}^2\approx4\widehat{P}_{\rm{c}}/3\sim \widehat{P}_{\rm{c}} \sim\mathcal{O}(10^{-k})$ with $k\gtrsim4\mbox{-}5$, we then have $a_2=b_2/s_{\rm{c}}^2\approx-1/6s_{\rm{c}}^2\sim\mathcal{O}(10^{k-1})$ and $a_4\lesssim a_2/\widehat{R}^2\sim\mathcal{O}(10^{k+1})$ (via $a_2\widehat{R}^2\gtrsim a_4\widehat{R}^4$ from the perturbative expansion of $\widehat{\varepsilon}$).
Thus $12a_4s_{\rm{c}}^4$ scales as $\mathcal{O}(10^{2-k})$.
Similarly, we have the scalings $144a_4^2s_{\rm{c}}^6\sim18a_6s_{\rm{c}}^4\lesssim\mathcal{O}(10^{5-k}),(62/35)a_4s_{\rm{c}}^2\lesssim\mathcal{O}(10^1)$ as well as $1/60s_{\rm{c}}^2\lesssim\mathcal{O}(10^{k-2})$.
Therefore, the $-4/15$ and $1/60s_{\rm{c}}^2$ terms denominate over others in Eq.\,(\ref{for_s2Newt}).
Consequently, Eq.\,(\ref{for_s2Newt}) could be approximated as
\begin{equation}\label{for_s2Newt-1}
s^2\approx s_{\rm{c}}^2-\frac{4}{15}\widehat{r}^2+\frac{1}{60s_{\rm{c}}^2}\widehat{r}^4.
\end{equation}
Consequently, the $s^2$ takes its minimum as $s_{\min}^2=-{s_{\rm{c}}^2}/{15}$ at $\widehat{r}^2_{\min}=8s_{\rm{c}}^2$.
The vanishing of $s^2$ gives two special $\widehat{r}^2$ values, namely $6s_{\rm{c}}^2$ and $10s_{\rm{c}}^2$.
Since the stability condition requires $s^2\geq0$, we find $s^2$ is a decreasing function of $\widehat{r}\leq\widehat{R}$ with $\widehat{R}^2=6s_{\rm{c}}^2$, see FIG.\,\ref{fig_s2New_r4} for the sketch of $s^2$ in the Newtonian limit.
The decreasing feature of $s^2$ under Newtonian limit could also be seen from Eq.\,(\ref{zf-1}) since $s^2/\widehat{P}_{\rm{c}}\approx 4/3+32\mu/15$ for $\widehat{P}_{\rm{c}}\approx0$, which decreases with decreasing $\widehat{\varepsilon}=\mu+1$ (outward from the center).

\begin{figure}[h!]
\centering
\includegraphics[width=7.5cm]{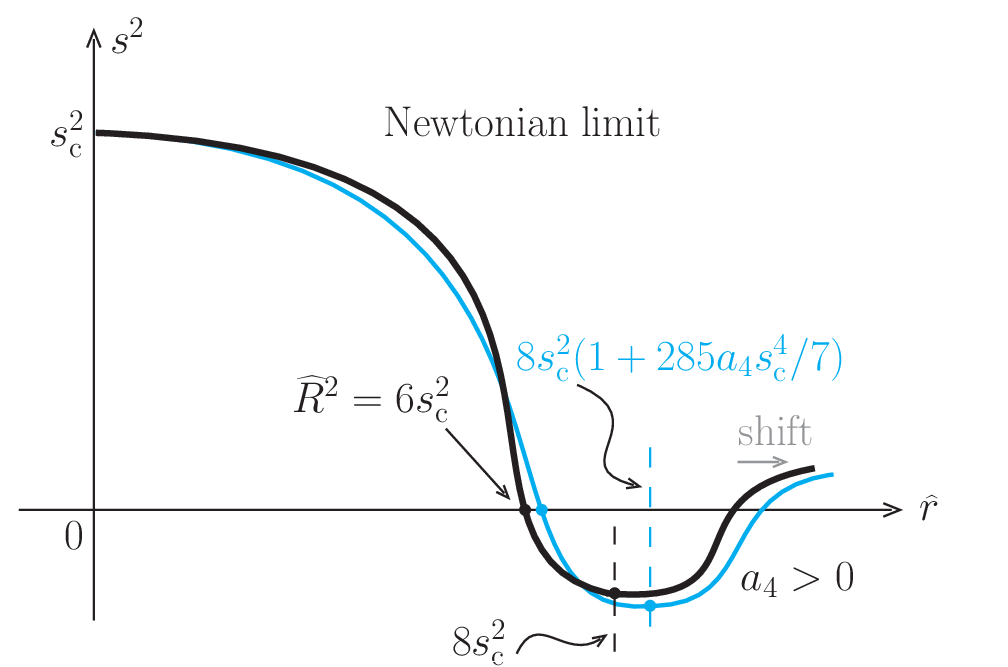}
\caption{(Color Online). Sketch of $s^2$ in the Newtonian limit when expanding $s^2$ to order $\widehat{r}^4$ (black), see Eq.\,(\ref{for_s2Newt-1}).
The correction from a positive $a_4$ is also shown (light-blue), see Eq.\,(\ref{for_s2_a6}).
}\label{fig_s2New_r4}
\end{figure}

The coefficient $l_4^{\rm{N}}\approx 1/60s_{\rm{c}}^2$ in Eq.\,(\ref{for_s2Newt-1}) (originated from $b_4$ of Eq.\,(\ref{def-b4})) is fundamental for explaining the radial dependence of $s^2$ in the Newtonian limit.
This is because even when the next high-order term $l_6^{\rm{N}}\widehat{r}^6$ is included,  the conclusion will not change qualitatively.
Specifically, we have under the Newtonian limit that $s^2\approx s_{\rm{c}}^2+l_2^{\rm{N}}\widehat{r}^2+l_4^{\rm{N}}\widehat{r}^4+l_6^{\rm{N}}\widehat{r}^6$:
\begin{align}\label{for_s2_a6}
s^2\approx& s_{\rm{c}}^2+\left(12a_4s_{\rm{c}}^4-\frac{4}{15}\right)\widehat{r}^2\notag\\
&+
\left(
144a_4^2s_{\rm{c}}^6+18a_6s_{\rm{c}}^4-\frac{62}{35}a_4s_{\rm{c}}^2+\frac{1}{60s_{\rm{c}}^2}\right)
\widehat{r}^4\notag\\
&+\left[
1728a_4^3s_{\rm{c}}^8+432a_4a_6s_{\rm{c}}^6+\left(24a_8-\frac{744}{35}a_4^2\right)s_{\rm{c}}^4\right.\notag\\
&\hspace*{2.cm}
\left.-\frac{52}{15}a_6s_{\rm{c}}^2+\frac{1}{35}a_4
\right]\widehat{r}^6\notag\\
\approx&s_{\rm{c}}^2-\frac{4}{15}\widehat{r}^2+\frac{1}{60s_{\rm{c}}^2}\widehat{r}^4
+\frac{1}{35}a_4\widehat{r}^6,
\end{align}
which becomes exact as $s_{\rm{c}}^2\to0$.
{Here the term $a_4\widehat{r}^6/35$ denominates at order $\widehat{r}^6$ using the same order-of-magnitude estimates given above.}
Notice that there are no terms inversely proportional to $s_{\rm{c}}^2$ appear in the coefficient $l_6^{\rm{N}}$ (which is different from $l_4^{\rm{N}}$ in this aspect).
Using the full form of $s^2$ in Eq.\,(\ref{for_s2_a6}), we obtain the location $\widehat{r}_{\min}^2$, the SSS at this location and the radius square $\widehat{R}^2$ as,
\begin{align}
\widehat{r}_{\min}^2\approx&8s_{\rm{c}}^2\left(1+{285}a_4s_{\rm{c}}^4/7+1416a_6s_{\rm{c}}^6\right),\\
s_{\min}^2\approx&-\frac{s_{\rm{c}}^2}{15}\left(1+{288}a_4s_{\rm{c}}^4/7+9344a_6s_{\rm{c}}^6\right),\\
\widehat{R}^2\approx& 6s_{\rm{c}}^2\left(1+{144}a_4s_{\rm{c}}^4/7+1620a_6s_{\rm{c}}^6\right).
\end{align}
The corrections in the brackets are perturbations (compared with the leading ``1''), e.g., $285a_4s_{\rm{c}}^4/7\lesssim\mathcal{O}(10^{3-k})\ll1$ and $1416a_6s_{\rm{c}}^6\lesssim\mathcal{O}(10^{7-2k})\ll1$ for $k\gtrsim4\mbox{-}5$, {based on the order-of-magnitude estimates given before Eq.\,(\ref{for_s2Newt-1}).}
The case of $a_4>0$ is shown in FIG.\,\ref{fig_s2New_r4} by the light-blue line, from which we find that the overall shape (such as the monotonicity) does not change.
Furthermore,  the contribution $l_8^{\rm{N}}\widehat{r}^8$ and beyond all contain no term inversely proportional to $s_{\rm{c}}^2$. Therefore, the expansion of $s^2$ converges and our analysis of its behavior at the Newtonian limit is stable.

Actually, the feature of $s^2$ for Newtonian stars near $\widehat{\varepsilon}\approx1$ could be extracted straightforwardly from their structure equations.
Starting directly from Eq.\,(\ref{s2-N}), we obtain,
\begin{align}\label{prf}
\left(\frac{\d s^2}{\d\widehat{\varepsilon}}\right)_{\rm{N}}
=&-\frac{3\widehat{\varepsilon}}{\widehat{r}^3}\left(\frac{\d\widehat{\varepsilon}}{\d\widehat{r}}\right)^2\left(\frac{\widehat{r}^3\widehat{\varepsilon}}{3}-\widehat{M}\right),~~\rm{with\;}\widehat{M}=\int\d\widehat{r}\widehat{r}^2\widehat{\varepsilon}\notag\\
&+
\frac{\widehat{\varepsilon}\widehat{M}}{\widehat{r}^2}
\left(\frac{\d\widehat{\varepsilon}}{\d\widehat{r}}\right)^{-3}
\left[\frac{\d^2\widehat{\varepsilon}}{\d\widehat{r}^2}
-\frac{1}{\widehat{r}}\frac{\d\widehat{\varepsilon}}{\d\widehat{r}}
\left(1+\frac{\widehat{r}}{\widehat{\varepsilon}}\frac{\d\widehat{\varepsilon}}{\d\widehat{r}}\right)
\right].
\end{align}
Since $
[{\d}/{\d\widehat{r}}]({\widehat{r}^3\widehat{\varepsilon}}/{3}-\widehat{M})=3^{-1}\widehat{r}^3{\d\widehat{\varepsilon}}/{\d\widehat{r}}<0$ (notice $\d\widehat{M}/\d\widehat{r}=\widehat{r}^2\widehat{\varepsilon}$ and $\d\widehat{\varepsilon}/\d\widehat{r}<0$),
i.e., $\widehat{r}^3\widehat{\varepsilon}/3-\widehat{M}$ decreases with increasing of $\widehat{r}$ (starting from 0), therefore we have $\widehat{r}^3\widehat{\varepsilon}/3-\widehat{M}<[\widehat{r}^3\widehat{\varepsilon}/3-\widehat{M}]_{\widehat{r}=0}=0$. Moreover,  the factor in square bracket of Eq.\,(\ref{prf}) is negative:
\begin{equation}
\frac{\d^2\widehat{\varepsilon}}{\d\widehat{r}^2}
-\frac{1}{\widehat{r}}\frac{\d\widehat{\varepsilon}}{\d\widehat{r}}
\left(1+\frac{\widehat{r}}{\widehat{\varepsilon}}\frac{\d\widehat{\varepsilon}}{\d\widehat{r}}\right)
\approx-4a_2^2\widehat{r}^2<0,
\end{equation}
one then has $(\d s^2/\d\widehat{\varepsilon})_{\rm{N}}>0$ definitely (near $\widehat{r}=0$), i.e.,  $s^2$ is an increasing function of $\widehat{\varepsilon}$ near $\widehat{\varepsilon}\approx1$.
{This is equivalent to saying that $l_2<0$ for Newtonian stars. In the next section, we shall find under certain conditions the coefficient $l_2$ can become positive for NSs.}

\section{Emerging Peaks in $s^2$ profiles near $\widehat{r}=0$ with Strong GR Effects in Massive NSs}\label{SEC_GR}

In this section, we analyze in details the emergence of a peak in $s^2$ profile near $\widehat{r}=0$ in the GR case.
Some quantitative features of the peaked $s^2$ profiles are also given and discussed.
Then we give an example of the radial profile of $s^2$ by fixing $M_{\rm{NS}}^{\max}=2M_{\odot}$ but allowing its radius to vary.
Finally, we discuss results for stable NSs along the M-R curve instead of the maximum-mass configuration $M_{\rm{NS}}^{\max}$.

\subsection{Geometry-matter Corrections/Couplings to $s^2$}\label{sub1}

Compared to the Newtonian case discussed above, the strong GR effects bring two modifications to $s^2$: (a) $\widehat{P}_{\rm{c}}$ can be sizable $\gtrsim\mathcal{O}(0.1)$ in NSs; and (b) the expression for $s^2$ is changed via the GR structure equations (of Eq.\,(\ref{st_s2})).
Parallel to Eq.\,(\ref{prf}), we obtain $\d s^2/\d\widehat{\varepsilon}$ with GR as,
\begin{align}\label{PM}
\frac{\d s^2}{\d\widehat{\varepsilon}}
=&\frac{Y}{1-2\widehat{M}/\widehat{r}}\left\{
\left(\frac{\d s^2}{\d\widehat{\varepsilon}}\right)_{\rm{N}}
-\frac{\widehat{\varepsilon}\widehat{M}}{1-2\widehat{M}/\widehat{r}}\frac{2}{\widehat{r}^4}
\left(\frac{\d\widehat{r}}{\d\widehat{\varepsilon}}\right)^2\left(\widehat{r}^3\widehat{\varepsilon}-\widehat{M}\right)\right.\notag\\
&\hspace*{0.5cm}-\frac{\widehat{\varepsilon}}{\widehat{M}}\frac{\d\widehat{r}}{\d\widehat{\varepsilon}}
\left(1+\frac{\widehat{r}^3\widehat{P}}{\widehat{M}}\right)^{-1}
\left[
\widehat{r}\widehat{M}s^2+\widehat{P}\frac{\d\widehat{r}}{\d\widehat{\varepsilon}}\left(3\widehat{M}-\widehat{r}^3\widehat{\varepsilon}\right)
\right]
\notag\\
&\hspace*{0.5cm}\left.-\frac{\widehat{M}}{\widehat{r}^2}\frac{\d\widehat{r}}{\d\widehat{\varepsilon}}
\left(1+\frac{\widehat{P}}{\widehat{\varepsilon}}\right)^{-1}\left(s^2-\frac{\widehat{P}}{\widehat{\varepsilon}}\right)
\right\},
\end{align}
where $Y=(1+\widehat{P}/\widehat{\varepsilon})(1+\widehat{r}^3\widehat{P}/\widehat{M})$ and $(\d s^2/\d\widehat{\varepsilon})_{\rm{N}}>0$ is given by Eq.\,(\ref{prf}). 
In Eq.\,(\ref{PM}),  besides $3\widehat{M}-\widehat{r}^3\widehat{\varepsilon}>0$ and $s^2-\widehat{P}/\widehat{\varepsilon}=s^2-P/\varepsilon=s^2-\phi>0$ is the derivative part of $s^2$\,\cite{Fuji22}, we also have (for small $\widehat{r}$),
\begin{equation}
\widehat{r}^3\widehat{\varepsilon}-\widehat{M}
\approx\frac{2\widehat{r}^3}{3}\left(1+\frac{6}{5}a_2\widehat{r}^2\right)>0.
\end{equation}
This means they contribute to $\d s^2/\d\widehat{\varepsilon}$ with definite signs.

The two terms  in Eq.\,(\ref{PM}) contributing negatively to the derivative $\d s^2/\d\widehat{\varepsilon}$ can be combined as,
\begin{align}\label{g-1}
-\frac{Y}{1-2\widehat{M}/\widehat{r}}\left(\frac{\d\widehat{r}}{\d\widehat{\varepsilon}}\right)^2&\left[\frac{\widehat{\varepsilon}\widehat{M}}{1-2\widehat{M}/\widehat{r}}\frac{2}{\widehat{r}^4}
\left(\widehat{r}^3\widehat{\varepsilon}-\widehat{M}\right)\right.\notag\\
&\hspace*{-0.5cm}\left.+\frac{\widehat{P}\widehat{\varepsilon}}{\widehat{M}}\left(1+\frac{\widehat{r}^3\widehat{P}}{\widehat{M}}\right)^{-1}\left(3\widehat{M}-\widehat{r}^3\widehat{\varepsilon}\right)
\right]<0.
\end{align}
The first term in (\ref{g-1}) originated from ``$-2\widehat{M}/\widehat{r}$'' in Eq.\,(\ref{st_s2}) survives even $\widehat{P}=0$ is considered (as it is a geometric correction),  while the second term is the geometry-matter coupling (from ``$\widehat{r}^3\widehat{P}/\widehat{M}$'' in Eq.\,(\ref{st_s2})), which disappears if $\widehat{P}\approx0$.
This demonstrates clearly that GR geometrical effects, the special relativity corrections for dense matter and/or the matter-geometry couplings can all effectively modify the $\widehat{\varepsilon}$-dependence (or equivalently the $\widehat{r}$-dependence) of $s^2$ compared with its Newtonian counterpart.
Specifically, they all tend to make the $s^2$ decrease with increasing $\widehat{\varepsilon}$ (i.e. $\d s^2/\d\widehat{\varepsilon}<0$).

On the other hand,  $s^2>0$ itself contributes positively to the derivative $\d s^2/\d\widehat{\varepsilon}$ in Eq. (\ref{PM}),  namely
{\begin{align}\label{g-2}
-\frac{Y}{1-2\widehat{M}/\widehat{r}}&\left[\frac{\widehat{\varepsilon}}{\widehat{M}}\frac{\d\widehat{r}}{\d\widehat{\varepsilon}}
\left(1+\frac{\widehat{r}^3\widehat{P}}{\widehat{M}}\right)^{-1}
\widehat{r}\widehat{M}s^2
+\frac{\widehat{M}}{\widehat{r}^2}\frac{\d\widehat{r}}{\d\widehat{\varepsilon}}
\left(1+\frac{\widehat{P}}{\widehat{\varepsilon}}\right)^{-1}s^2\right]\notag\\
=-&\frac{\widehat{\varepsilon}\widehat{r} s^2}{1-2\widehat{M}/\widehat{r}}\left(\frac{\d\widehat{r}}{\d\widehat{\varepsilon}}\right)\left(1+\frac{2\widehat{P}}{\widehat{\varepsilon}}+\frac{\widehat{M}}{\widehat{\varepsilon}\widehat{r}^3}\right)>0.
\end{align}}This means $s^2$ itself tends to remove the peak, i.e., it tends to make $\d s^2/\d\widehat{\varepsilon}>0$ near $\widehat{r}=0$.
{Only the first term in the curry bracket of Eq.\,(\ref{PM}), namely $(\d s^2/\d\widehat{\varepsilon})_{\rm{N}}$ survives in the Newtonian limit, and all the other terms (proportional to $(\d\widehat{r}/\d\widehat{\varepsilon})^2$ and $\d\widehat{r}/\d\widehat{\varepsilon}$) disappear.
Therefore the correction (\ref{g-2}) also disappears in $\d s^2/\d\widehat{\varepsilon}$ for Newtonian stars.}
The final sign of $\d s^2/\d\widehat{\varepsilon}$ is the result of a balance/competition between Eqs.\,(\ref{g-1}) and (\ref{g-2}). It may depend on the EOS model, and therefore not every dense matter EOS could induce a peaked $s^2$ profile.

\subsection{Expressions for $s^2$ Profiles Including GR Effects}

A formula for $s^2\approx s_{\rm{c}}^2+l_2\widehat{r}^2+l_4\widehat{r}^4$ similar to the Eq.\,(\ref{for_s2Newt}) for Newtonian stars could be obtained for the GR case too:
\begin{align}
s^2\approx&s_{\rm{c}}^2+\frac{2s_{\rm{c}}^2}{b_2}\left(b_4-a_4s_{\rm{c}}^2\right)\widehat{r}^2
\notag\\
&+\frac{s_{\rm{c}}^2}{b_2^2}
\left[
4a_4s_{\rm{c}}^2\left(a_4s_{\rm{c}}^2-b_4\right)-3b_2\left(a_6s_{\rm{c}}^2-b_6\right)
\right]\widehat{r}^4,\label{k-0}
\end{align}
and this is a  general expression for $s^2$ to order $\widehat{r}^4$.
{By putting the expressions for $b_2=-6^{-1}(1+3\widehat{P}_{\rm{c}}^2+4\widehat{P}_{\rm{c}})$, $a_2=b_2/s_{\rm{c}}^2$,  $b_4$ (of Eq.\,(\ref{def-b4})) as well as $b_6$\,\cite{CLZ23-a},
\begin{align}\label{def_b6}
b_6=&-\frac{1}{216}\left(1+9\widehat{P}_{\rm{c}}^2\right)\left(1+3\widehat{P}_{\rm{c}}^2+4\widehat{P}_{\rm{c}}\right)-\frac{a_2^2}{30}\notag\\
&+\left(\frac{2}{15}\widehat{P}_{\rm{c}}^2+\frac{1}{45}\widehat{P}_{\rm{c}}-\frac{1}{54}\right)a_2-\frac{5+12\widehat{P}_{\rm{c}}}{63}a_4,
\end{align}
 into the coefficients $l_2$ and $l_4$, and expanding terms to linear order of $\widehat{P}_{\rm{c}}$ (while the $s_{\rm{c}}^2$ is still kept without being expanded over $\widehat{P}_{\rm{c}}$),  we obtain,}
\begin{align}
s^2\approx&s_{\rm{c}}^2+\left[\left(12a_4s_{\rm{c}}^4-\frac{4}{15}\right)
-\left(48a_4s_{\rm{c}}^4+s_{\rm{c}}^2+\frac{3}{5}\right)\widehat{P}_{\rm{c}}
\right]\widehat{r}^2\notag\\
&+
\left[\left(
144a_4^2s_{\rm{c}}^6+18a_6s_{\rm{c}}^4-\frac{62}{35}a_4s_{\rm{c}}^2+\frac{1}{60s_{\rm{c}}^2}
+\frac{1}{12}s_{\rm{c}}^2-\frac{1}{18}
\right)\right.\notag\\
&
+\left(
\frac{1}{15s_{\rm{c}}^2}+\frac{1}{15}
-12a_4s_{\rm{c}}^4-72a_6s_{\rm{c}}^4\right.\notag\\
&\hspace*{1.cm}\left.\left.
-1152a_4^2s_{\rm{c}}^6+\frac{116}{35}a_4s_{\rm{c}}^2
\right)\widehat{P}_{\rm{c}}
\right]
\widehat{r}^4.\label{k-1}
\end{align}
Expanding $s_{\rm{c}}^2$ of Eq.\,(\ref{def_sc2}) over $\widehat{P}_{\rm{c}}$ as $s_{\rm{c}}^2\approx 4\widehat{P}_{\rm{c}}/3+4\widehat{P}_{\rm{c}}^2/3+\cdots$ further approximates (\ref{k-1}) as,
\begin{align}
s^2\approx& \frac{4}{3}\widehat{P}_{\rm{c}}+\frac{4}{3}\widehat{P}_{\rm{c}}^2
+\left[-\frac{4}{15}-\frac{3}{5}\widehat{P}_{\rm{c}}
+\left(\frac{64a_4}{3}-\frac{4}{3}\right)\widehat{P}_{\rm{c}}^2\right]\widehat{r}^2\notag\\
&+\left[
\frac{1}{80\widehat{P}_{\rm{c}}}-\frac{13}{720}+
\left(\frac{229}{1440}-\frac{248a_4}{105}\right)\widehat{P}_{\rm{c}}\right.\notag\\
&\hspace*{1.cm}\left.
+\left(\frac{157}{360}+\frac{72a_4}{35}+32a_6\right)\widehat{P}_{\rm{c}}^2
\right]\widehat{r}^4.
\label{for_1}
\end{align}

\begin{figure}[h!]
\centering
\includegraphics[width=6.5cm]{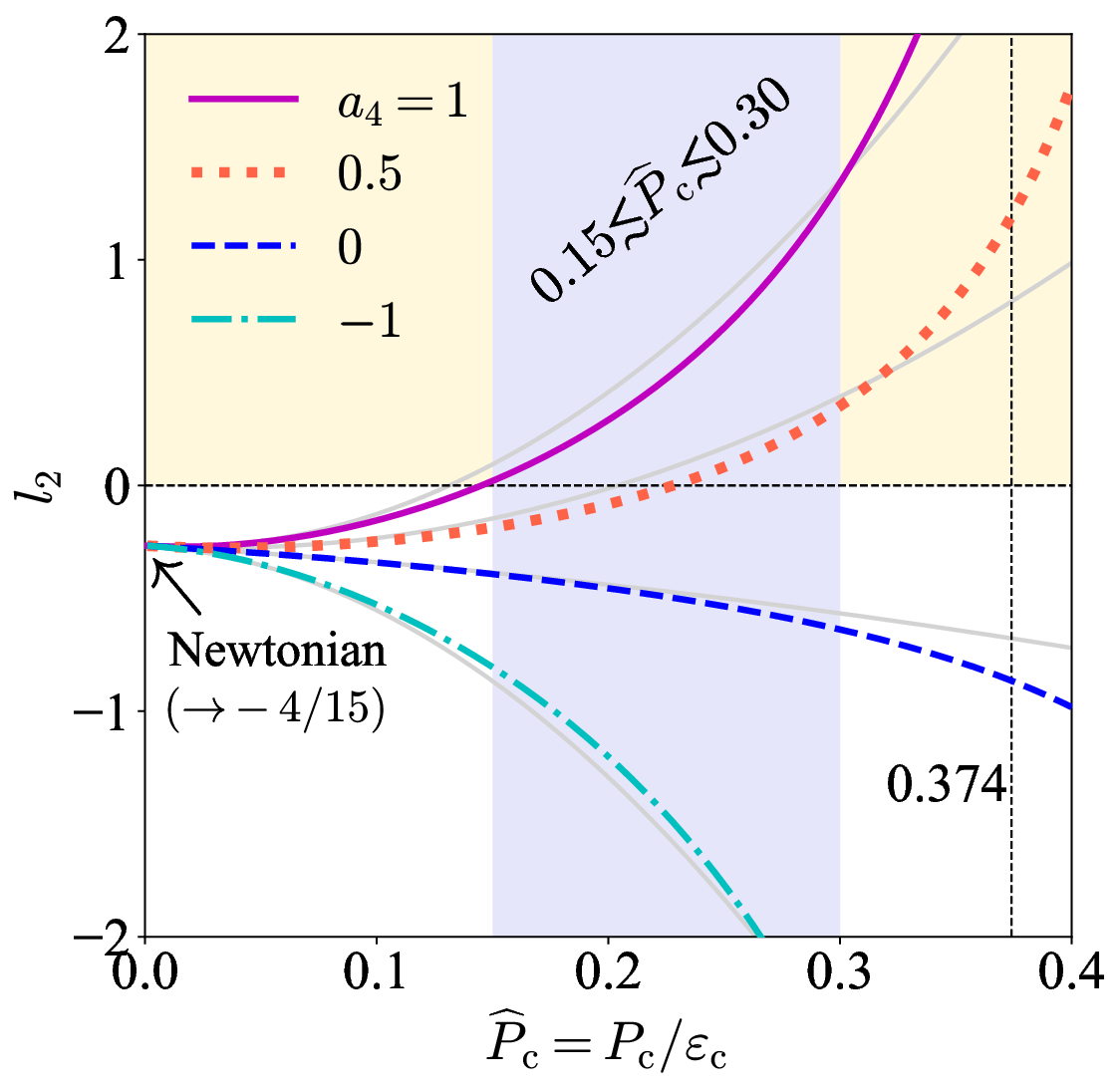}
\caption{(Color Online). The coefficient $l_2$ as a function of $\widehat{P}_{\rm{c}}$ adopting different $a_4$ values, the grey line near each colored curve is for the approximated $l_2\approx-4/15-3\widehat{P}_{\rm{c}}/5+(64a_4/3-4/3)\widehat{P}_{\rm{c}}^2$ to order $\widehat{P}_{\rm{c}}^2$.
The light-yellow background shows the necessary condition for a peaked SSS profile, i.e., $l_2>0$.
}\label{fig_s2_GR}
\end{figure}

We find that Eq.\,(\ref{k-1}) with GR is very different from Eq.\,(\ref{for_s2Newt}) for Newtonian stars even if $\widehat{P}_{\rm{c}}\approx0$ is taken (this is due to the geometrical correction ``$-2\widehat{M}/\widehat{r}$''), e.g., the terms ``$s_{\rm{c}}^2/12$'' and ``$-1/18$'' at order $\widehat{r}^4\widehat{P}_{\rm{c}}^0=\widehat{r}^4$ are new compared with Eq.\,(\ref{for_s2Newt}).
Similarly, new terms may emerge at other orders of $\widehat{P}_{\rm{c}}$.
Considering Eq.\,(\ref{for_1}), e.g., we find $a_4$ starts to operate at order $\widehat{P}_{\rm{c}}^2$ {(in the form of $64a_4\widehat{P}_{\rm{c}}^2/3$).}
Then for finite $\widehat{P}_{\rm{c}}\sim\mathcal{O}(0.1)$, the coefficient $l_2$ may take positive values.
For example, taking $a_4\approx1$ ($a_4\approx0.5$) and requiring $l_2>0$ leads to $\widehat{P}_{\rm{c}}\gtrsim0.15$ ($\widehat{P}_{\rm{c}}\gtrsim0.23$), which are typical values of $\widehat{P}_{\rm{c}}$ in massive NSs.
Thus, a sizable $\widehat{P}_{\rm{c}}$ is necessary for inducing a peaked $s^2$ profile, i.e.,
\begin{equation}
l_2>0\rm{\;needs\;}a_4>0\rm{\;as\;well\;as\;}\widehat{P}_{\rm{c}}\gtrsim\mathcal{O}(0.1).
\end{equation}
However,  $a_4>0$ in the Newtonian limit could still not generally make $l_2>0$ (see Section \ref{SEC_NEWT} and especially Eq.\,(\ref{for_s2_a6})),  we thus see clearly that $\widehat{P}_{\rm{c}}\gtrsim\mathcal{O}(0.1)$ is fundamental. FIG.\,\ref{fig_s2_GR} gives an illustration on the $\widehat{P}_{\rm{c}}$-dependence of $l_2$ (adopting four different $a_4$'s), where $0.15\lesssim\widehat{P}_{\rm{c}}\lesssim0.30$ is marked as the range for $\widehat{P}_{\rm{c}}$ in typical massive NSs.
The grey line near each colored curve is for the approximated $l_2\approx-4/15-3\widehat{P}_{\rm{c}}/5+(64a_4/3-4/3)\widehat{P}_{\rm{c}}^2$. Obviously, $l_2^{\rm{N}}=-4/15$ (see Eq.\,(\ref{for_s2Newt-1})).

\begin{figure}[h!]
\centering
\includegraphics[width=8.6cm]{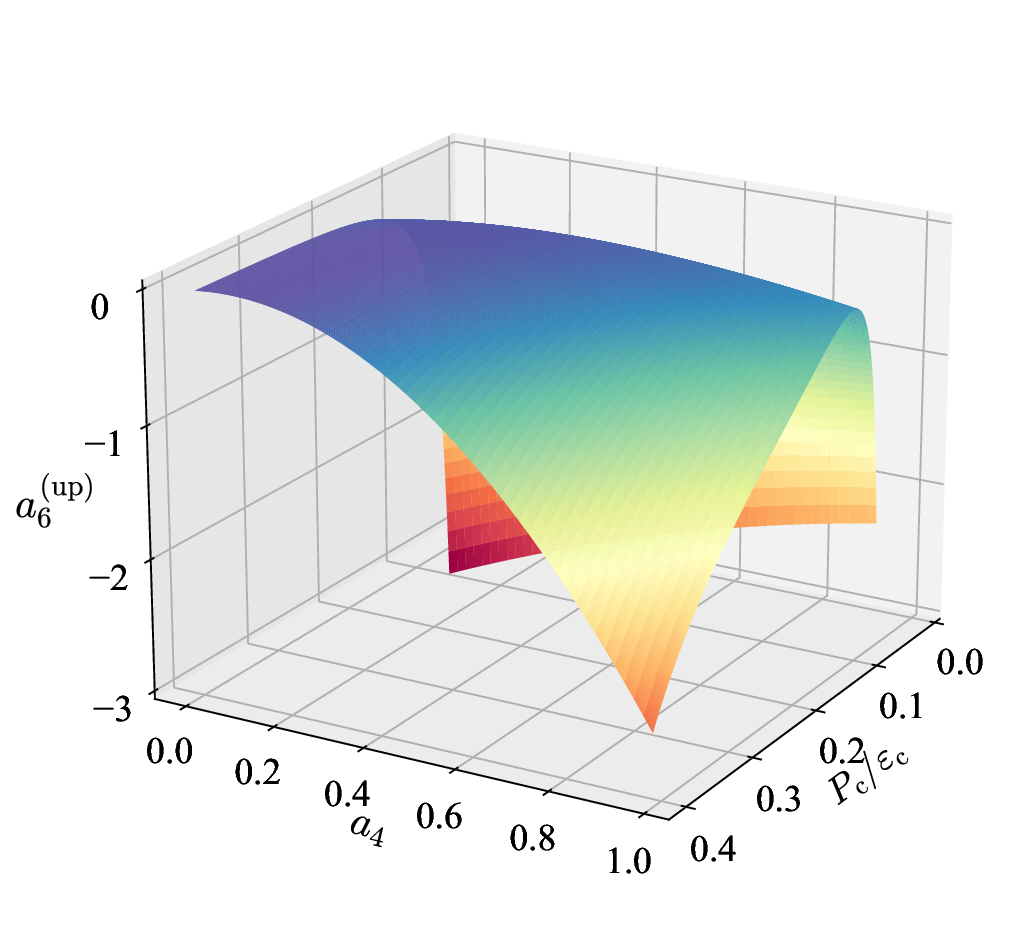}
\caption{(Color Online). The surface of the upper limit $a_6^{\rm{(up)}}$ as a function of $a_4$ and $\widehat{P}_{\rm{c}}$ in order to make the coefficient $l_4$ negative.
}\label{fig_s2_GR_a6surf}
\end{figure}

Once the coefficient $l_2$ becomes positive, there would unavoidably be a peak in $s^2$ profile at some finite distance $\widehat{r}<\widehat{R}$ regardless of the higher-order coefficients (since near the surface the $s^2\to0$).
One can demonstrate that the coefficient $a_6$ needs to be negative if $l_4<0$ is required.
The resulted condition for $a_6$ could be obtained from Eq.\,(\ref{k-0}), namely $l_4<0$,
or equivalently,
{\begin{equation}
a_6<\frac{b_6}{s_{\rm{c}}^2}+\frac{4}{3}\frac{a_4}{b_2}\left(s_{\rm{c}}^2a_4-b_4\right)\equiv a_6^{\rm{(up)}},
\end{equation}}where all the expressions (for $b_2,b_4$ and $b_6$ as well as $s_{\rm{c}}^2$) are available.
The $a_4$-dependence of $a_6^{\rm{(up)}}$ is shown in FIG.\,\ref{fig_s2_GR_a6surf}, from which
we find that $a_6<0$ and notice that both $a_4$ and $a_6$ are $\sim\mathcal{O}(1)$, e.g.,
for $\widehat{P}_{\rm{c}}\approx0.2$ and $a_4\approx1$, we then have $a_6\lesssim-1$.
If the coefficient $l_4$ is also positive, then an analysis of the even higher-order terms becomes necessary, e.g., the coefficient $l_6$, etc.
Nonetheless, the peak will emerge at some finite $\widehat{r}$.
In this work (see the next subsection), we mainly focus on the situation in which $l_2>0$ and $l_4<0$, and point out the extension if the $l_6$-term is included when necessary.

When the $b_6$- and $a_6$-terms are included, the core EOS of Eq.\,(\ref{zf-1}) should be improved.
The leading-order contribution from $a_6$ appears at the order $\mu^3\widehat{P}_{\rm{c}}^3$,
\begin{align}\label{zf-3}
\widehat{P}/\widehat{P}_{\rm{c}}\approx&1+\frac{4}{3}\mu+\frac{16}{15}\mu^2+\frac{4}{15}\mu^3
+\left(\frac{4}{3}-\frac{4}{5}\mu-\frac{268}{135}\mu^2\right)\mu\widehat{P}_{\rm{c}}\notag\\
&+\left[2+\left(\frac{28}{3}-\frac{256a_4}{3}\right)\mu
+\left(\frac{370}{27}-\frac{30208}{315}a_4\right)\mu^2
\right]\mu\widehat{P}_{\rm{c}}^2\notag\\
&+\left[4+\left(\frac{1280a_4}{3}-\frac{262}{15}\right)\mu
+\left(\frac{68608}{105}a_4+\frac{2048}{3}a_6\right.\right.\notag\\
&\hspace*{1.cm}\left.\left.-\frac{1496}{27}\right)\mu^2
\right]\mu\widehat{P}_{\rm{c}}^3+\mathcal{O}(\widehat{P}_{\rm{c}}^4,\mu^4),
\end{align}
where a new term $-30208a_4\mu^3\widehat{P}_{\rm{c}}^2/315$ proportional to $a_4$ at order $\mu^3\widehat{P}_{\rm{c}}^2$ also emerges.
We have $\mu^3\widehat{P}_{\rm{c}}^3\sim-10^{-5}$ for $\widehat{P}_{\rm{c}}\sim0.2$ and $\mu\sim-0.1$, implying $a_6$ has small effect on the core EOS near $\widehat{r}=0$, namely $|2048a_6\mu^3\widehat{P}_{\rm{c}}^3/3|\lesssim1\%$ by considering $a_6\sim\mathcal{O}(1)$.
Based on Eq.\,(\ref{zf-3}), we shall obtain the SSS for finite $\mu$ straightforwardly as $s^2/\widehat{P}_{\rm{c}}=\d(\widehat{P}/\widehat{P}_{\rm{c}})/\d\mu$, here the $\mu^3$-terms are new compared with Eq.\,(\ref{zf-1}).
Then for Newtonian stars (neglecting the finite-$\widehat{P}_{\rm{c}}$ corrections), we have $\widehat{P}_{\rm{c}}^{-1}\d s^2/\d\mu=\d^2(\widehat{P}/\widehat{P}_{\rm{c}})/\d\mu^2=32/15+8\mu/5>0$ for $-1\leq\mu\leq0$, i.e., the $s^2$ is a monotonically increasing function of $\widehat{\varepsilon}$, as expected.

\begin{figure}[h!]
\centering
\includegraphics[width=6.6cm]{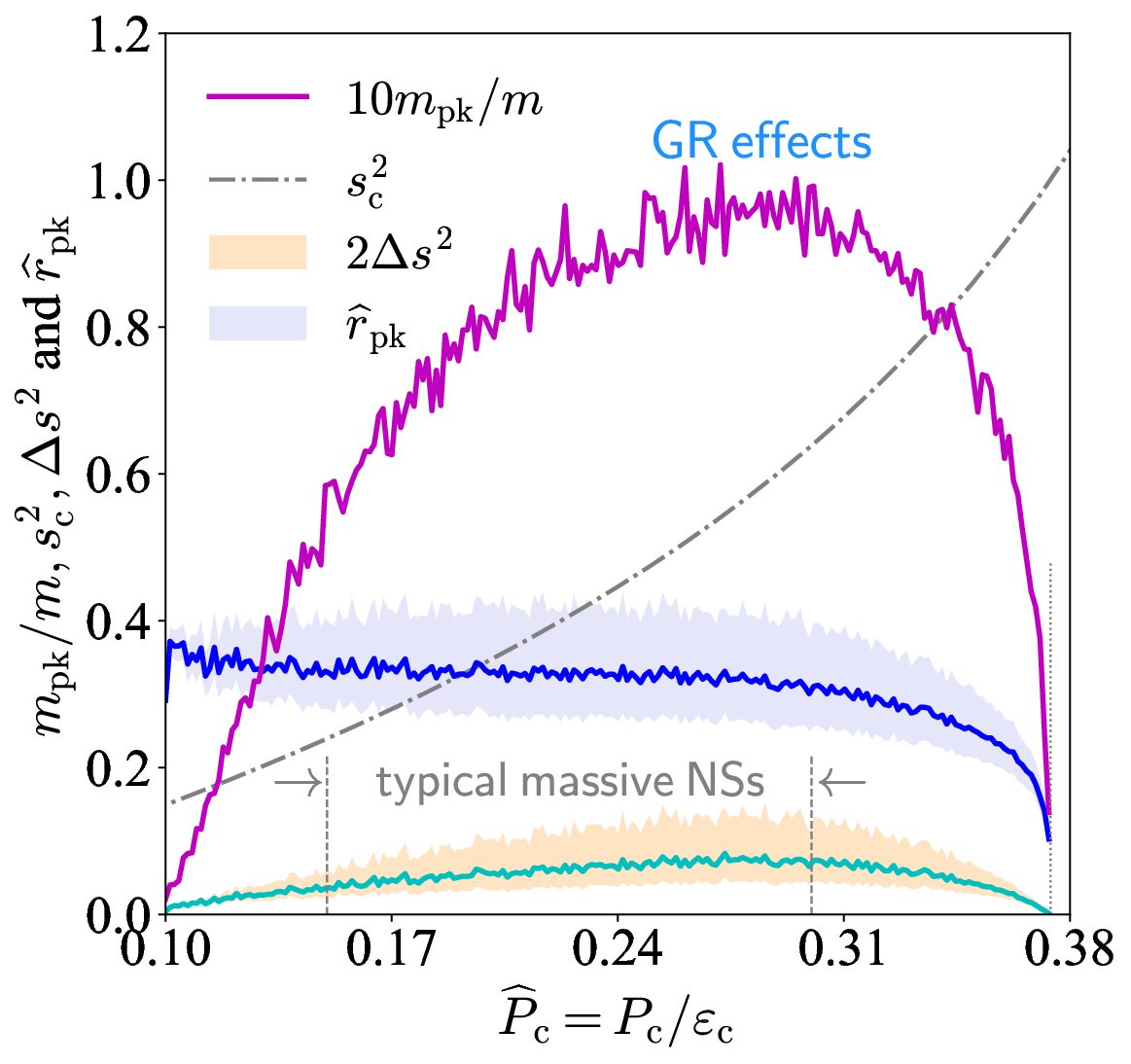}
\caption{(Color Online).  {Probability of the occurrence of a peak in $s^2$ profile (magenta line),  the position of the peak $\widehat{r}_{\rm{pk}}=\sqrt{-l_2/2l_4}$ of $s^2$ (blue line), and (two times) the enhancement $\Delta s^2\equiv s_{\max}^2/s_{\rm{c}}^2-1=-l_2^2/4l_4s_{\rm{c}}^2$ on $s_{\rm{c}}^2$ (green line).
The background bands on $\widehat{r}_{\rm{pk}}$ (lavender) and $2\Delta s^2$ (orange) represent their 1$\sigma$ uncertainties.  The $\widehat{P}_{\rm{c}}$-dependence of $s_{\rm{c}}^2$ is also shown (grey dash-dotted line).}
}\label{fig_s2_peak_stat_prop}
\end{figure}

\begin{figure}[h!]
\centering
\includegraphics[width=6.4cm]{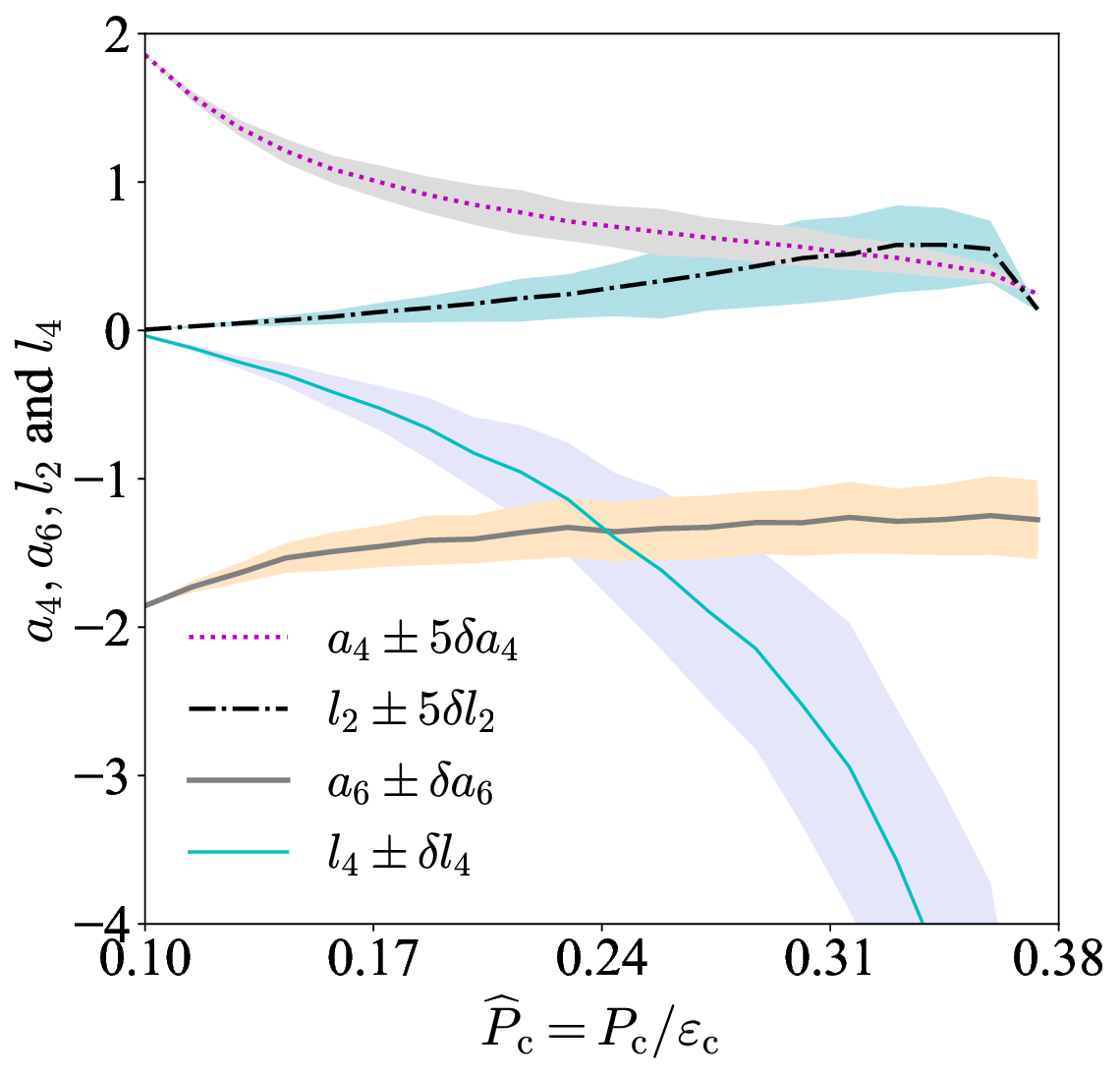}
\caption{(Color Online). Coefficients $a_4,a_6,l_2$ and $l_4$ as functions of $\widehat{P}_{\rm{c}}$.
The uncertainties on $a_4$ and $l_2$ are amplified by 5 times for a more clear visualization due to their relative smallness.
}\label{fig_s2_peak_stat_a4a6}
\end{figure}

{It is now useful to point out that one could directly work out the dependence of $s^2$ on $\phi=P/\varepsilon$.
Starting from $\widehat{P}/\widehat{P}_{\rm{c}}\approx1+4\mu/3+16\mu^2/15+(4/3-4\mu/5)\mu\widehat{P}_{\rm{c}}$ to order $\mu^2\widehat{P}_{\rm{c}}$ (see Eq.\,(\ref{zf-3})), for example, we obtain 
\begin{align}
\phi\approx&\frac{\widehat{P}_{\rm{c}}}{1+\mu}\left[1+\frac{4\mu}{3}+\frac{16\mu^2}{15}+\left(
\frac{4}{3}
-\frac{4\mu}{5}\right)\mu\widehat{P}_{\rm{c}}\right]\notag\\
\approx&\widehat{P}_{\rm{c}}\left[1+\frac{\mu}{3}\left(4\widehat{P}_{\rm{c}}+1\right)+\frac{\mu^2}{15}\left(11-32\widehat{P}_{\rm{c}}\right)\right],
\end{align}
which gives inversely that 
\begin{equation}
\mu\approx \frac{3}{4\widehat{P}_{\rm{c}}+1}\left(\frac{\phi}{\widehat{P}_{\rm{c}}}-1\right)+
\frac{9}{5}\frac{32\widehat{P}_{\rm{c}}-11}{4\widehat{P}_{\rm{c}}+1}
\left(\frac{\phi}{\widehat{P}_{\rm{c}}}-1\right)^2.
\end{equation}
 Putting this $\mu=\mu(\phi)$ back into $s^2/\widehat{P}_{\rm{c}}\approx4/3+32\mu/15+4\mu^2/5+(4/3-8\mu/5-268\mu^2/45)\widehat{P}_{\rm{c}}$ (via Eq.\,(\ref{zf-3})) gives the $s^2$ profile near the center to order $(\phi/\phi_{\rm{c}}-1)^2$ as (with $\phi/\phi_{\rm{c}}\lesssim1$),
\begin{align}
s^2/\phi_{\rm{c}}\approx&\frac{4}{3}+\frac{32}{5}\left(1-\frac{19}{4}\widehat{P}_{\rm{c}}\right)\left(\frac{\phi}{\phi_{\rm{c}}}-1\right)\notag\\
&-\frac{876}{25}\left(1-\frac{3439}{219}\widehat{P}_{\rm{c}}\right)\left(\frac{\phi}{\phi_{\rm{c}}}-1\right)^2,~~
\phi_{\rm{c}}\equiv \widehat{P}_{\rm{c}}.
\end{align}
We then find $[\phi_{\rm{c}}^{-1}\d s^2/\d \phi]_{\phi=\phi_{\rm{c}}}=(32/5)(1-19\widehat{P}_{\rm{c}}/4)$,  which may either be positive (Newtonian limit $\widehat{P}_{\rm{c}}\to0$) or negative,  depending on the magnitude of $\widehat{P}_{\rm{c}}$. In fact, all the descriptions, namely the $\widehat{r}$-dependence, the $\mu$-dependence and the $\phi$-dependence, are equivalent on the peaked profile of $s^2$. Thus, in the following we may
adopt these representations interchangeably.}

\subsection{Peak Position $\widehat{r}_{\textmd{pk}}$ and Enhancement $\Delta s^2$ of $s_{\textmd{c}}^2$}\label{sub2}

\begin{figure*}
\centering
\includegraphics[width=5.4cm]{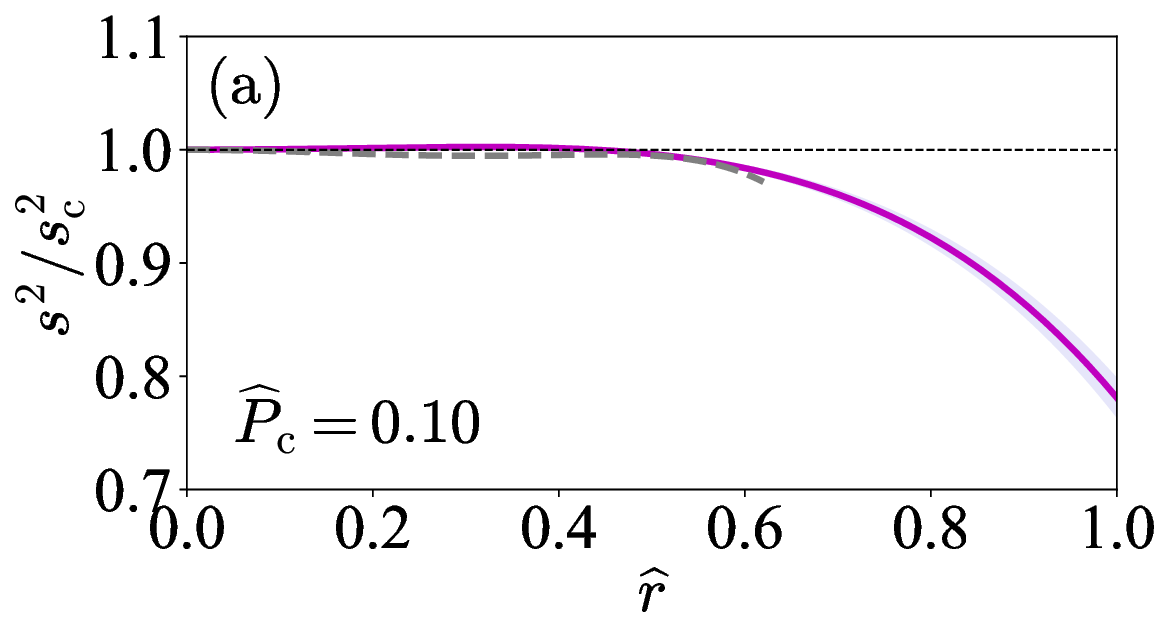}
\includegraphics[width=5.4cm]{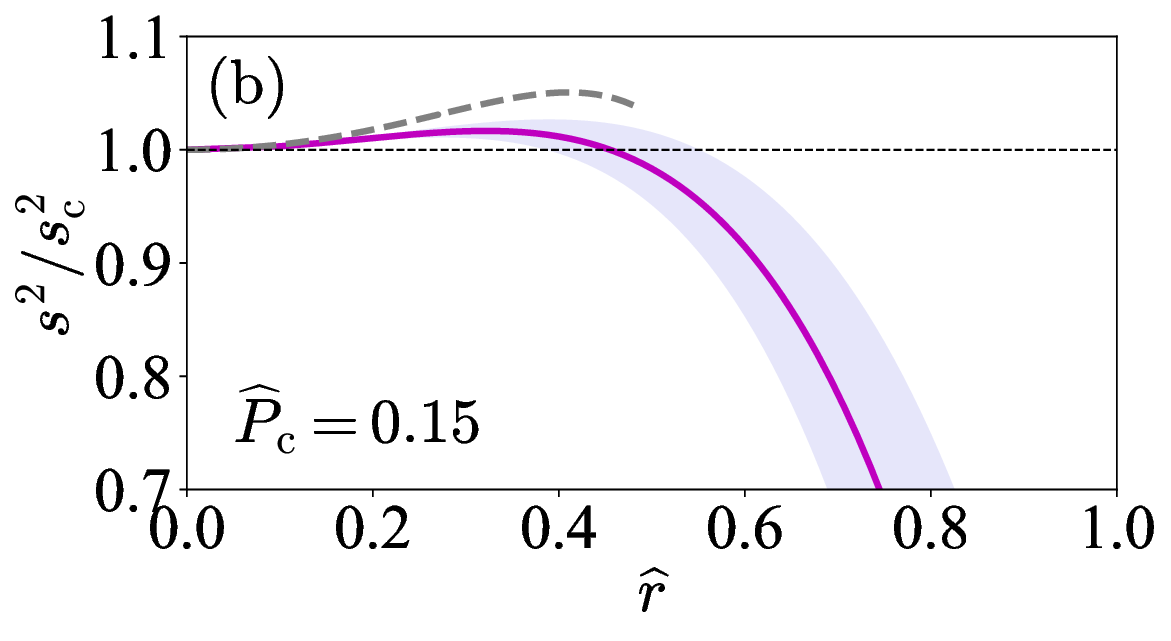}
\includegraphics[width=5.4cm]{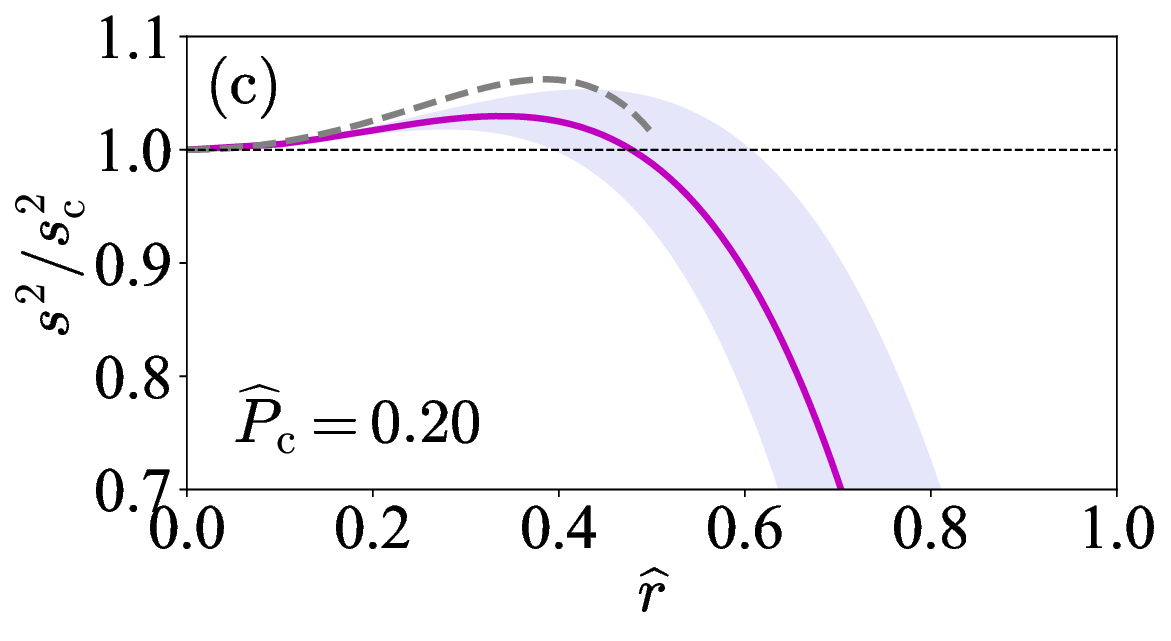}\\
\includegraphics[width=5.4cm]{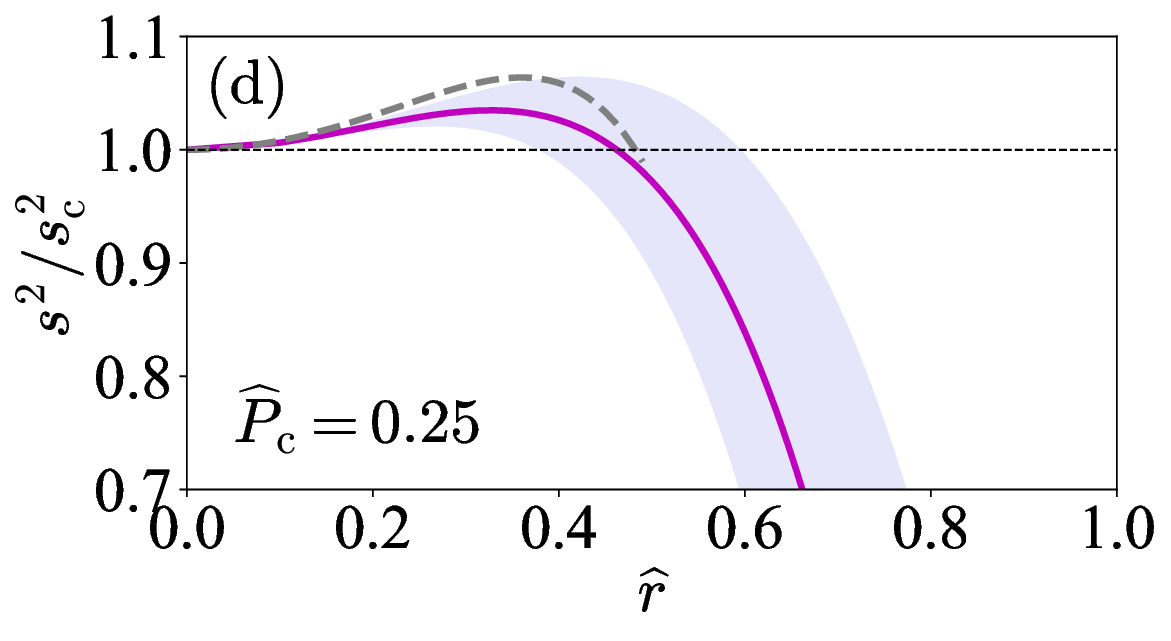}
\includegraphics[width=5.4cm]{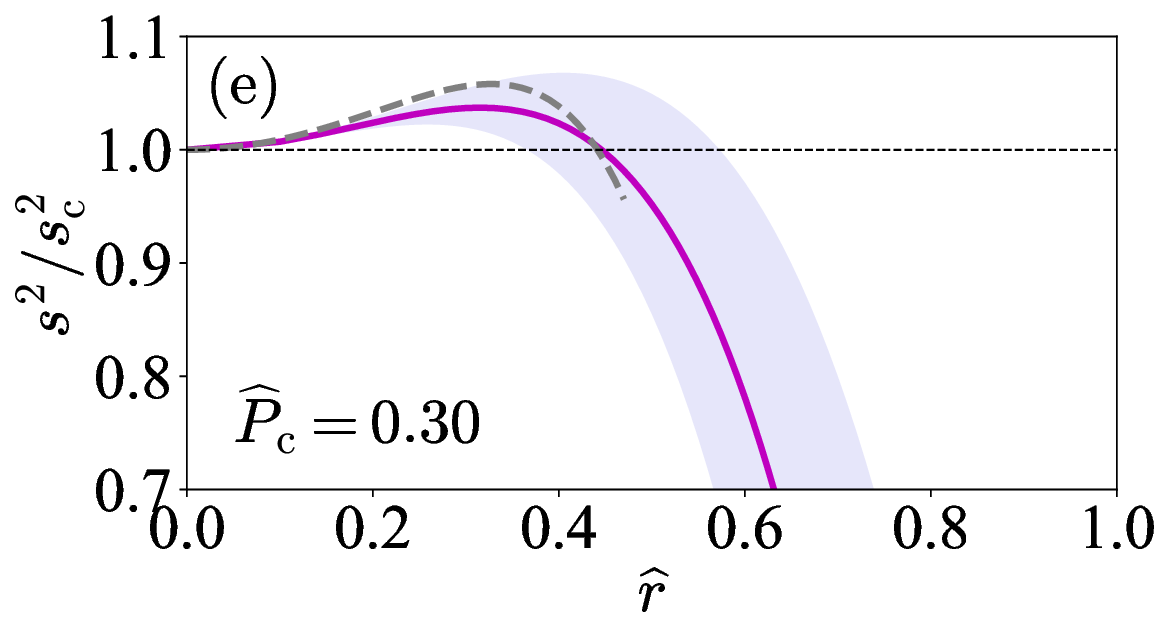}
\includegraphics[width=5.4cm]{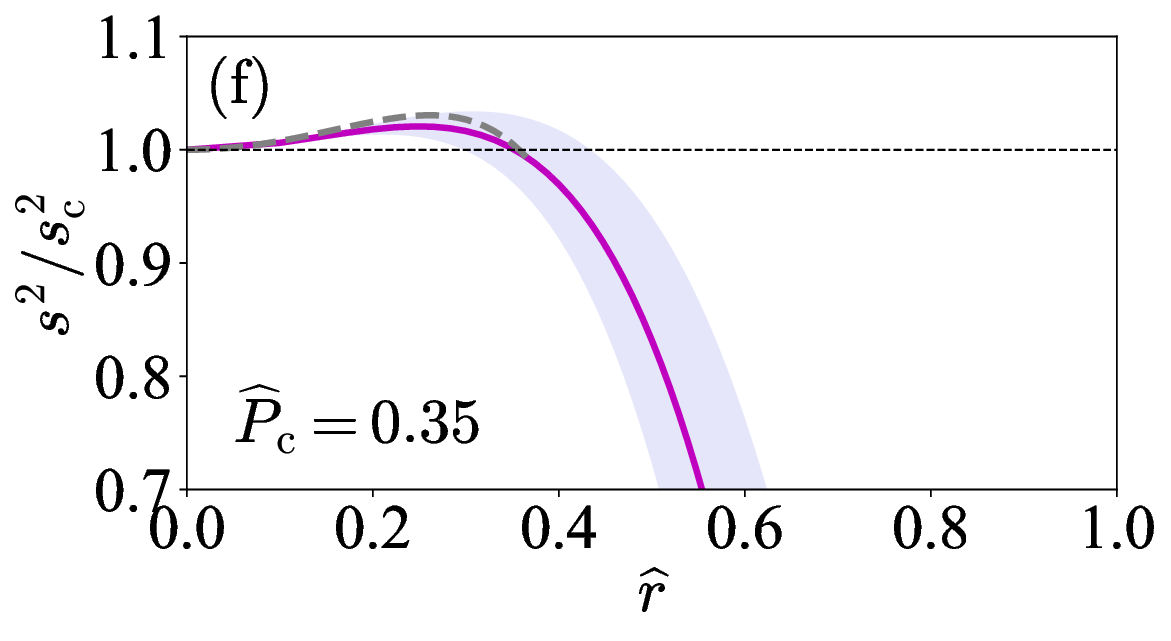}\\
\includegraphics[width=5.4cm]{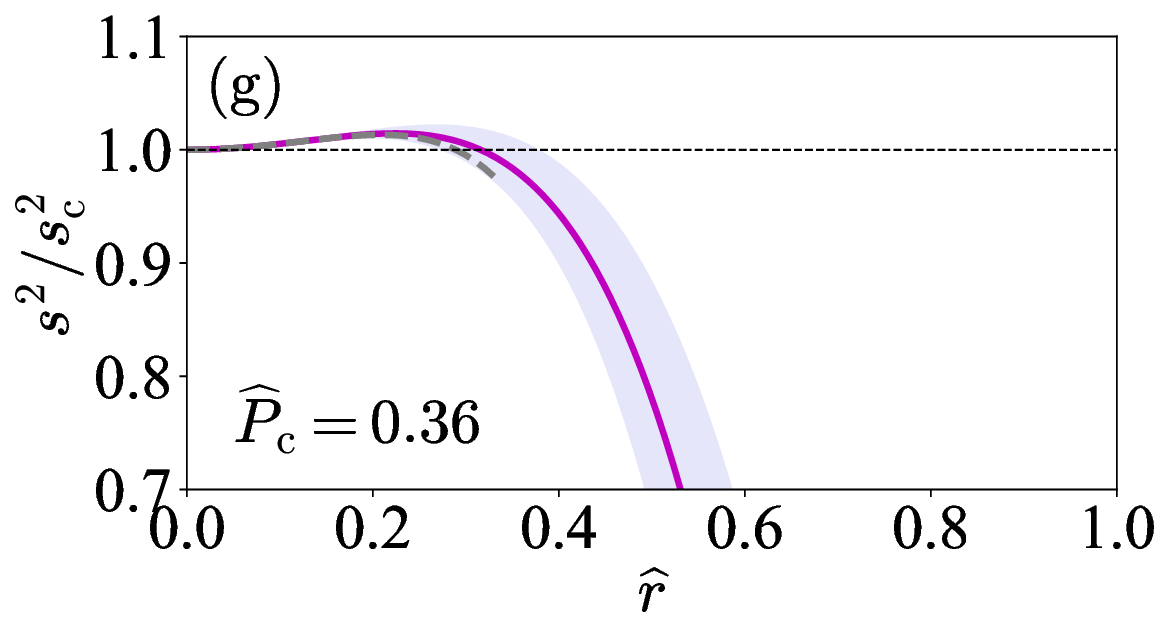}
\includegraphics[width=5.4cm]{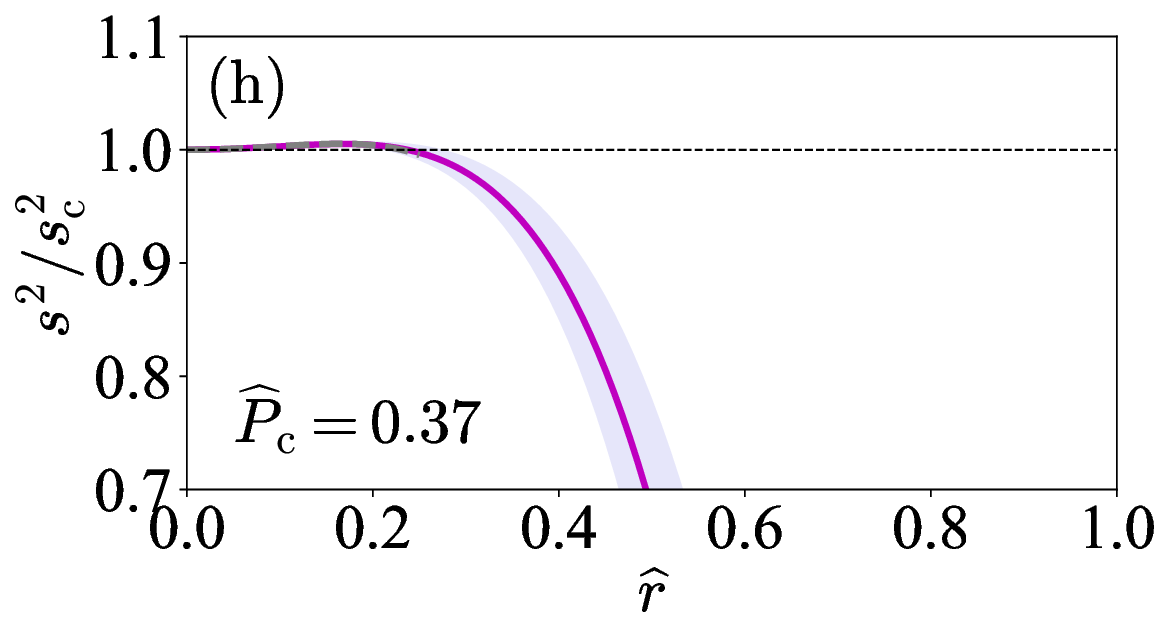}
\includegraphics[width=5.4cm]{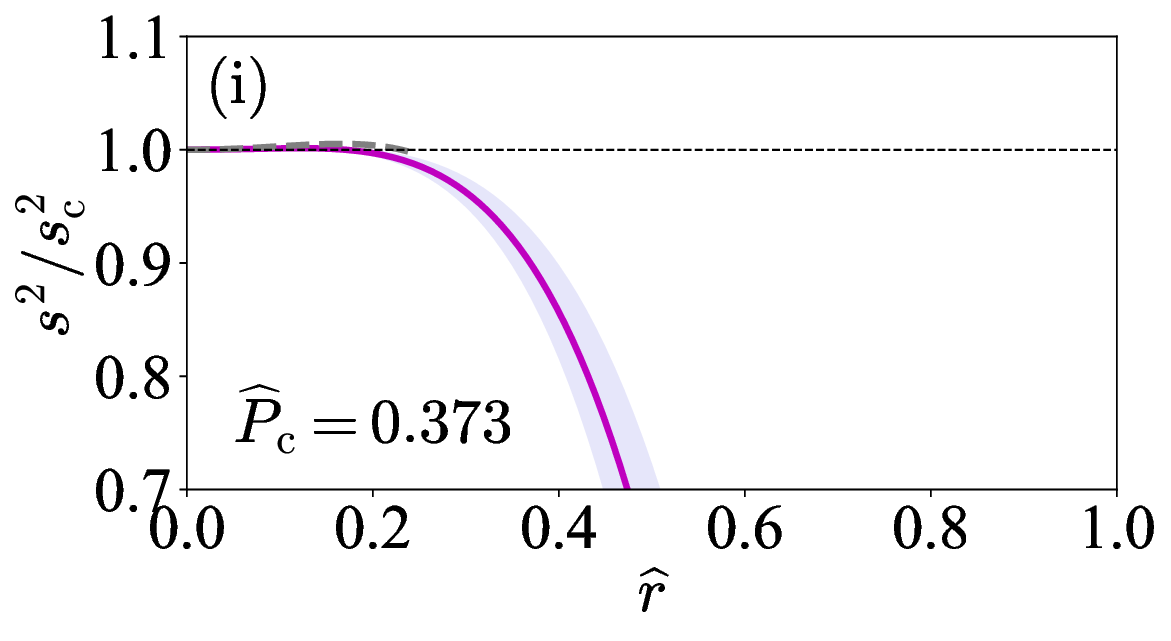}
\caption{(Color Online). $s^2/s_{\rm{c}}^2$ as a function of $\widehat{r}$ {with increasing values of $\widehat{P}_{\rm{c}}$ and with 1$\sigma$ uncertainties,  the grey dashed line is the one including contribution of $l_6\widehat{r}^6$, namely $s^2\approx s_{\rm{c}}^2+l_2\widehat{r}^2+l_4\widehat{r}^4+l_6\widehat{r}^6$.}
}\label{fig_s2_peak_stat-af}
\end{figure*}

\begin{figure*}
\centering
\includegraphics[width=5.4cm]{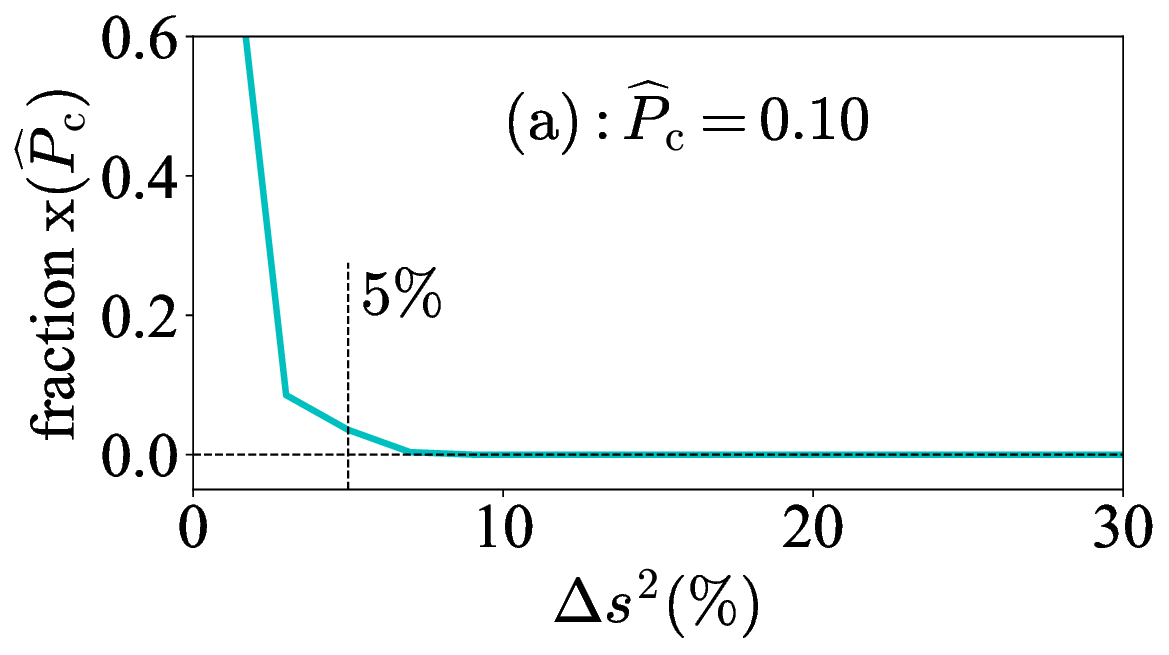}
\includegraphics[width=5.4cm]{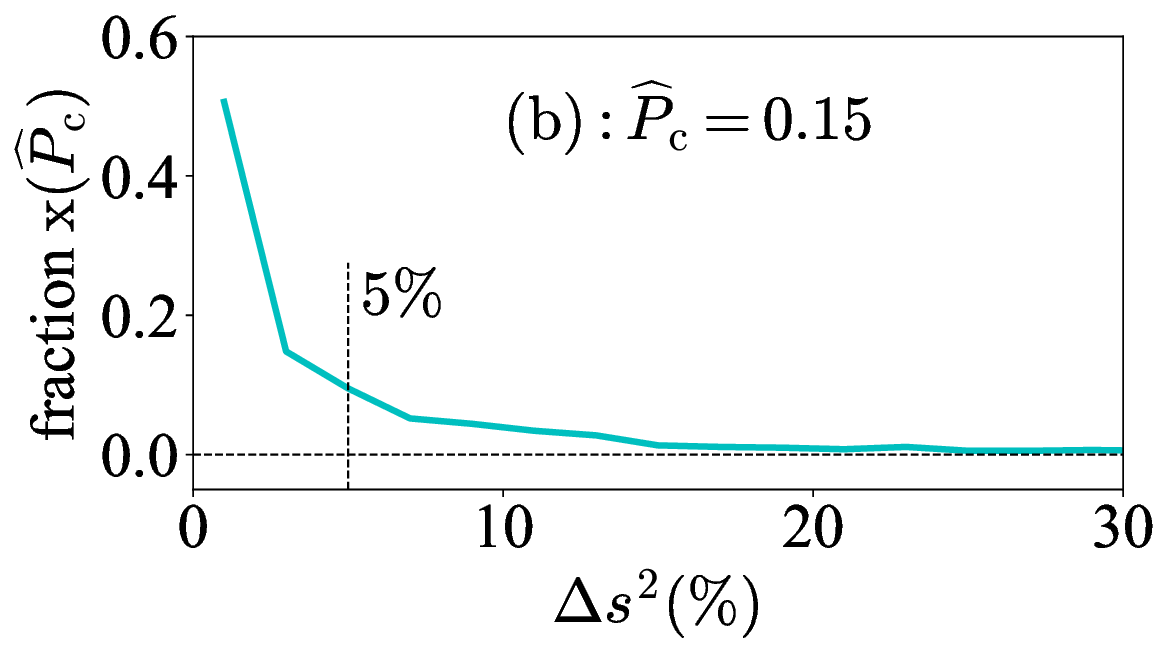}
\includegraphics[width=5.4cm]{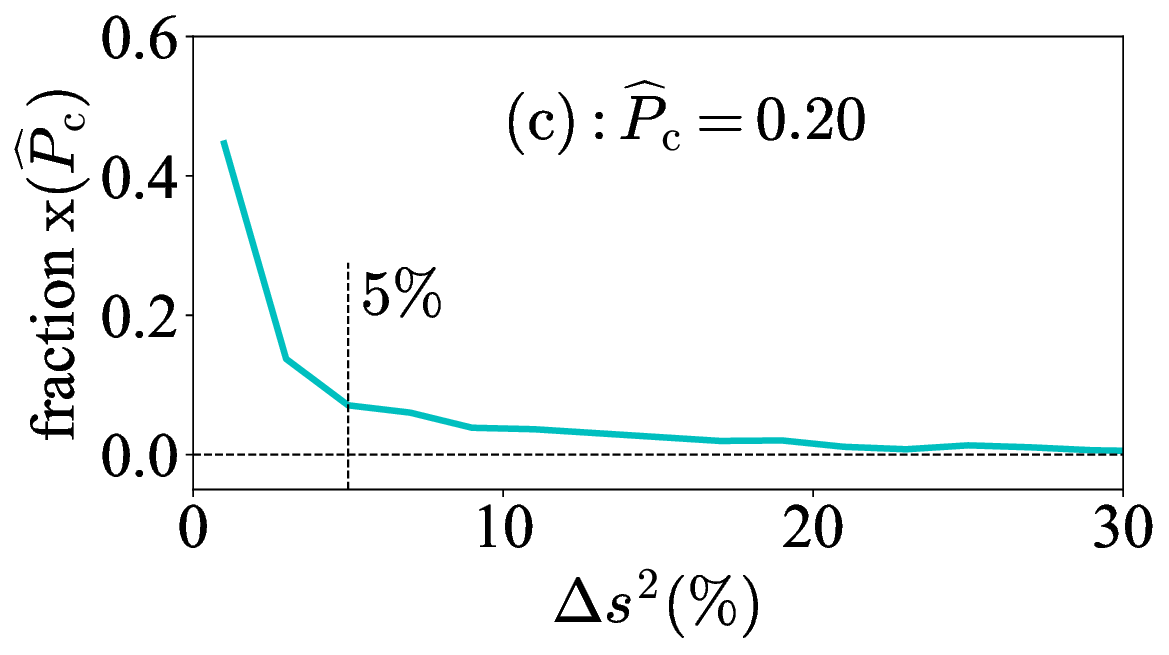}\\
\includegraphics[width=5.4cm]{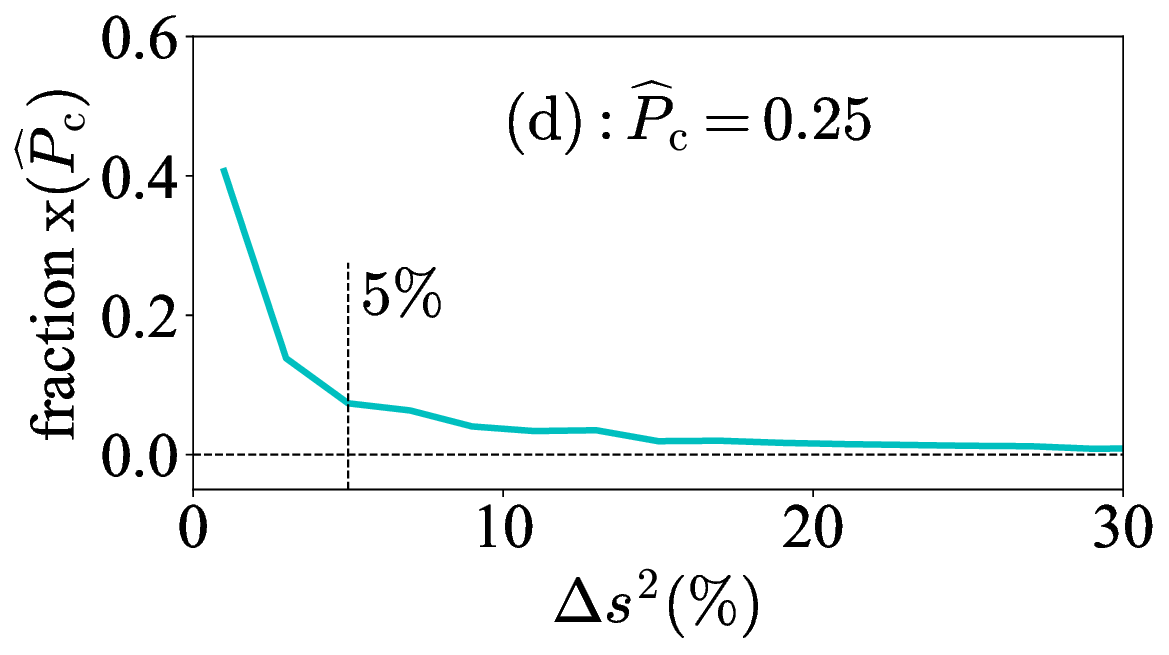}
\includegraphics[width=5.4cm]{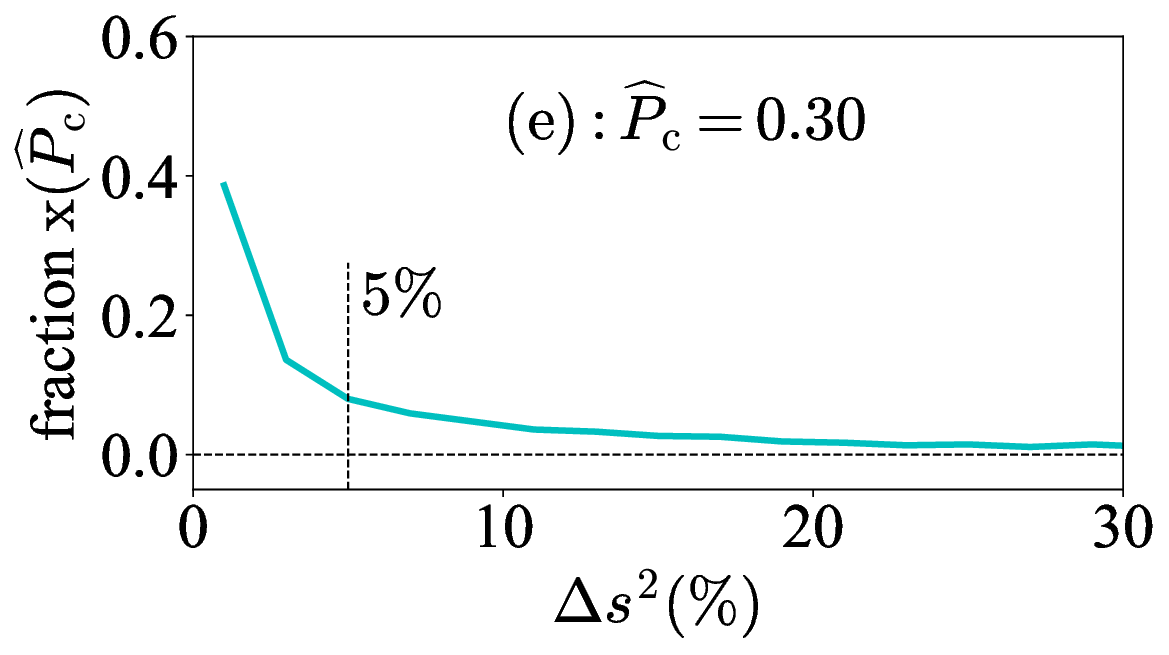}
\includegraphics[width=5.4cm]{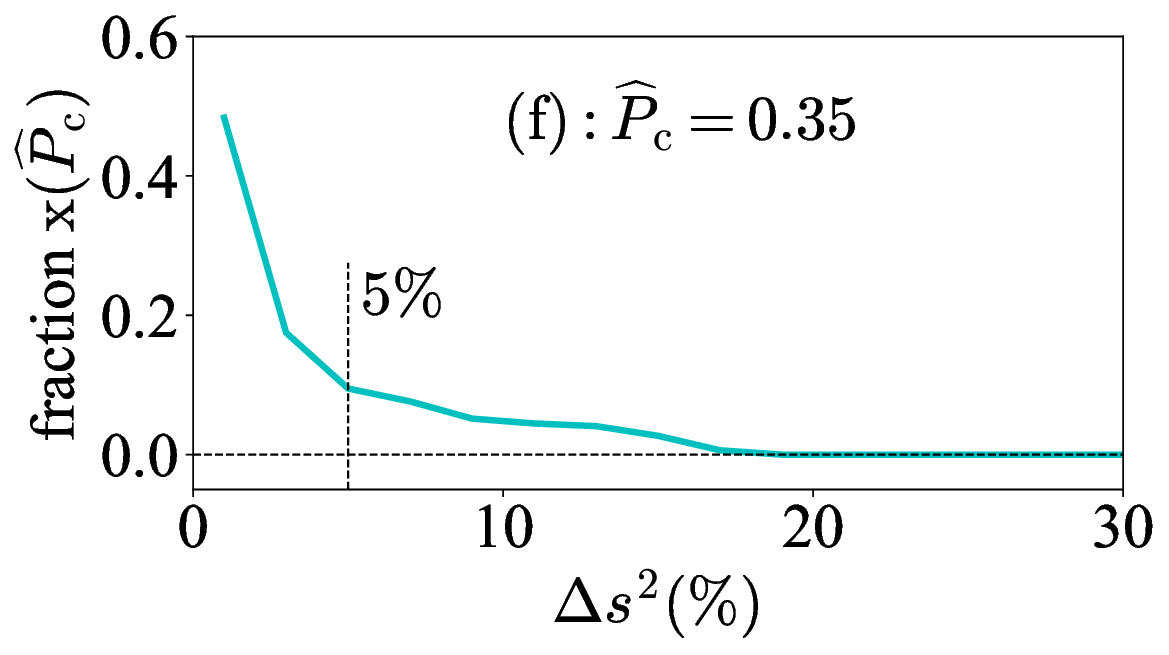}\\
\includegraphics[width=5.4cm]{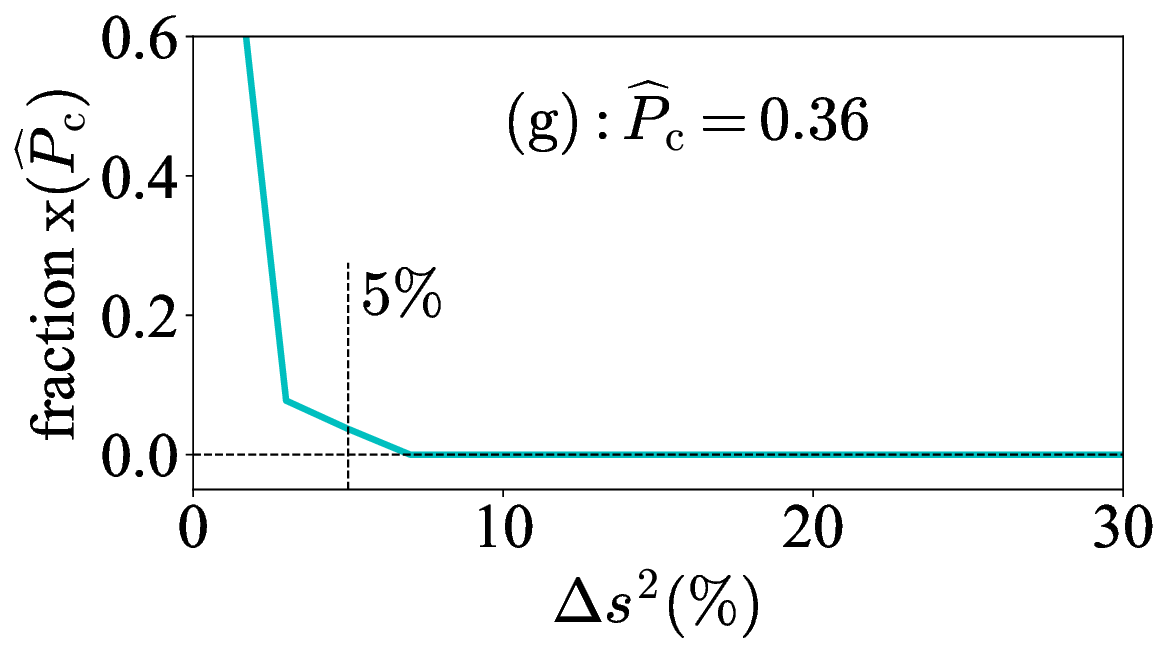}
\includegraphics[width=5.4cm]{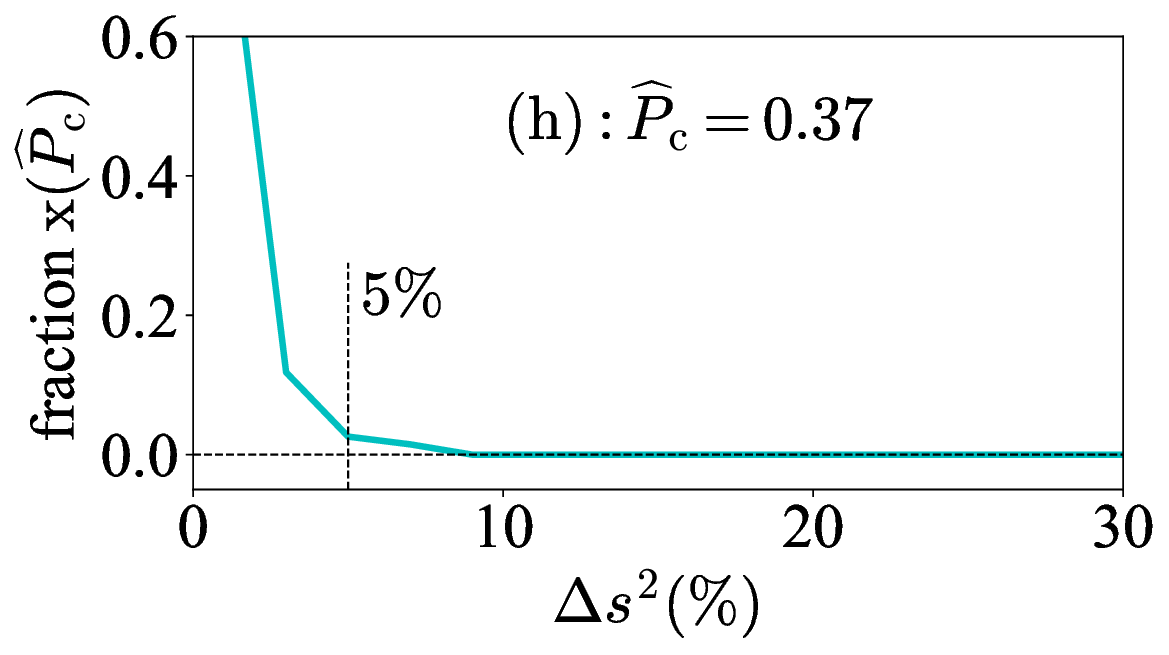}
\includegraphics[width=5.4cm]{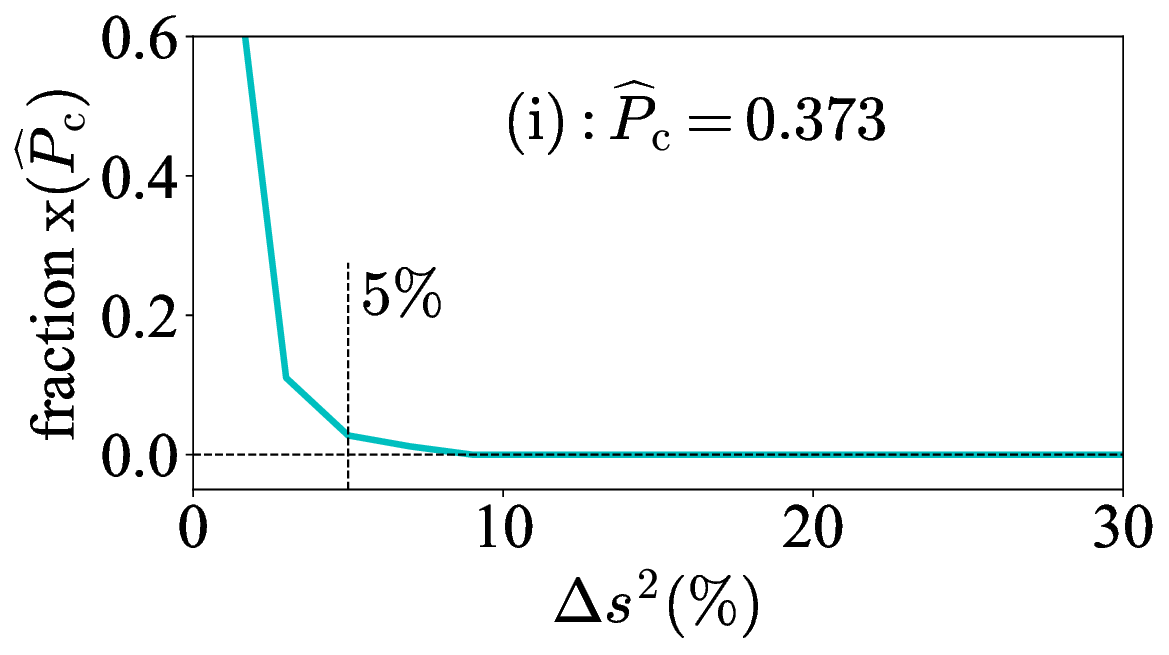}
\caption{(Color Online). Same as FIG.\,\ref{fig_s2_peak_stat-af} but for the fraction $\rm{x}(\widehat{P}_{\rm{c}})$ for the enhancement $\Delta s^2$.
}\label{fig_s2_peak_stat-3af}
\end{figure*}

\begin{figure*}
\centering
\includegraphics[width=5.4cm]{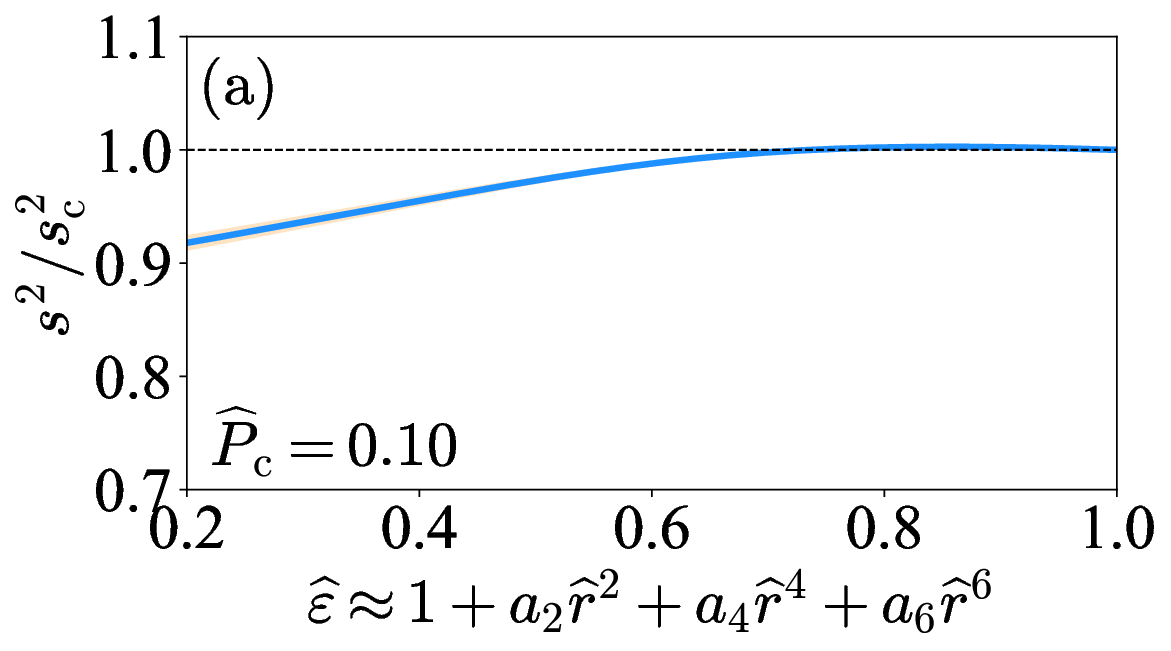}
\includegraphics[width=5.4cm]{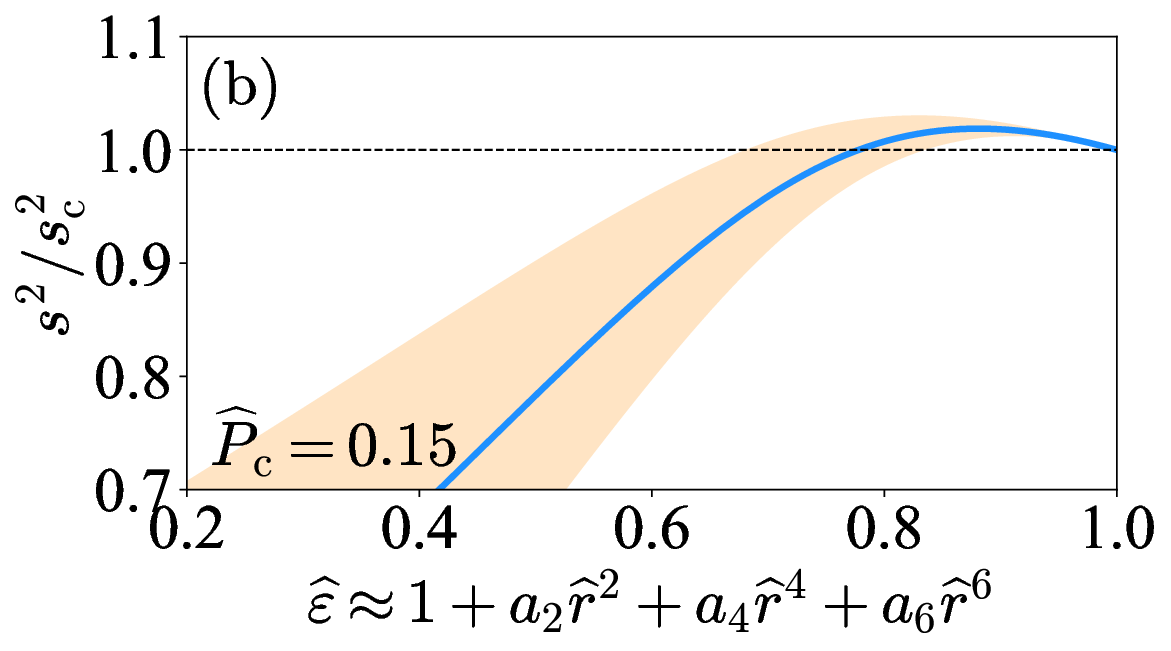}
\includegraphics[width=5.4cm]{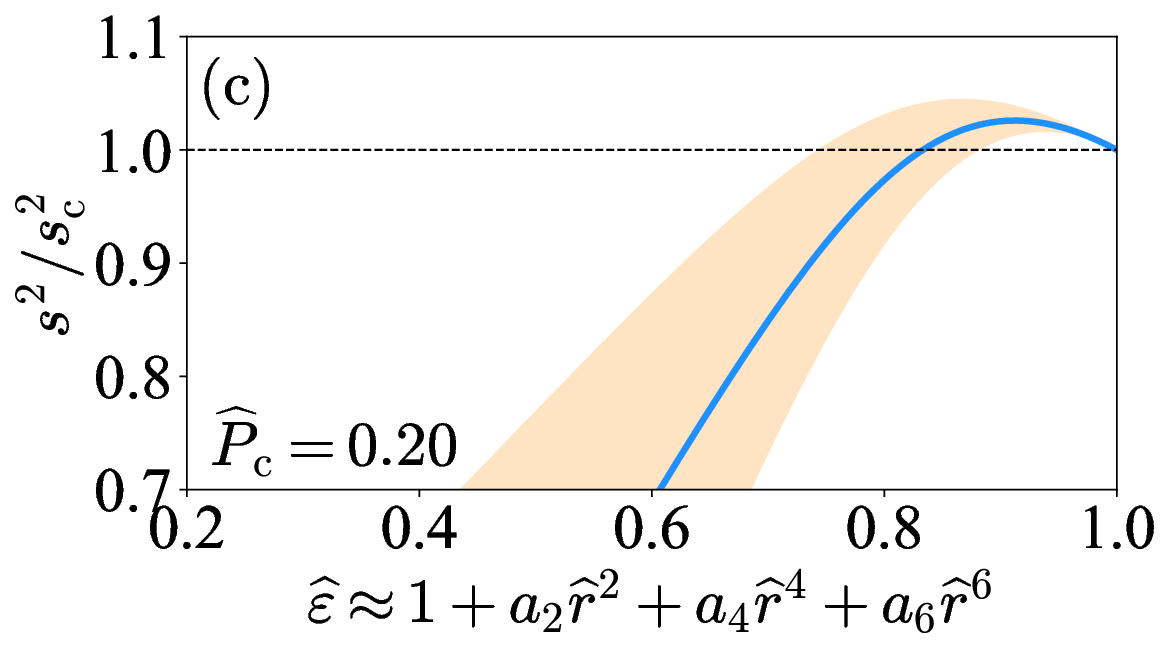}\\
\includegraphics[width=5.4cm]{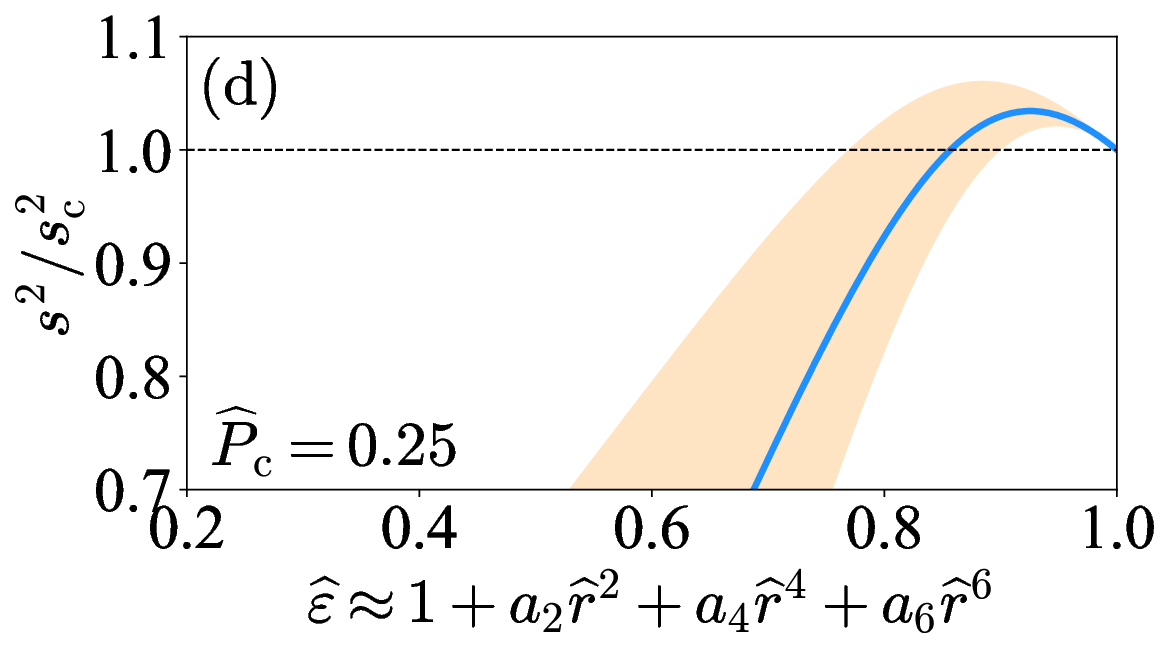}
\includegraphics[width=5.4cm]{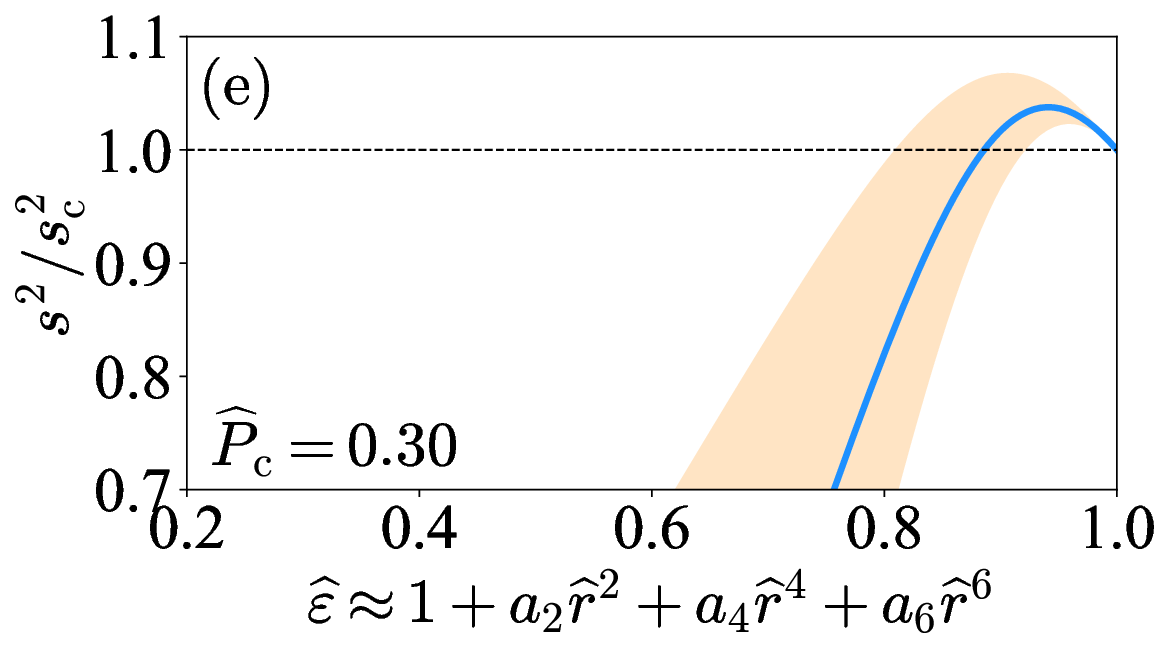}
\includegraphics[width=5.4cm]{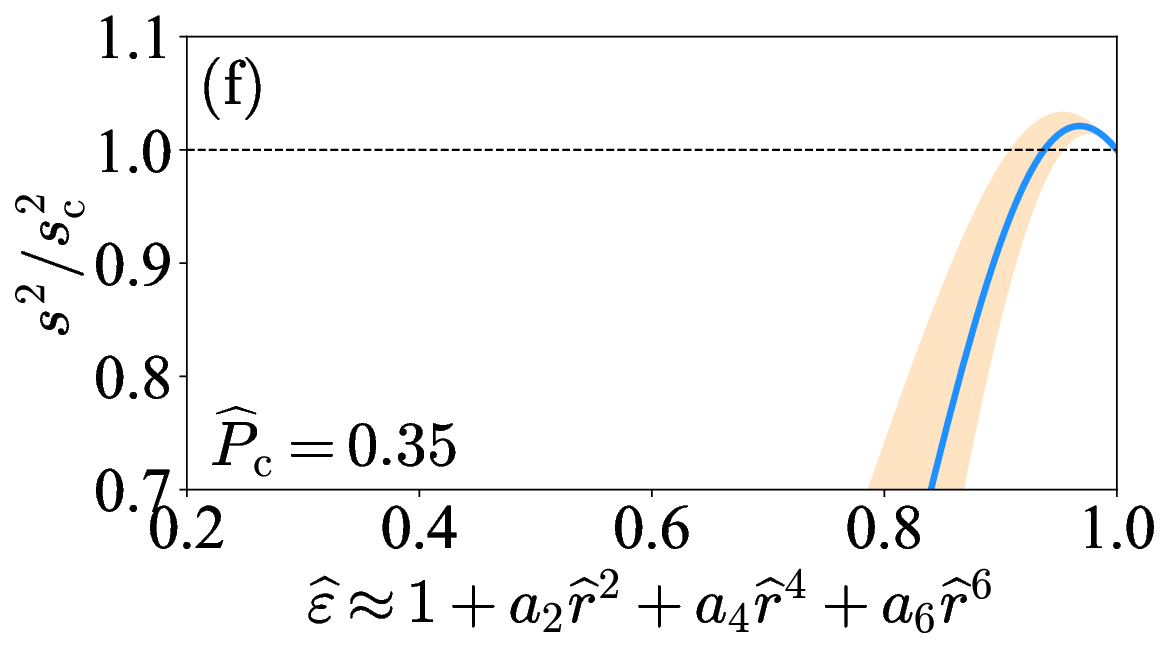}\\
\includegraphics[width=5.4cm]{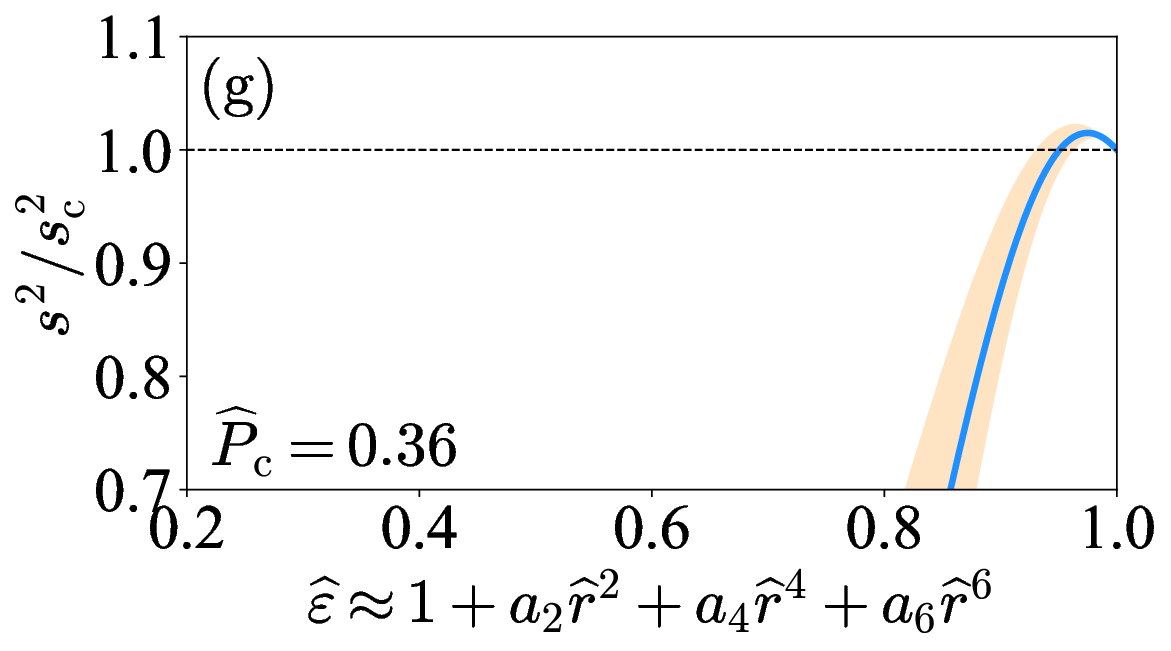}
\includegraphics[width=5.4cm]{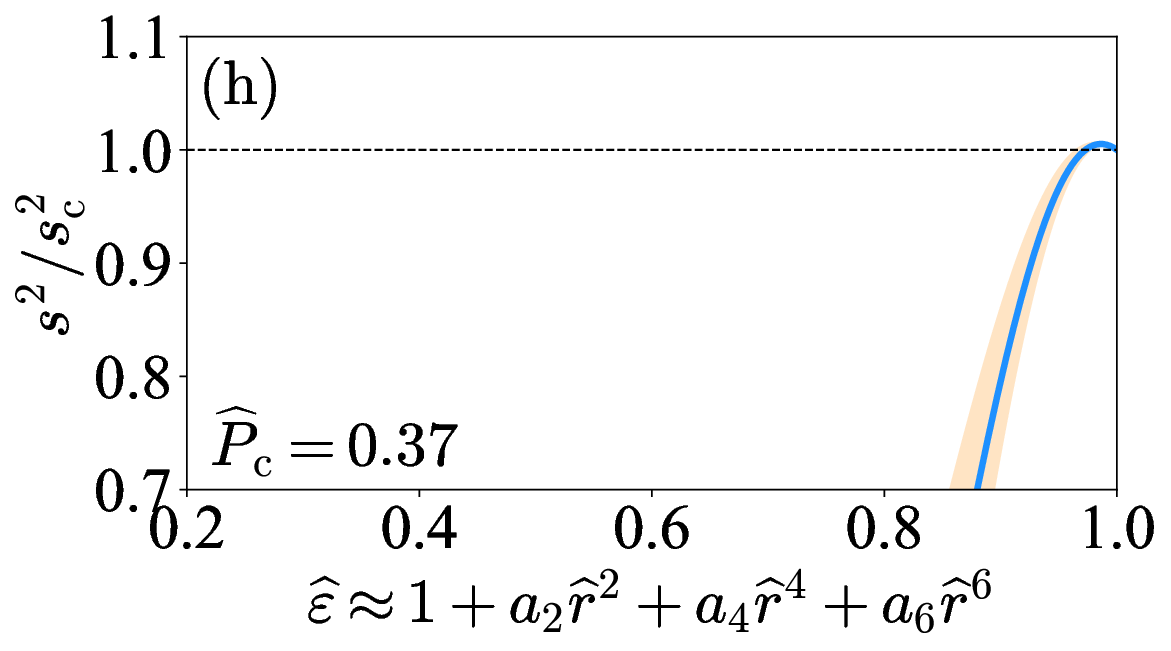}
\includegraphics[width=5.4cm]{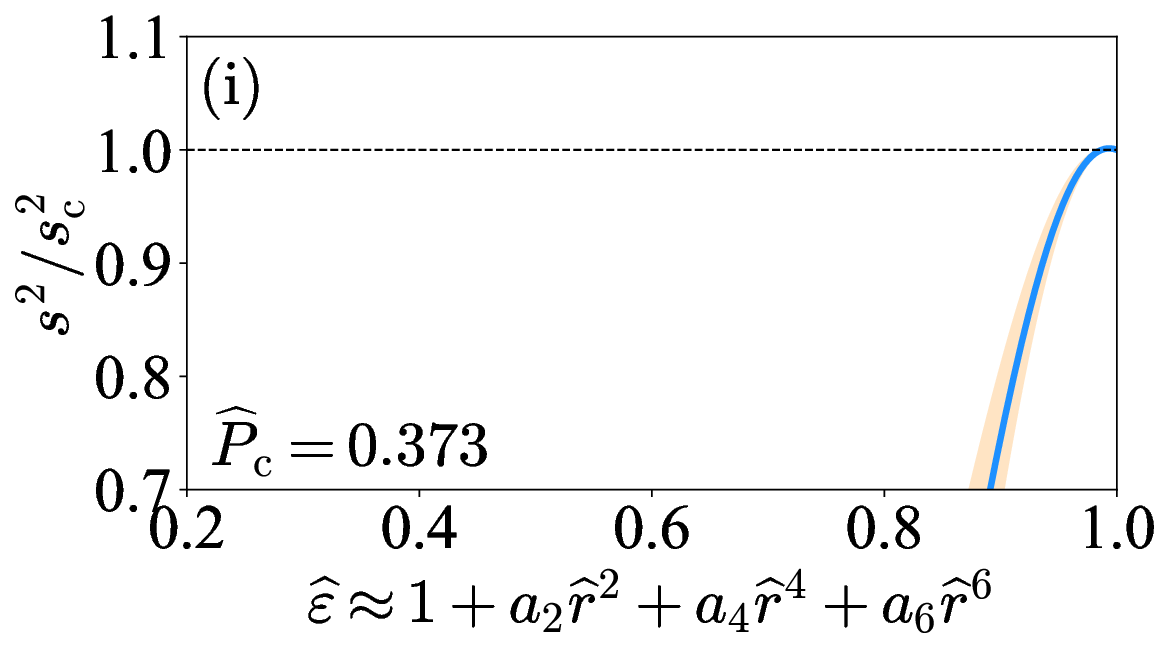}
\caption{(Color Online). Same as FIG.\,\ref{fig_s2_peak_stat-af} but for the $s^2/s_{\rm{c}}^2$ as a function of $\widehat{\varepsilon}$ (near NS centers).
}\label{fig_s2_peak_stat-1af}
\end{figure*}

\begin{figure*}
\centering
\includegraphics[width=5.4cm]{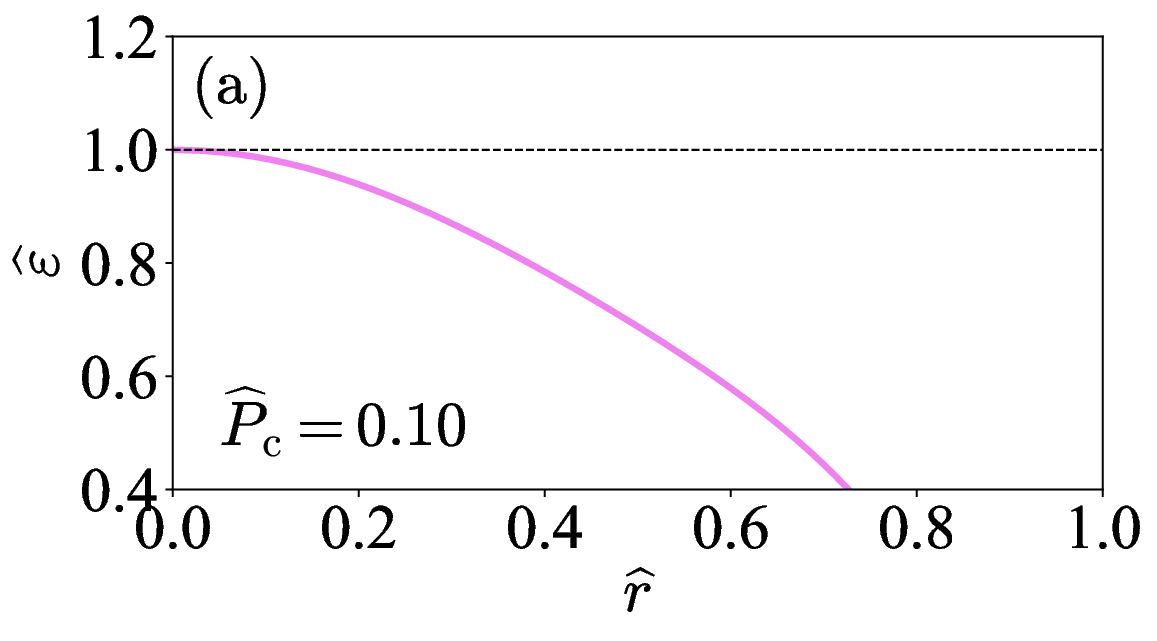}
\includegraphics[width=5.4cm]{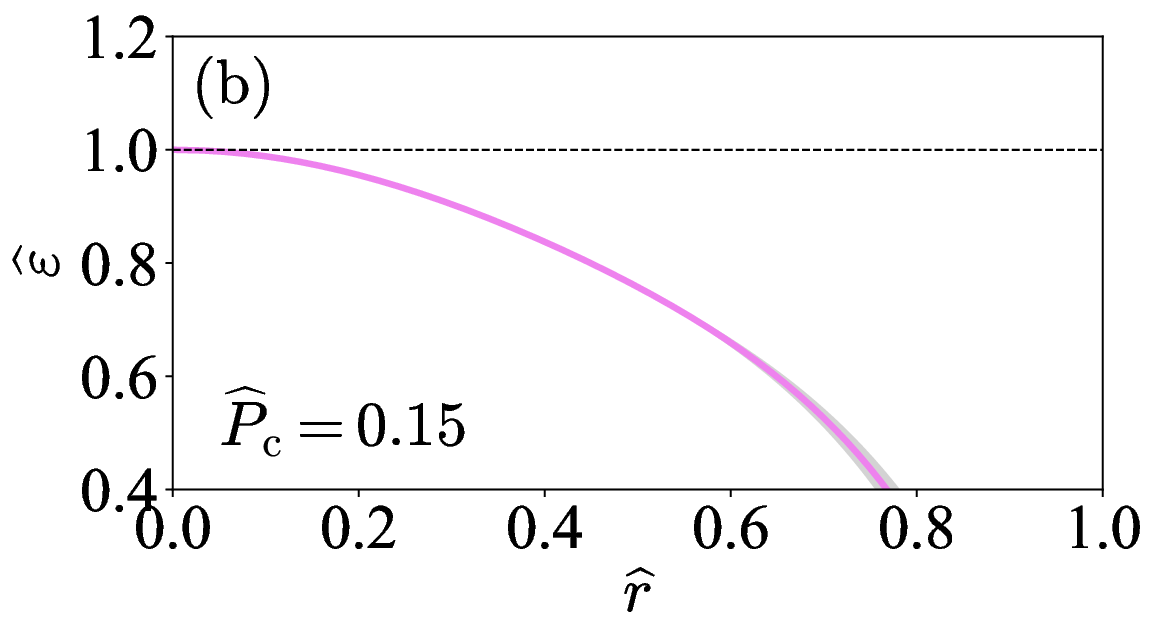}
\includegraphics[width=5.4cm]{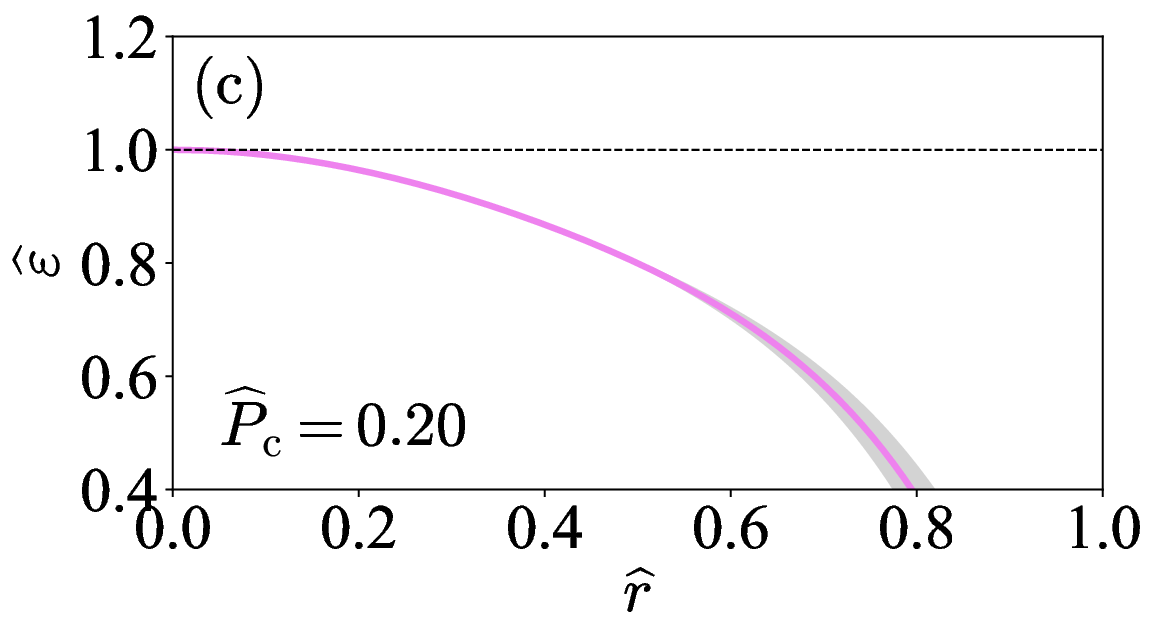}\\
\includegraphics[width=5.4cm]{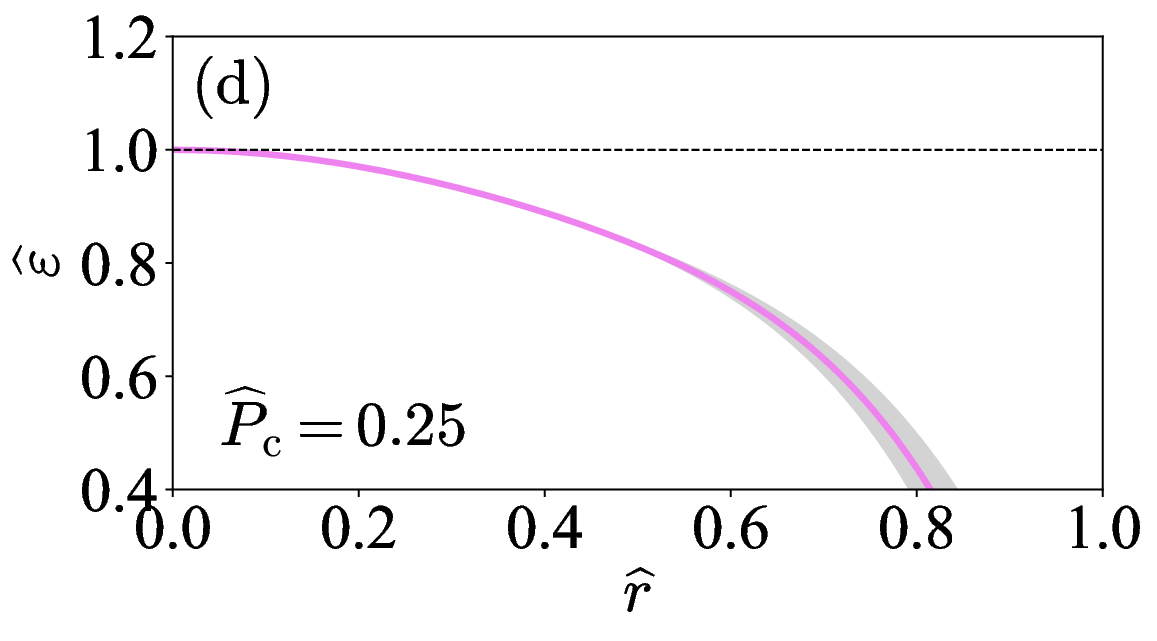}
\includegraphics[width=5.4cm]{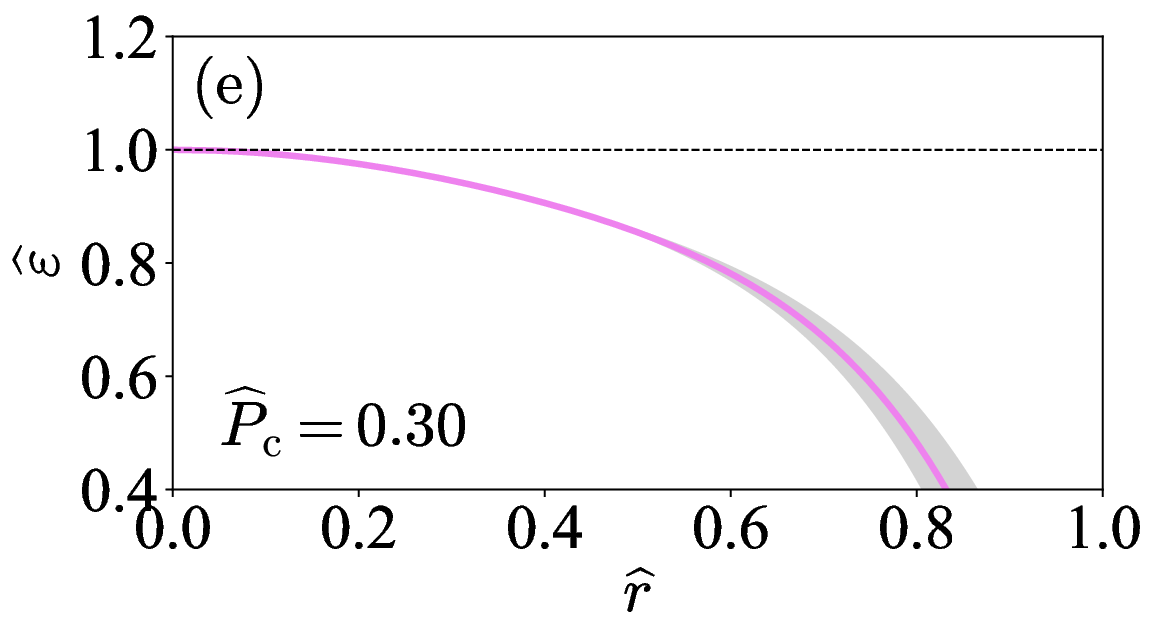}
\includegraphics[width=5.4cm]{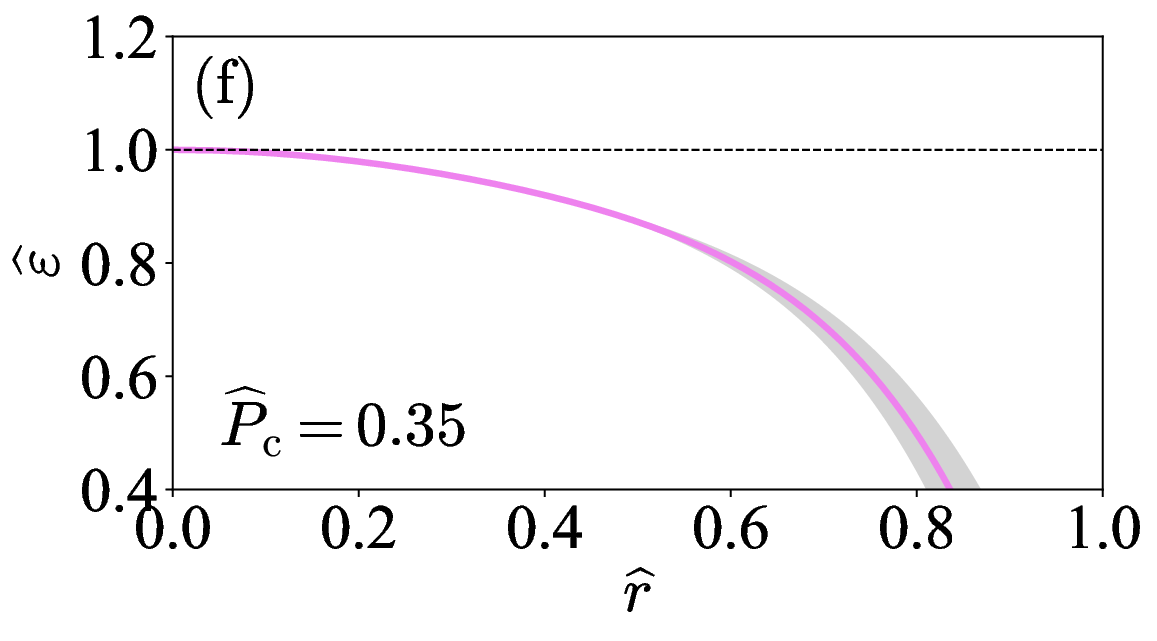}\\
\includegraphics[width=5.4cm]{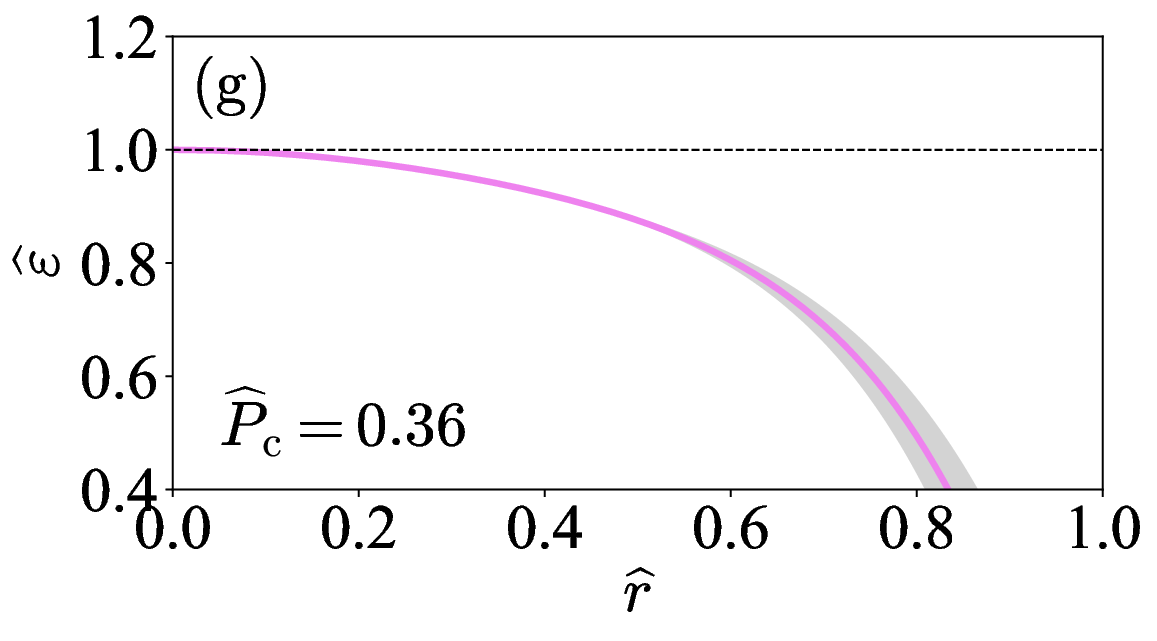}
\includegraphics[width=5.4cm]{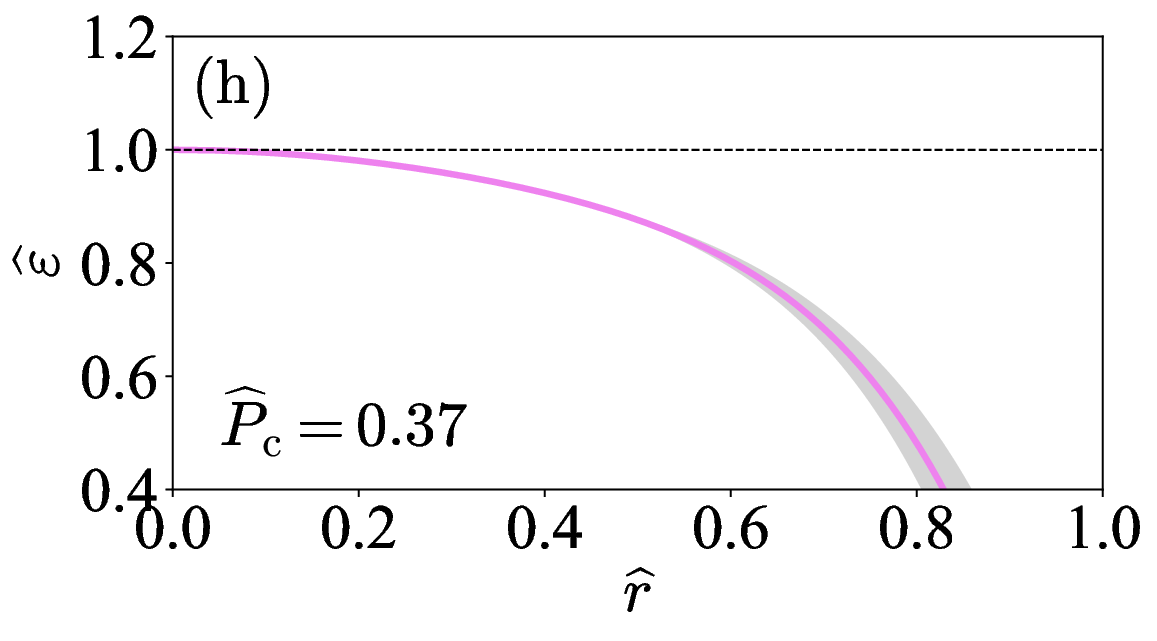}
\includegraphics[width=5.4cm]{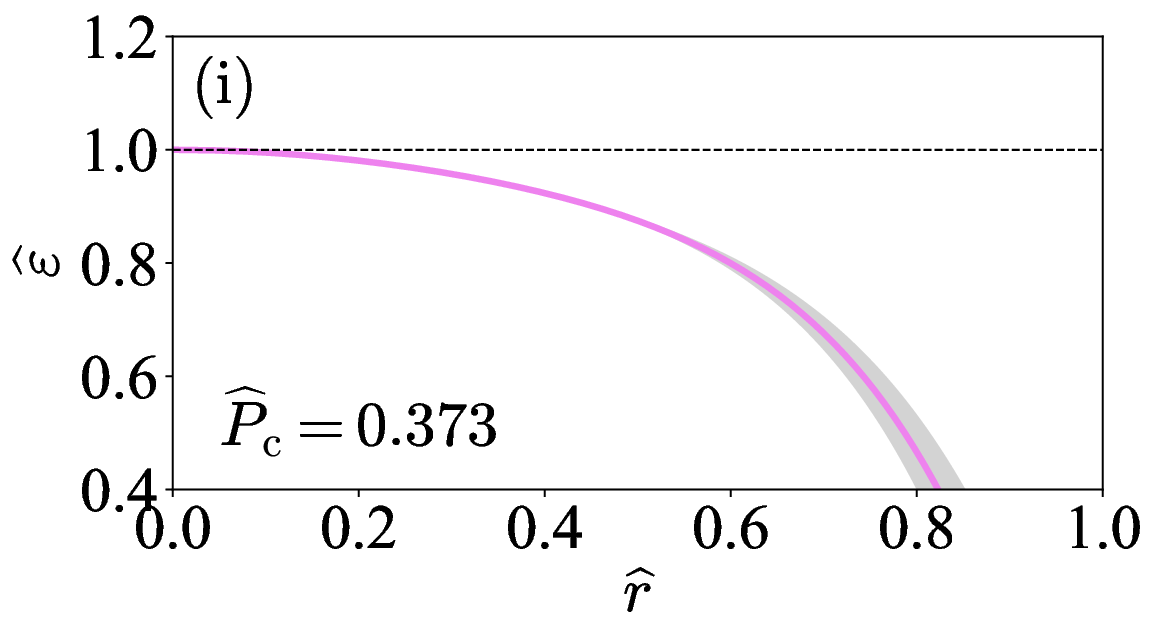}
\caption{(Color Online). Same as FIG.\,\ref{fig_s2_peak_stat-af} but for $\widehat{\varepsilon}\approx 1+a_2\widehat{r}^2+a_4\widehat{r}^4+a_6\widehat{r}^6$ as a function of $\widehat{r}$.
}\label{fig_s2_peak_stat-2af}
\end{figure*}

We estimate here the occurring probability of a peak in the $s^2$ profile near $\widehat{r}=0$ for $l_2>0$ and $l_4<0$, where the peak is at 
\begin{equation}
\widehat{r}_{\rm{pk}}=\sqrt{-l_2/2l_4}.
\end{equation}
For each $\widehat{P}_{\rm{c}}$, we randomly sample the coefficients $a_4\sim\rm{Unif}[0,2]$ and $a_6\sim\rm{Unif}[-2,2]$ by $m$ times, and count the number $m_{\rm{pk}}$ of samples where there is a peak in the $s^2$ profile while satisfying the causality condition $s_{\max}^2\leq1$, here
\begin{equation}\label{def-Ds2}
s_{\max}^2\equiv s^2(\widehat{r}_{\rm{pk}})=s_{\rm{c}}^2-l_2^2/4l_4.
\end{equation}
The probability is estimated as $m_{\rm{pk}}/m$, see FIG.\,\ref{fig_s2_peak_stat_prop} for the simulated results.
For a relatively small $\widehat{P}_{\rm{c}}\approx0.1$, the probability is also small since the $\widehat{P}_{\rm{c}}$ is not large enough to make $l_2$ positive (even if $a_4$ is positive). The probability eventually increases with  increasing $\widehat{P}_{\rm{c}}$, but it becomes small again when the $\widehat{P}_{\rm{c}}$ is close to the causality limit about 0.374. This is because the matter is too stiff to be further compressed by gravity, i.e.,  there is no space for $s^2(\widehat{r})>s_{\rm{c}}^2$ since $s_{\rm{c}}^2$ itself is $\to1$\,\cite{CLZ23-b}.
Combining the above information with what we learned from FIG.\,\ref{fig_a4peak}, we find that the probability is relatively larger (compared with its surroundings) for $0.2\lesssim\widehat{P}_{\rm{c}}\lesssim0.35$, where the strong gravitational force effectively bends down the $s^2$.

Interestingly, $\widehat{P}_{\rm{c}}\approx0.24_{-0.07}^{+0.05}$\,\cite{CLZ23-a} for PSR J0740+6620 extracted directly from its observed mass and radius\,\cite{Fon21,Miller21,Riley21,Salmi22} is quite close to this region, indicating massive NSs with radii $\approx12\mbox{-}14\,\rm{km}$ are excellent objects to study the peaked $s^2$ profile.
The resulted $a_4$ and $a_6$ for $\widehat{P}_{\rm{c}}\approx0.24$ are found respectively to be about $a_4\approx0.71\pm0.03$ and $a_6\approx-1.37\pm0.19$, both are $\sim\mathcal{O}(1)$.
See FIG.\,\ref{fig_s2_peak_stat_a4a6} for the general $\widehat{P}_{\rm{c}}$-dependence of the coefficients $a_4,a_6,l_2$ and $l_4$, where the coefficient $l_4$ (compared with $a_4,a_6$ and $l_2$) may take some large values (for large $\widehat{P}_{\rm{c}}$).
Furthermore, the coefficient $l_2$ starting from about zero at $\widehat{P}_{\rm{c}}\approx0.1$ implies that for even smaller $\widehat{P}_{\rm{c}}$ values there would be no peaked feature in the $s^2$ profile.

Also shown in FIG.\,\ref{fig_s2_peak_stat_prop} is the peak position $\widehat{r}_{\rm{pk}}$ which is located at about $0.35\lesssim\widehat{r}_{\rm{pk}}\lesssim0.4$ for a wide range of $\widehat{P}_{\rm{c}}$.
Moreover,  we show in FIG.\,\ref{fig_s2_peak_stat_prop} the (averaged) enhancement of $s_{\rm{c}}^2$, namely (see Eq.\,(\ref{def-Ds2})),
 \begin{equation}
\Delta s^2\equiv s_{\max}^2/s_{\rm{c}}^2-1=-l_2^2/4l_4s_{\rm{c}}^2,
\end{equation}
which is generally small about $\lesssim5\%$.  

FIG.\,\ref{fig_s2_peak_stat-af} shows the radial profiles of $s^2$ adopting nine different $\widehat{P}_{\rm{c}}$ values specified in the plots.
The $s^2$ is probably a monotonically decreasing function of $\widehat{r}$ for small $\widehat{P}_{\rm{c}}\lesssim0.1$ (panel (a) of FIG.\,\ref{fig_s2_peak_stat-af}), the peak eventually emerges/develops as $\widehat{P}_{\rm{c}}$ increases to about $\widehat{P}_{\rm{c}}\approx0.30$ (panels (b) to (e)) and then tends to disappear for even large $\widehat{P}_{\rm{c}}\gtrsim0.35$ (panels (f)-(i)).
In FIG.\,\ref{fig_s2_peak_stat-3af}, we show the dependence of the fraction $\rm{x}(\widehat{P}_{\rm{c}})$ on the enhancement $\Delta s^2$. It is interesting to see that at the two limiting sides of $\widehat{P}_{\rm{c}}$ (i.e., small $\widehat{P}_{\rm{c}}\lesssim\mathcal{O}(0.1)$ and $\widehat{P}_{\rm{c}}\gtrsim0.35$) the fraction $\rm{x}(\widehat{P}_{\rm{c}})$ for large enhancement $\Delta s^2(\gtrsim15\%)$ is extremely small,
and most of the fraction is accumulated at $\Delta s^2\lesssim5\%$.

If the contribution $l_6\widehat{r}^6$ (which involves the coefficient $a_8$) is included as $s^2\approx s_{\rm{c}}^2+l_2\widehat{r}^2+l_4\widehat{r}^4+l_6\widehat{r}^6$,
then the equation determining the peak position $\widehat{r}_{\rm{pk}}$ becomes ($l_2>0$),
\begin{equation}\label{kk-1}
l_2+2l_4\widehat{r}^2+3l_6\widehat{r}^4=0.
\end{equation}
The expression for $l_6$ could be obtained straightforwardly (we do not give it explicitly here due to its length).
Using Eq.\,(\ref{kk-1}), we immediately obtain the peak position $\widehat{r}_{\rm{pk}}$ as $\widehat{r}_{\rm{pk}}^2=[-l_4+(l_4^2-3l_2l_6)^{1/2}]/3l_6$ if $l_4>0$ and $\widehat{r}_{\rm{pk}}^2=[-l_4-(l_4^2-3l_2l_6)^{1/2}]/3l_6$ for $l_4<0$ (both can fulfill the conditions $l_6\leq l_4^2/3l_2$ and $\widehat{r}_{\rm{pk}}^2>0$).  
By sampling $a_8\sim\rm{Unif}[-2,2]$, one can similarly analyze the peaked behavior of $s^2$, and the results are shown by the dashed grey lines in FIG.\,\ref{fig_s2_peak_stat-af}.
It is seen that the correction $l_6\widehat{r}^6$ slightly increases the enhancement $\Delta s^2$ while has almost no effect on the location of the peak $\widehat{r}_{\rm{pk}}$.

Moreover, using $\widehat{\varepsilon}\approx 1+a_2\widehat{r}^2+a_4\widehat{r}^4+a_6\widehat{r}^6$, one finds that $85\%\lesssim\widehat{\varepsilon}_{\rm{pk}}\lesssim95\%$  for $0.15\lesssim\widehat{P}_{\rm{c}}\lesssim0.35$.
In FIG.\,\ref{fig_s2_peak_stat-1af},  we show the $s^2/s_{\rm{c}}^2$ as a function of $\widehat{\varepsilon}$.
The peak location $\widehat{\varepsilon}_{\rm{pk}}$ in energy density is consistent with our previous studies, i.e., $\widehat{\rho}\equiv
\rho/\rho_{\rm{c}}\approx
\widehat{\varepsilon}-\mu(1+4\mu/3)\widehat{P}_{\rm{c}}(1-\widehat{P}_{\rm{c}})
$, from which we inferred that $\rho_{\rm{pk}}/\rho_{\rm{c}}\approx\widehat{\varepsilon}_{\rm{pk}}\lesssim95\%$\,\cite{CLZ23-b}.
This is also consistent with findings in a few recent studies by others, see, e.g., Ref.\,\cite{Mro23} predicted that the SSS peak if exists would be very close to the NS center.
Similarly, Ref.\,\cite{Cao23} predicted that the peak in $s^2$ profile for a $2M_{\odot}$ NS is at $\rho_{\rm{pk}}\approx0.55\,\rm{fm}^{-3}$ while its central density is about $\rho_{\rm{c}}\approx0.56\,\rm{fm}^{-3}$ and thus $\rho_{\rm{pk}}/\rho_{\rm{c}}\approx98\%$.
Finally, we show in FIG.\,\ref{fig_s2_peak_stat-2af} the radial profile of the energy density, i.e., 
the $\widehat{r}$-dependence of $\widehat{\varepsilon}\approx 1+a_2\widehat{r}^2+a_4\widehat{r}^4+a_6\widehat{r}^6$. It is seen that they have a direct correspondence 
with little uncertainty especially for $\widehat{r}\approx0$, consistent with that shown in FIG.\,\ref{fig_s2_prep_r}.

\subsection{Example: the $s^2$ Profile of a NS with $M_{\textmd{NS}}^{\max}=2M_{\odot}$}\label{sub3}

Consider a NS with mass $M_{\rm{NS}}^{\max}=2M_{\odot}$ fixed but its radius $R$ could vary.
The central EOS ${P}_{\rm{c}}$-$\varepsilon_{\rm{c}}$ is determined by the correlation (\ref{sdd-1}).
Inversely, it gives,
\begin{equation}
P_{\rm{c}}(\varepsilon_{\rm{c}})\approx u^{2/3}\varepsilon_{\rm{c}}^{4/3}\left(1+4u^{2/3}\varepsilon_{\rm{c}}^{1/3}+19u^{4/3}\varepsilon_{\rm{c}}^{2/3}+\cdots\right),
\end{equation}
here $u/[\rm{fm}^{3/2}/\rm{MeV}^{1/2}]=(M_{\rm{NS}}^{\max}/M_{\odot}+0.106)/1730$\,\cite{CLZ23-a} depends on $M_{\rm{NS}}^{\max}$.
See the light-blue line shown in FIG.\,\ref{fig_EOS_sc2}.
The $\widehat{P}_{\rm{c}}$ eventually decreases with increasing radius $R$  (e.g., from 11\,km to 17\,km) since the NS becomes less compact.
The resulted $\widehat{P}_{\rm{c}}$'s are also shown in FIG.\,\ref{fig_EOS_sc2}, e.g., for $R\approx13\,\rm{km}$ we have $\widehat{P}_{\rm{c}}\approx0.20$, using the correlation (\ref{sdd-2}).

\begin{figure}[h!]
\centering
\includegraphics[width=8.6cm]{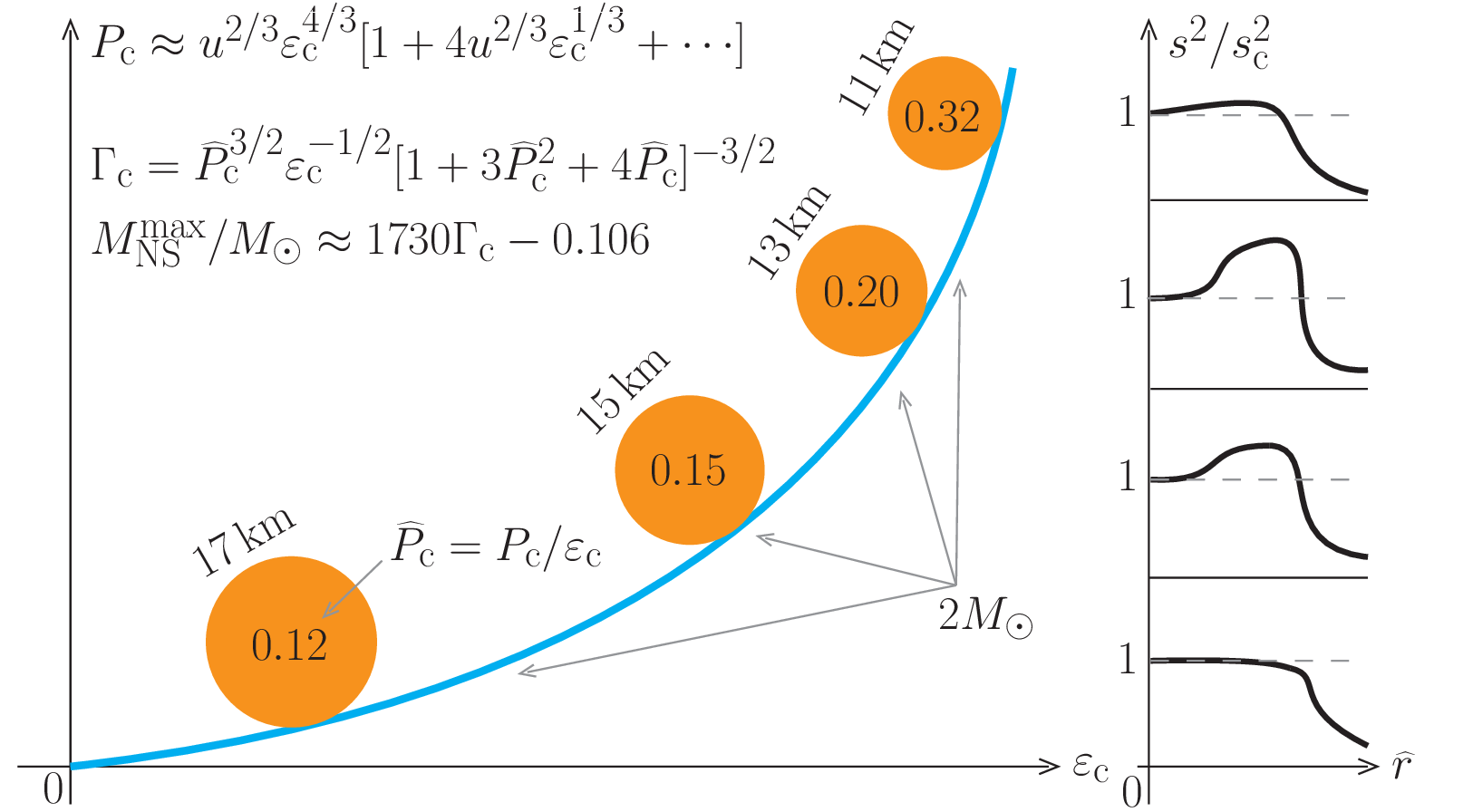}
\caption{(Color Online). Central EOS of a $M_{\rm{NS}}^{\max}=2M_{\odot}$ NS and the sketch of variation of the peaked behavior of $s^2$ near $\widehat{r}=0$, here $u/[\rm{fm}^{3/2}/\rm{MeV}^{1/2}]=(M_{\rm{NS}}^{\max}/M_{\odot}+0.106)/1730$\,\cite{CLZ23-a}.
}\label{fig_EOS_sc2}
\end{figure}

\begin{figure}[h!]
\centering
\includegraphics[width=7.cm]{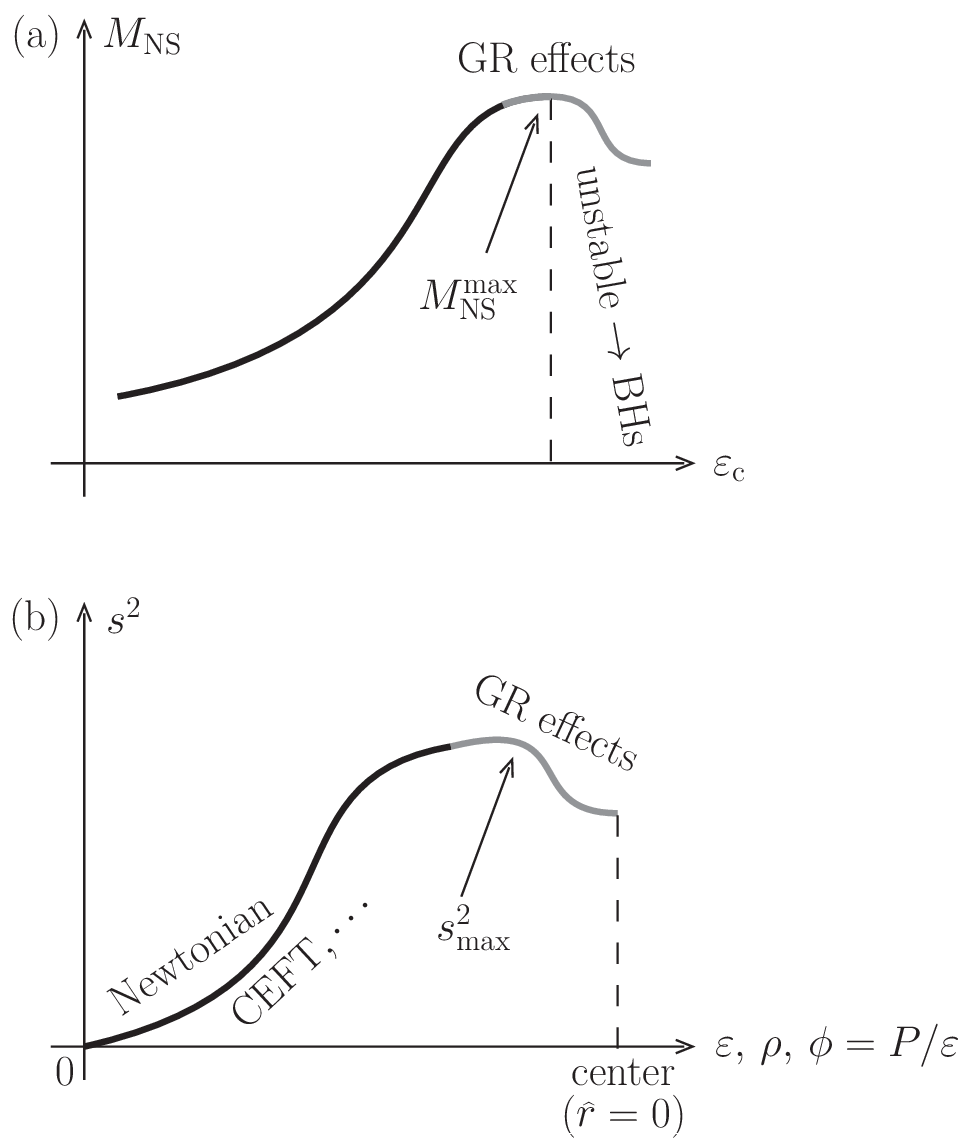}
\caption{(Color Online). Sketch of the $\varepsilon_{\rm{c}}$-dependence of NS mass $M_{\rm{NS}}$ (upper panel) and the energy density $\varepsilon$-,  the density $\rho$- or the $\phi$-dependence of $s^2$ (lower panel).
}\label{fig_s2peak_sk}
\end{figure}

We now have the following coherent picture for the peaked profile of $s^2$ in massive NSs.
There may exist a peak in $s^2$ near $\widehat{r}=0$ when $R$ is relatively small (e.g., 11\,km) and the peak becomes enhanced when $R$ increases a little further (e.g., to 13\,km or 15\,km). As a comparison, we have $\widehat{P}_{\rm{c}}\approx0.11\mbox{-}0.12$ for a 12-13\,km canonical NS of mass about $1.4M_{\odot}$. 
Our analysis indicates that there tends to be no peak in the $s^2$ profile near $\widehat{r}=0$ for these canonical NSs.
This is consistent with the finding of a recent analysis\,\cite{Ecker22} using as input in solving the TOV equations in the traditional approach the EOS from
nuclear theory and pQCD and imposing observational
constraints on NS masses, radii, and tidal deformabilities. In essence, the profile of $s^2$ is analogous to the M-R curve where the GR leads to a maximum-mass point beyond which the NSs are unstable against collapsing into black holes (BHs), sketched in the upper panel of FIG.\,\ref{fig_s2peak_sk}.
The strong-gravitational force makes $\phi=P/\varepsilon$ sizable $\gtrsim\mathcal{O}(0.1)$ and extrudes a peak in the $s^2$ profile, as sketched in the lower panel of FIG.\,\ref{fig_s2peak_sk} (and numerically demonstrated in FIG.\,\ref{fig_s2_peak_stat-af}).
If $R$ is extra-ordinarily large (e.g., 17\,km and correspondingly $\widehat{P}_{\rm{c}}\approx0.12$ from Eqs.\,(\ref{sdd-1}) and (\ref{sdd-2})), the peak in $s^2$ may reduce and even disappear as illustrated in the right column of FIG.\,\ref{fig_EOS_sc2}. 

We emphasize that not every EOS can induce a peaked $s^2$ profile. It is similar to the fact that not every EOS could induce a maximum mass for NSs below the causality limit, e.g.,  the ultra-relativistic Fermi gas predicts a linear M-R relation $M_{\rm{NS}}=3R/14$\,\cite{Lightman1975}. It should be pointed out that the observed masses and radii of massive NSs as well as properties of post-merger remnants of NSs from future high-frequency GW observations are expected to be useful for inferring features of the $s^2$ profile especially near $\widehat{r}=0$.

\subsection{Normally Stable NSs along the M-R Curve}\label{sub4}

For normally stable NSs (instead of the maximum-mass configuration $M_{\rm{NS}}^{\max}$) on the M-R curve, the expression for $b_2$ and the relation $a_2=b_2/s_{\rm{c}}^2$ do not change. However, the expression for $s_{\rm{c}}^2$ is modified from Eq.\,(\ref{def_sc2}) to\,\cite{CLZ23-b},
\begin{equation}\label{def_sc2_Psi}
s_{\rm{c}}^2=\widehat{P}_{\rm{c}}\left(1+\frac{1+\Psi}{3}\frac{1+3\widehat{P}_{\rm{c}}^2+4\widehat{P}_{\rm{c}}}{1-3\widehat{P}_{\rm{c}}^2}\right),
\end{equation}
where $
\Psi={2\varepsilon_{\rm{c}}}{M_{\rm{NS}}^{-1}}{\d M_{\rm{NS}}}/{\d\varepsilon_{\rm{c}}}
=2\d\ln M_{\rm{NS}}/\d\ln\varepsilon_{\rm{c}}>0$\,\cite{CLZ23-a}.
Taking $\varepsilon_{\rm{c}}\approx900\,\rm{MeV}/\rm{fm}^3$ and $\varepsilon_{\rm{c}}\approx400\,\rm{MeV}/\rm{fm}^3$ for a $2M_{\odot}$ NS and a canonical NS, respectively,  we find approximately $\Psi\approx0.88$.

\begin{figure}[h!]
\centering
\includegraphics[width=6.5cm]{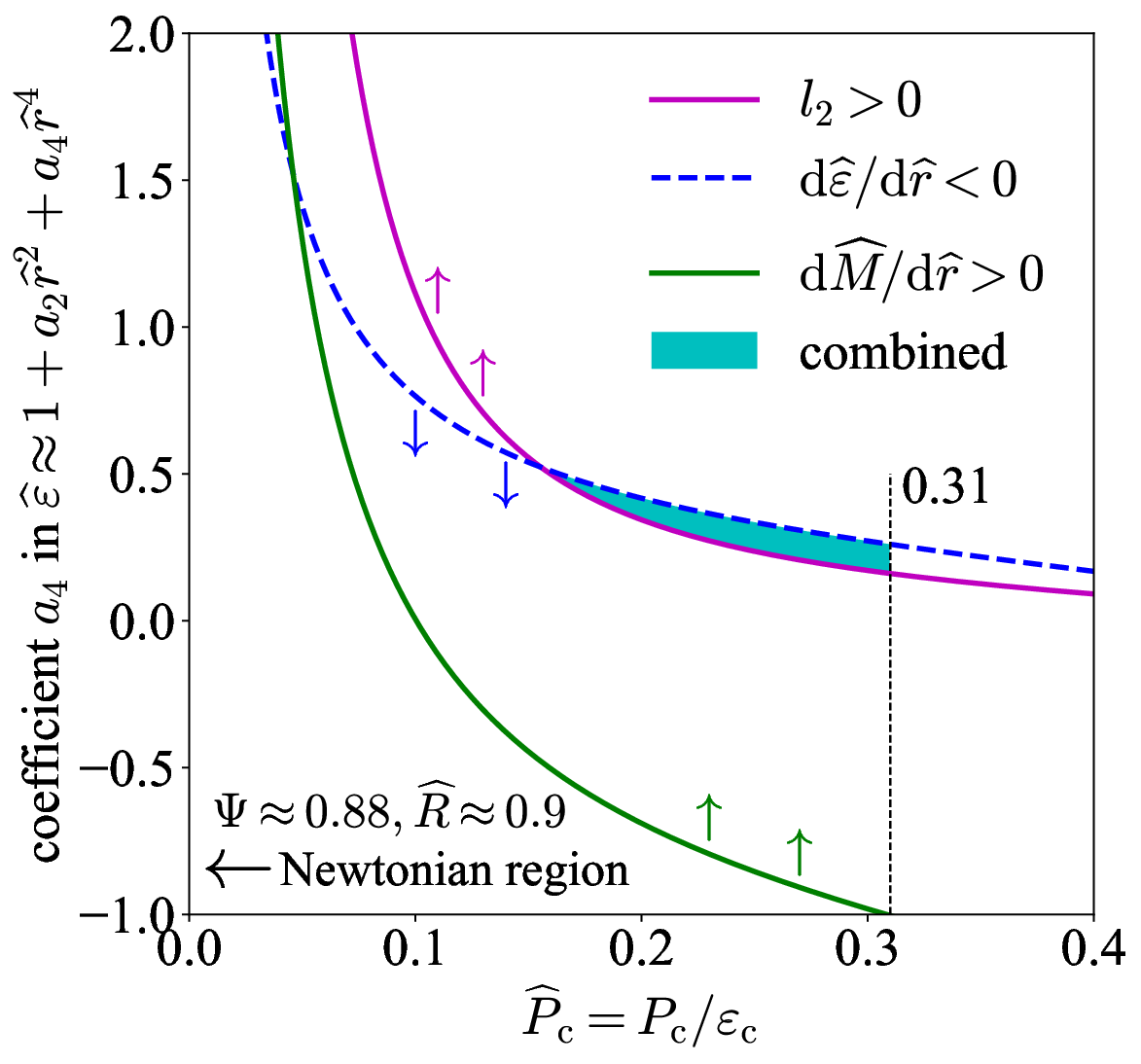}
\caption{(Color Online). Same as FIG.\,\ref{fig_a4peak} but for NS configurations climbing the M-R curve (instead of being at $M_{\rm{NS}}^{\max}$), here $\Psi=0.88$ is adopted.
}\label{fig_a4peak-psi}
\end{figure}

\begin{figure}[h!]
\centering
\includegraphics[width=7.6cm]{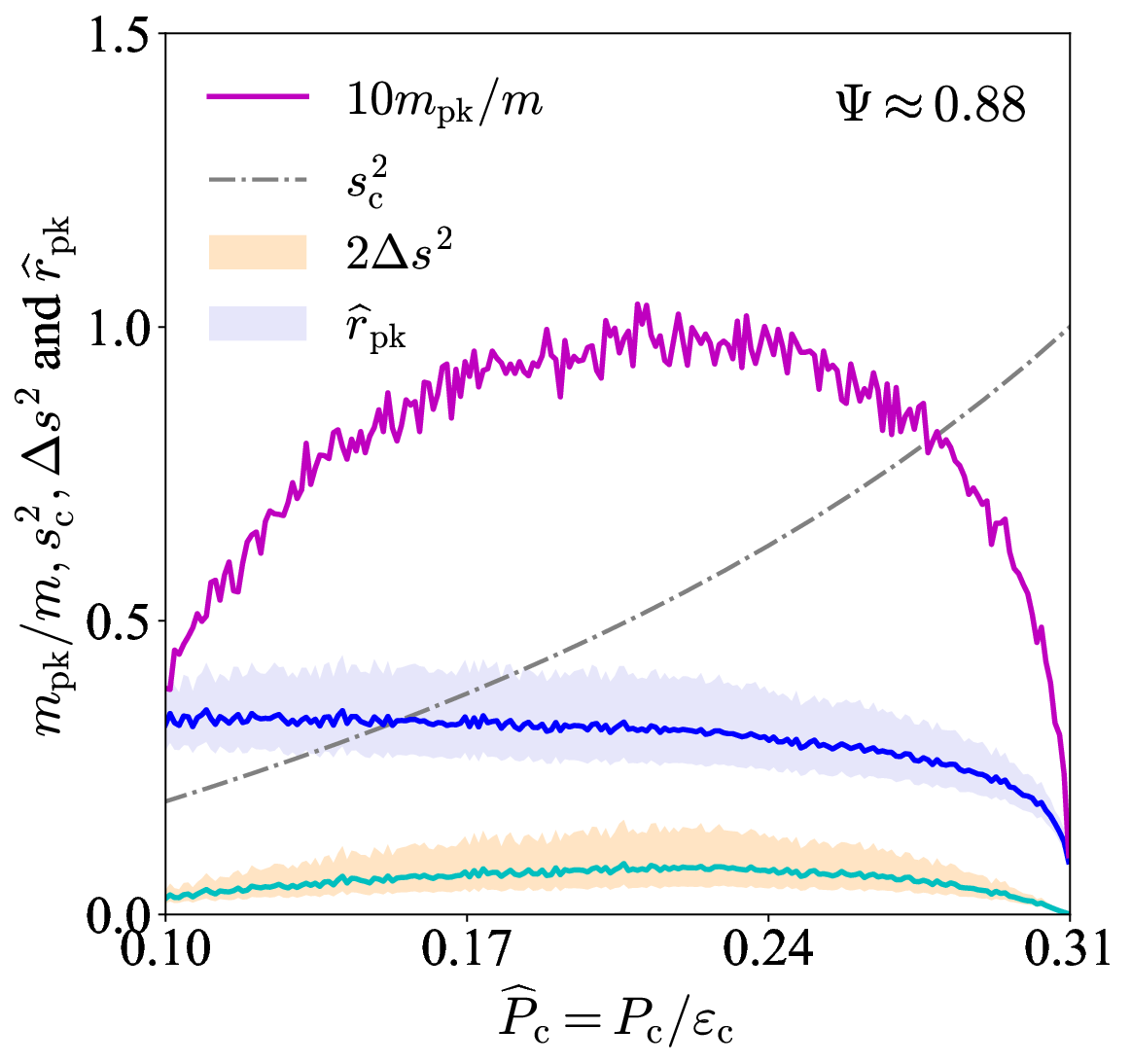}
\caption{(Color Online). Same as FIG.\,\ref{fig_s2_peak_stat_prop} but with $\Psi=0.88$.
}\label{fig_s2_peak_stat_prop-psi}
\end{figure}

Adopting the same criteria of Eq.\,(\ref{def_a4ineq}),  $a_4<-a_2/2\widehat{R}^2$ and $a_4\widehat{R}^4>-1-a_2\widehat{R}^2$,  we can analyze the allowed region for $a_4$ for the onset of a peak in $s^2$ near $\widehat{r}=0$.
The result is shown in FIG.\,\ref{fig_a4peak-psi} with $\Psi=0.88$.
We find that the combined region for $a_4$ is similar to the one shown in FIG.\,\ref{fig_a4peak} for $M_{\rm{NS}}^{\max}$.
In this case,  one has $\widehat{P}_{\rm{c}}\lesssim0.31$ by requiring $s_{\rm{c}}^2\leq1$.
Similarly, we could analyze the probability for the $s^2$ profile to have a peak near $\widehat{r}=0$ as done for the $M_{\rm{NS}}^{\max}$ configuration in subsection \ref{sub2}, and the result is shown in FIG.\,\ref{fig_s2_peak_stat_prop-psi}.
We find that the probability is relatively large for $0.17\lesssim\widehat{P}_{\rm{c}}\lesssim0.25$ (compared with its low-$\widehat{P}_{\rm{c}}$ surroundings), while on the other hand the location of the peak $\widehat{r}_{\rm{pk}}$ (lavender band) and the enhancement $\Delta s^2$ (tan band) are very similar to those of FIG.\,\ref{fig_s2_peak_stat_prop}.
Combining the results of FIG.\,\ref{fig_a4peak-psi} and FIG.\,\ref{fig_s2_peak_stat_prop-psi} and 
considering $\widehat{P}_{\rm{c}}\lesssim0.14$\,\cite{Brandes23-a} for a canonical NS,  we see that the probability of occurrence of a peak in $s^2$ for such NSs is probably very low.
These features are consistent with the findings of a recent analysis in Ref.\,\cite{Ecker22}, which predicted that the $s^2$ in canonical NSs is a monotonically decreasing function of $\widehat{r}$.

\section{Connections with Some Other Predictions in the Literature}\label{SEC_COM}

We discuss here some possible relations with other alternatives on the peaked $s^2$ profiles predicted in the literature.
If the asymptotic pQCD constraint $s^2\leq1/3$\,\cite{Fuji22} at extremely large densities $\rho_{\rm{pQCD}}\sim40\rho_{\rm{sat}}$\,\cite{Bjorken83,Kur10,Gorda21,Gorda23-a} is considered, then there would essentially be a peak in $s^2(\varepsilon)$ if the $s^2\gtrsim1/3$\,\cite{Bed15} in the cores of massive NSs where $\rho_{\rm{c}}\approx\rho_{\max}\approx (4\mbox{-}8)\rho_{\rm{sat}}$, in order to extrapolate together the two $s^2$'s. Of course, this does not necessarily mean that the $s^2(\widehat{r})$ profile may have a peaked structure in NSs\,\cite{Mro23}.
Our analysis above shows that even without considering the pQCD prediction, there would be a possible intrinsic peak in $s^2$ in massive NS cores, fundamentally due to the strong GR effects. FIG.\,\ref{fig_s2pQCD} illustrates a few possible ways to extrapolate the $s^2$ in NSs predicted by the TOV equations to the asymptotic pQCD limit. The double circles indicate the central $s^2$ in NSs. Here the peaked behavior in $s^2$ of case (a) is due to the desire to meet the pQCD prediction (i.e., no peak in $s^2$ at densities within NSs), whiles in cases (b) and (c) the peak is originated from the GR effects and is already located somewhere within NSs. The actual outcome depends on the EOS model. Case (d) may correspond to a low-mass NS (and correspondingly $\widehat{P}_{\rm{c}}$ is small) whose $s^2$ is always smaller than 1/3.
See Ref.\,\cite{Tews18} for a similar discussion.

\begin{figure}[h!]
\centering
\includegraphics[width=8.0cm]{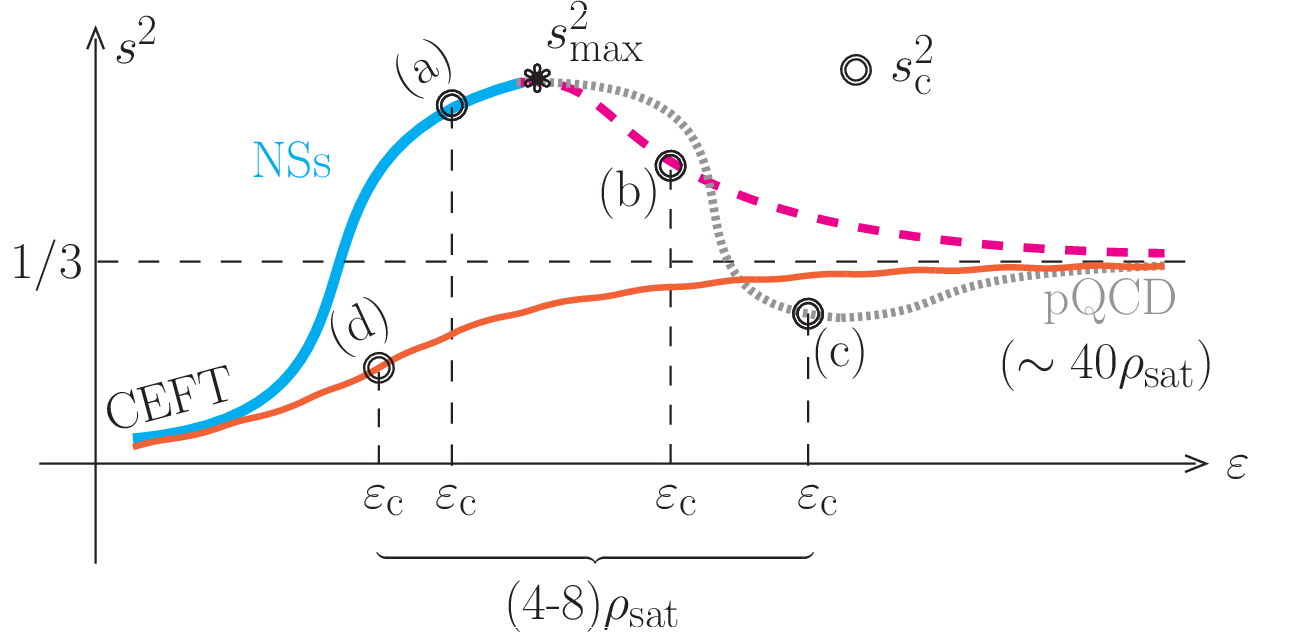}
\caption{(Color Online). Different patterns of extrapolating the $s^2$ in the core of NSs with $\rho_{\rm{c}}\approx\rho_{\max}\approx(4\mbox{-}8)\rho_{\rm{sat}}$ to high densities $\sim40\rho_{\rm{sat}}$ (pQCD),  and the double-circle on each line denotes its $s_{\rm{c}}^2$.
{Here,  the $s^2$ in patterns (a) and (d) shows a monotonic behavior and the difference lies in whether $s_{\rm{c}}^2$ is greater (smaller) than $1/3$; the $s^2$ in patterns (b) and (c) shows a peak at densities smaller than the central density of NSs where the $s_{\rm{c}}^2$ in pattern (b) (pattern (c)) is larger (smaller) than $1/3$.
Both pattern (b) and (c) indicate a continuous crossover behavior near the NS center.}
Other possible nontrivial features in $s^2$ (like plateau, spike or bump, etc.) are not sketched in the figure.
}\label{fig_s2pQCD}
\end{figure}

As we study the peaked structure in the $s^2$ profile near $\widehat{r}=0$ or $\widehat{\varepsilon}=1$ in this work, we would like to point out that other types of structures such as plateau, bump and spike, etc., may also exist in the $s^2(\rho)$ profiles of NSs\,\cite{Tan22,Tan22-a}, see the very recent Ref.\,\cite{Mro23} for relevant discussions. Moreover,  nontrivial features of $s^2$ are predicted to emerge in relativistic heavy-ion collision experiments\,\cite{Olii23,Kut22} where similarly high densities as in NSs are reached but at high temperatures and with little gravity effect. Moreover, a peaked profile of $s^2$ was also predicted recently in isospin-asymmetric strongly interacting matter simulated on a lattice\,\cite{Bra23,Aya23}. In this case, $\varepsilon\sim\mathcal{O}(10^{1\mbox{-}2})\,\rm{MeV}/\rm{fm}^3$ is an order-of-magnitude smaller than that in NSs. Nevertheless,  $\phi=P/\varepsilon$ is sizable $\gtrsim\mathcal{O}(0.1)$ in these systems, e.g.,  $P/\varepsilon\gtrsim0.2$ at the peak position of $s^2$\,\cite{Bra23}.  This means that the ``effective gravitational force'' is made strong enough by compressing the matter into a tiny volume. 

These examples and our analysis above all point to the same fundamental Hamilton's variational principle at work on the total action of any system under consideration\,\cite{NAP2011}. It would thus be exciting to investigate whether and how different ingredients with various forms of the system including space, strong-field gravity (with possibly modified gravity besides Einstein's GR) and matter (including nuclear/hyper/quark matter, dark matter and energy) as well as their couplings could influence the system's $s^2$ profile. Supermassive NSs are among the possible sites for such studies, see, e.g. Refs.\,\cite{Soren2023,He14,Li19} and references therein.

\section{Summary and conclusions}\label{SEC_SUM}

In summary, we investigated systematically the $s^2$ profile especially the nature of its possible peak in massive NSs in a new approach that is independent of the nuclear EOS model and without any presumption about the NS structure and/or composition. In terms of the small quantities $\widehat{r}$ (or equivalently $\mu=\widehat{\varepsilon}-1$) near NS centers  and $\widehat{P}_{\rm{c}}\equiv P_{\rm{c}}/\varepsilon_{\rm{c}}$, we performed the double-element perturbative expansions in solving perturbatively the scaled TOV equations and analyzing the $s^2$ profiles of NSs from the Newtonian limit to the GR case. The results obtained are intrinsic properties of the GR stellar structure equations. The relevant expansions developed in this work provide convergent and universal predictions independent of the still very uncertain EOS of supradense neutron-rich matter in massive NSs. 

Among the many interesting findings, the GR term in the TOV equations was found particularly to play a twofold role: it compresses NS matter and modifies $\widehat{P}_{\rm{c}}$ from small $\lesssim\mathcal{O}(10^{-4})$ in Newtonian stars to $\gtrsim\mathcal{O}(0.1)$ in massive NSs ($R\approx12\mbox{-}14\,\rm{km}$) and can generate a peak in $s^2$ profile near the centers of massive NSs, and eventually takes away the peak in extremely compact/massive NSs ($R\lesssim11\,\rm{km}$) whose $\widehat{P}_{\rm{c}}$ approaches the causality limit. {In essence, the strong-field gravity in massive NSs naturally extrudes peaks in the $s^2$ radial profiles. The nuclear EOSs are required to match these profiles to satisfy the TOV equations. Consequently, a continuous crossover characterized by a smooth reduction of $s^2$ is expected to occur in cores of massive NSs like PSR J0740+6620 with their radii being around 12-14\,km. Moreover, due to the facts that $s_{\rm{c}}^2\neq0$ and $\gamma_{\rm{c}}=s_{\rm{c}}^2\varepsilon_{\rm{c}}/{P}_{\rm{c}}\gtrsim1.33$, a sharp phase transition signaled by an abruptly vanishing of $s^2$ near the center is basically excluded and the dense matter in NS cores could hardly be conformal signaled by $\gamma_{\rm{c}}\to1$.}

\section*{Acknowledgements}We would like to thank James Lattimer and Zhen Zhang for helpful discussions.
This work is supported in part by the U.S. Department of Energy, Office of Science, under Award No. DE-SC0013702, the CUSTIPEN (China-U.S. Theory Institute for Physics with Exotic Nuclei) under
US Department of Energy Grant No. DE-SC0009971.

\appendix
\renewcommand\theequation{\Alph{section}\arabic{equation}}
\renewcommand\thefigure{\Alph{section}\arabic{figure}}

\section{Proof on Absence of Odd Terms in the Perturbative Expansions of  $\widehat{\varepsilon}$ and $\widehat{P}$ over $\widehat{r}$}\label{App0}
{
The evenness/oddness of reduced energy density,  pressure and mass as functions of the reduced radius were proved in Ref.\,\cite{CLZ23-a} by explicitly calculating individually the relevant expansion coefficients in perturbatively solving the dimensionless TOV equations using polynomials. Here we provide an alternative and more general mathematical proof by exploring properties of these functions under a coordinate transformation satisfying physical requirements.

First consider Newtonian stars.
The reduced mass $\widehat{M}$ as a function of radial distance $\widehat{r}$ from the center could be written as,
\begin{align}\label{ok-1}
\widehat{M}(\widehat{r})=\int_0^{\widehat{r}}\d xx^2\widehat{\varepsilon}(x),
\end{align}
here $x$ is an integration variable.
Let's make the coordinate transformation as follows,
\begin{equation}
x\to-x,
\end{equation}
then  the $\widehat{M}(\widehat{r})$ transforms as,
\begin{equation}\label{ok-2}
\widehat{M}(\widehat{r})=\int_0^{-\widehat{r}}(-\d x)(-x)^2\widehat{\varepsilon}(-x)
=-\int_0^{-\widehat{r}}\d xx^2\widehat{\varepsilon}(-x).
\end{equation}
On the other hand,  we have straightforwardly from Eq.\,(\ref{ok-1}) that,
\begin{equation}\label{ok-3}
\widehat{M}(-\widehat{r})=\int_0^{-\widehat{r}}\d xx^2\widehat{\varepsilon}(x).
\end{equation}
Similarly, starting from the pressure equation, namely
\begin{equation}
\widehat{P}(\widehat{r})=-\int_0^{\widehat{r}}\d x\frac{\widehat{\varepsilon}(x)\widehat{M}(x)}{x^2},
\end{equation}
we shall obtain
\begin{align}\label{oq-2}
\widehat{P}(\widehat{r})=&+\int_0^{-\widehat{r}}\d x\frac{\widehat{\varepsilon}(-x)\widehat{M}(-x)}{x^2},\\
\widehat{P}(-\widehat{r})=&-\int_0^{-\widehat{r}}\d x\frac{\widehat{\varepsilon}(x)\widehat{M}(x)}{x^2}.
\end{align}
In order that both Eq.\,(\ref{ok-2}) and Eq.\,(\ref{ok-3}) hold, only two possibilities exist:
\begin{enumerate}[leftmargin=*,label=(\alph*)]
\item[(a)] $\widehat{\varepsilon}(-x)=\widehat{\varepsilon}(x)$ and $\widehat{M}(-x)=-\widehat{M}(x)$,  we have in this case  $\widehat{P}(-x)=\widehat{P}(x)$, or
\item[(b)]$\widehat{\varepsilon}(-x)=-\widehat{\varepsilon}(x)$ and $\widehat{M}(-x)=\widehat{M}(x)$, now we still have $\widehat{P}(-x)=\widehat{P}(x)$.
\end{enumerate}
We therefore know at once that $\widehat{P}(\widehat{r})=\widehat{P}(-\widehat{r})$.
In addition, since we have the physical requirement that $\widehat{\varepsilon}(0)=1$ at $\widehat{r}=0$, only the possibility (a) above is allowed as the option (b) would lead to $\widehat{\varepsilon}(0)=0$ that is unphysical. This means that $\widehat{\varepsilon}(\widehat{r})=\widehat{\varepsilon}(-\widehat{r})$ and $\widehat{M}(\widehat{r})=-\widehat{M}(-\widehat{r})$.

For NSs, the mass evolution equation remains the same as Eq.\,(\ref{ok-1}), therefore we can similarly infer that $\widehat{\varepsilon}(\widehat{r})=\widehat{\varepsilon}(-\widehat{r})$ and $\widehat{M}(\widehat{r})=-\widehat{M}(-\widehat{r})$.
Looking back into the pressure evolution of Eqs.\,(\ref{def-1}),  we shall infer that $\widehat{P}(\widehat{r})$ has the same properties under the transformation $\widehat{r}\to-\widehat{r}$ as $\widehat{\varepsilon}(\widehat{r})$ or  as $\widehat{r}^3/\widehat{M}(\widehat{r})$, considering the factor $1+\widehat{P}(\widehat{r})/\widehat{\varepsilon}(\widehat{r})$ or factor $1+\widehat{r}^3\widehat{P}(\widehat{r})/\widehat{M}(\widehat{r})$, respectively.
Specifically,  we have
\begin{align}
\widehat{P}(\widehat{r})=-&\int_0^{-\widehat{r}}\d x\frac{\widehat{\varepsilon}(x)\widehat{M}(x)}{x^2}\notag\\
&\times
\frac{[1+\widehat{P}(-x)/\widehat{\varepsilon}(x)][1+x^3\widehat{P}(-x)/\widehat{M}(x)]}{1-2\widehat{M}(x)/x},\label{kq-1}\\
\widehat{P}(-\widehat{r})=-&\int_0^{-\widehat{r}}\d x\frac{\widehat{\varepsilon}(x)\widehat{M}(x)}{x^2}\notag\\
&\times\frac{[1+\widehat{P}(x)/\widehat{\varepsilon}(x)][1+x^3\widehat{P}(x)/\widehat{M}(x)]}{1-2\widehat{M}(x)/x},\label{kq-2}
\end{align}
where Eq.\,(\ref{kq-1}) is obtained by transforming $x\to-x$ in the integration of $\widehat{P}(\widehat{r})$ while Eq.\,(\ref{kq-2}) is the pressure at $-\widehat{r}$.
Thus $\widehat{P}(\widehat{r})=\widehat{P}(-\widehat{r})$ is inferred from Eqs.\,(\ref{kq-1}) and (\ref{kq-2}).

These analyses show that $\widehat{\varepsilon}$ and $\widehat{P}$ are even functions of $\widehat{r}$ and $\widehat{M}$ is an odd function of $\widehat{r}$, though physically $\widehat{r}$ is non-negative.
Consequently, no odd (even) terms in $\widehat{r}$ could appear in the expansions of $\widehat{\varepsilon}$ and $\widehat{P}$ (of $\widehat{M}$).}

\setcounter{figure}{0}
\section{Estimating the Maximum Size of $a_4$ Using Empirical Parameters of Nuclear EOS}\label{App}

We pointed out in the main text that the coefficient $a_4$ appearing in $\widehat{\varepsilon}\approx 1+a_2\widehat{r}^2+a_4\widehat{r}^4+\cdots$ encapsulates certain uncertainties of the dense matter EOS.
Once the EOS model is adopted (e.g., a nuclear matter model or a quark matter model), the $a_4$ could be worked out. There is no {\it a prior} that $a_4$ should be either positive or negative, though $a_4>0$ is a necessary condition for inducing a peaked $s^2$ near the center, see the discussion given around Eq.\,(\ref{s2app}).
Here we (approximately) express $a_4$ using a meta EOS model for NS matter and study under which circumstances $a_4$ shall become positive.
Our illustration is qualitative with the motivation to demonstrate how the nuclear parameters may affect $a_4$. Within the currently known uncertainty ranges of these EOS parameters, we then estimate the maximum size (absolute value) of $a_4$.

We start by considering the EOS of asymmetric nuclear matter (ANM) to order $\delta^2$\,\cite{LCK08},
\begin{equation}
E(\rho)\approx E_0(\rho)+E_{\rm{sym}}(\rho)\delta^2,
\end{equation}
where $\rho=\rho_{\rm{n}}+\rho_{\rm{p}}$ is the total density, and
\begin{equation}
\delta=(\rho_{\rm{n}}-\rho_{\rm{p}})/(\rho_{\rm{n}}+\rho_{\rm{p}}),
\end{equation} 
is the isospin asymmetry between neutrons ($\rho_{\rm{n}}$) and protons ($\rho_{\rm{p}}$). The EOS $E_0(\rho)$ of symmetric nuclear matter (SNM) and the symmetry energy $E_{\rm{sym}}(\rho)$ could further be expanded around the nuclear saturation density $\rho_{\rm{sat}}\equiv\rho_0$ as\,\cite{LCK08,ZLi23},
\begin{align}
E_0(\rho)\approx &E_0(\rho_0)+2^{-1}K_0\chi^2+6^{-1}J_0\chi^3,\\
E_{\rm{sym}}(\rho)\approx&S+L\chi+2^{-1}K_{\rm{sym}}\chi^2+6^{-1}J_{\rm{sym}}\chi^3.
\end{align}
Here $S\equiv E_{\rm{sym}}(\rho_0)\approx30\pm4\,\rm{MeV}$ is the magnitude of the symmetry at $\rho_{\rm{sat}}$,  $K_0\approx230\pm20\,\rm{MeV}$ is the incompressibility coefficient of SNM, $J_0\approx-300\pm300\,\rm{MeV}$ is the skewness parameter of SNM, $L\approx60\pm20\,\rm{MeV}$, $K_{\rm{sym}}\approx-100\pm200\,\rm{MeV}$ and $J_{\rm{sym}}\approx0\mbox{-}800\,\rm{MeV}$ are the slope, curvature and skewness of the symmetry energy\,\cite{LCXZ21}. Moreover, the dimensionless quantity $\chi$ is defined as,
\begin{equation}\label{ddef-chi}
\chi=({\rho-\rho_0})/{3\rho_0}
=({\widehat{\rho}-\widehat{\rho}_0})/{3\widehat{\rho}_0}.
\end{equation}
In the second relation, we introduce the reduced densities $\widehat{\rho}=\rho/\rho_{\rm{c}}$ and $\widehat{\rho}_0=\rho_0/\rho_{\rm{c}}$, with $\rho_{\rm{c}}$ the central density in NSs.
The reduced density $\widehat{\rho}$ is given by\,\cite{CLZ23-b},
\begin{align}
\widehat{\rho}\approx&1+\left(\frac{b_2/s_{\rm{c}}^2}{1+\widehat{P}_{\rm{c}}}\right)\widehat{r}^2+\frac{1}{1+\widehat{P}_{\rm{c}}}\left(a_4-\frac{{b_2^2}/{2s_{\rm{c}}^2}}{1+\widehat{P}_{\rm{c}}}\right)\widehat{r}^4\notag\\
\equiv& 1+\eta\widehat{r}^2+(\beta a_4-\xi)\widehat{r}^4
,\label{pk-4}
\end{align}
via the thermodynamic relation $\rho\partial\varepsilon/\partial\rho=P+\varepsilon$.
The second line defines the coefficients $\eta$, $\beta$ and $\xi$.

\begin{figure*}
\centering
\includegraphics[height=4.cm]{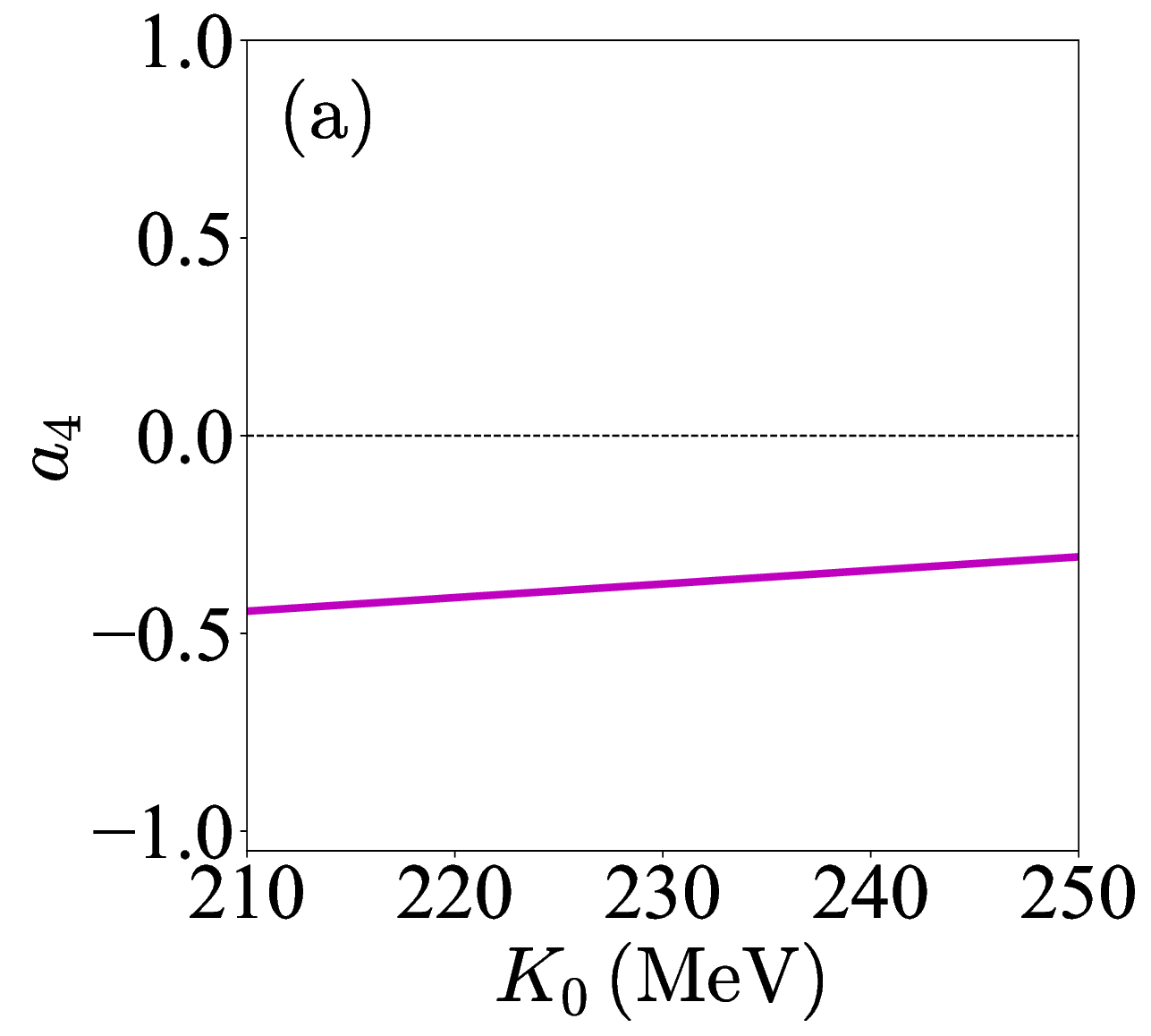}
\includegraphics[height=4.cm]{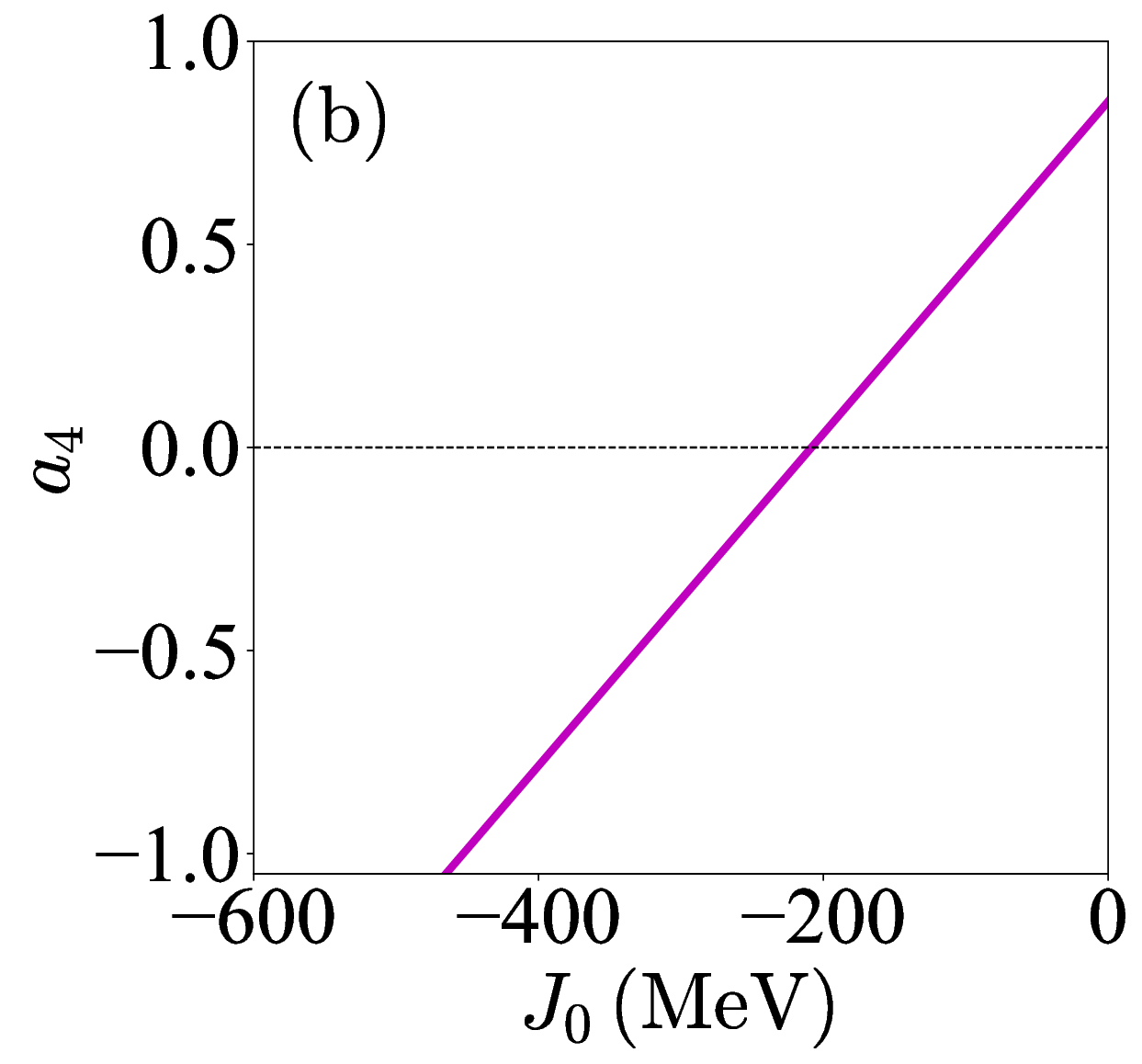}
\includegraphics[height=4.cm]{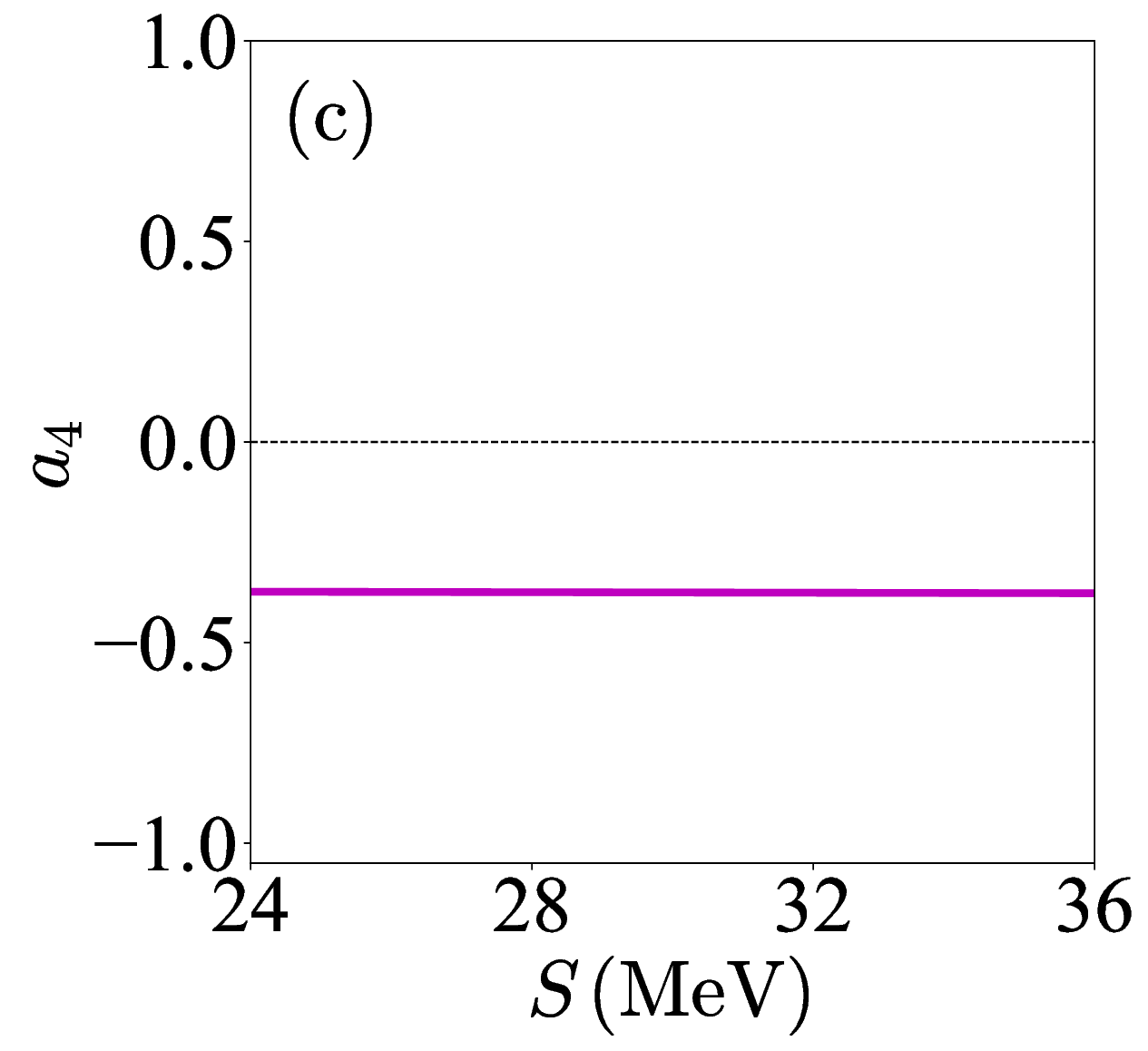}\\
\hspace*{0.0cm}
\includegraphics[height=4.cm]{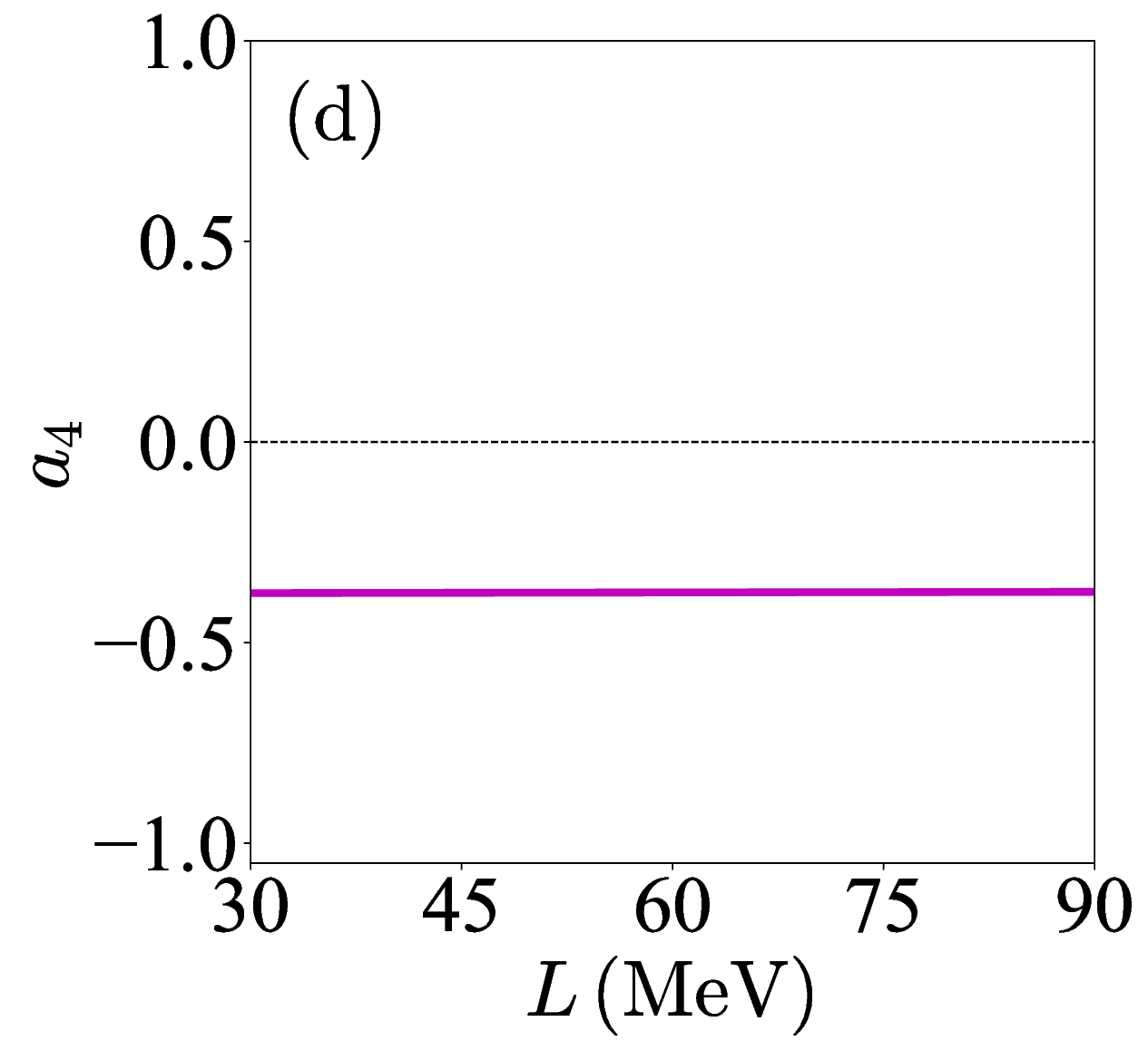}
\hspace*{0.0cm}
\includegraphics[height=4.cm]{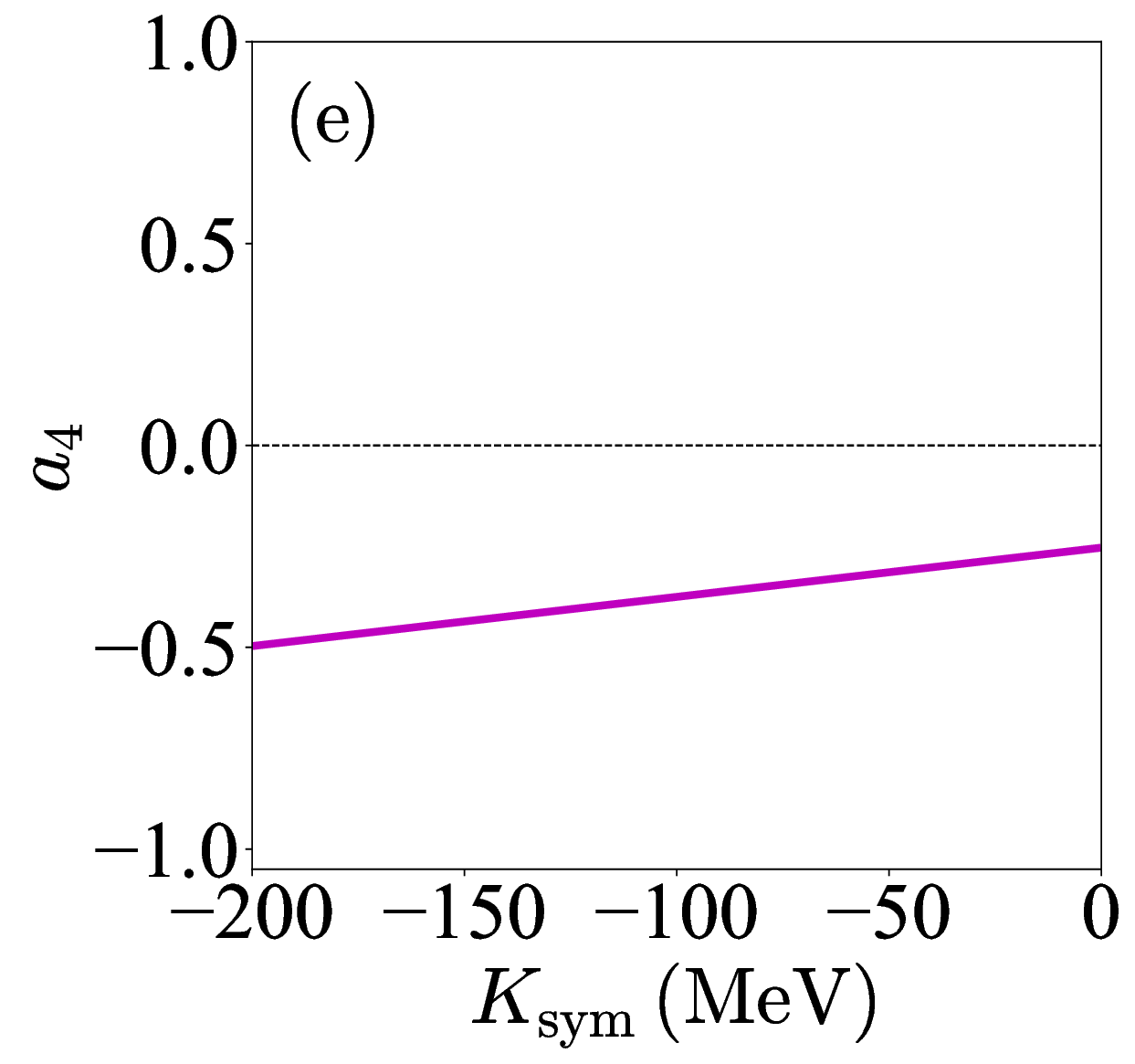}
\hspace*{0.0cm}
\includegraphics[height=4.cm]{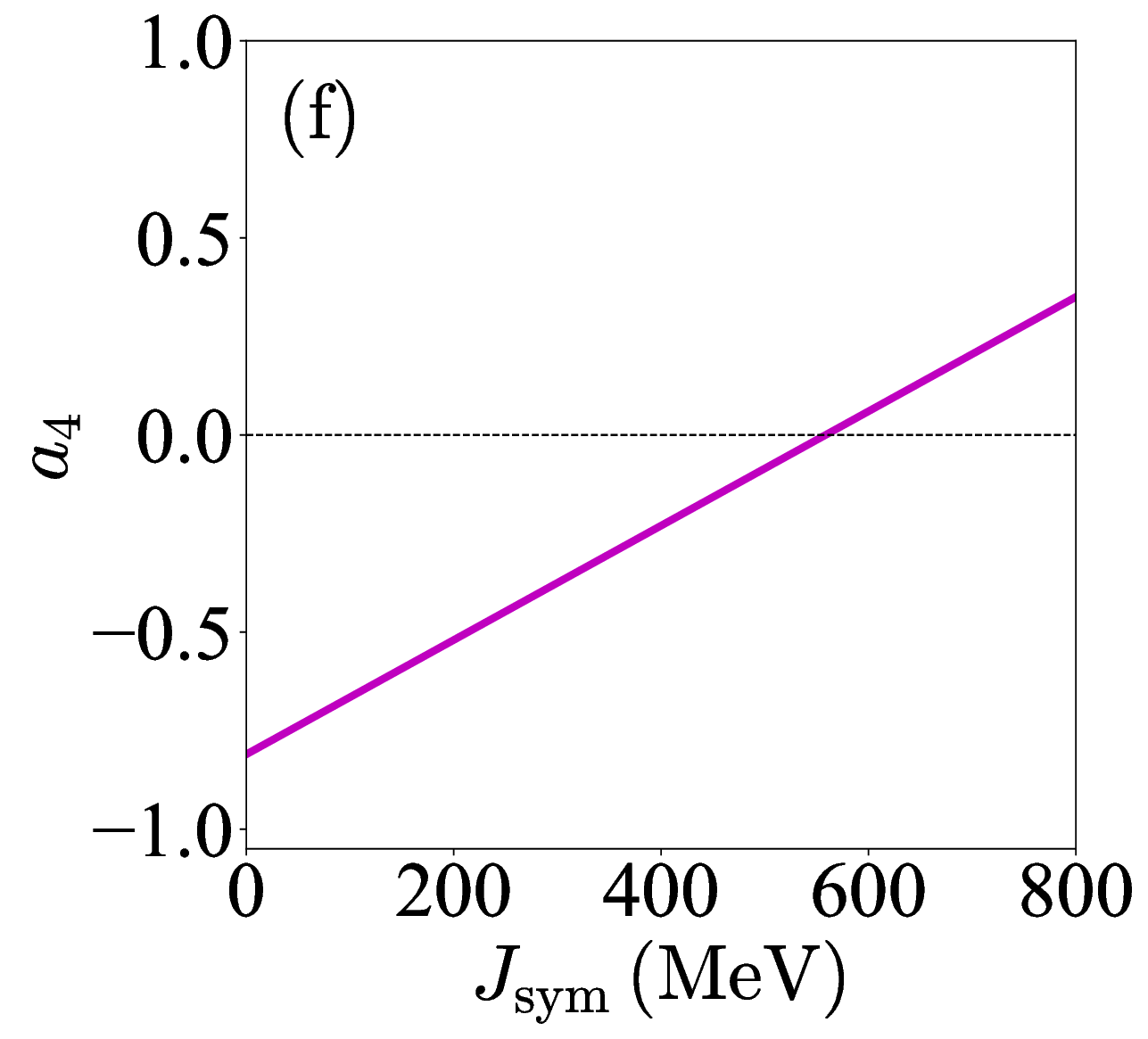}
\caption{(Color Online).  Dependence of $a_4$ on the empirical EOS parameters $K_0$, $J_0$, $S$, $L$, $K_{\rm{sym}}$ and $J_{\rm{sym}}$ according to Eq.\,(\ref{fff}),  see text for the details on the choice/determination of the parameters.
}\label{fig_a4_ll}
\end{figure*}

Based on these relations, we can express the energy density $\varepsilon=[E(\rho)+M_{\rm{N}}]\rho$, or its reduced form as
\begin{align}\label{ddef-eps}
\widehat{\varepsilon}
={\varepsilon}/{\varepsilon_{\rm{c}}}
=&[E(\rho)+M_{\rm{N}}]({\rho_{\rm{c}}}/{\varepsilon_{\rm{c}}})\widehat{\rho}.
\end{align}
Putting the $\widehat{\rho}$ of Eq.\,(\ref{pk-4}) into $\chi$ of Eq.\,(\ref{ddef-chi}) and further putting $\chi$ back into $E_0(\rho)$ as well as $E_{\rm{sym}}(\rho)$ gives the $\widehat{\varepsilon}$ of Eq.\,(\ref{ddef-eps}) as a function of $\widehat{r}^2$, i.e.,
\begin{align}\label{deff-2}
\widehat{\varepsilon}\approx& \left(\frac{\rho_{\rm{c}}}{\varepsilon_{\rm{c}}}\right)
\left[M_{\rm{N}}+E_0(\rho_0)+2^{-1}K_0\chi^2+6^{-1}J_0\chi^3\right.\notag\\
&\hspace*{1cm}
\left.+\left(S+L\chi+2^{-1}K_{\rm{sym}}\chi^2+6^{-1}J_{\rm{sym}}\chi^3\right)\delta^2\right].
\end{align}
For our purpose, we need the expression of the $\widehat{r}^4$-term in Eq.\,(\ref{deff-2}) which itself involves $a_4$ due to the expression for $\widehat{\rho}$ of Eq.\,(\ref{pk-4}).
Equaling it with $a_4\widehat{r}^4$ gives a self-consistent equation for $a_4$ (both sides of which involve $a_4$).

The expression for $a_4$ is quite complicated and since the magnitude of $\overline{M}_{\rm{N}}\equiv M_{\rm{N}}+E_0(\rho_0)\approx923\,\rm{MeV}$ (using $M_{\rm{N}}\approx939\,\rm{MeV}$ and $E_0(\rho_0)\approx-16\,\rm{MeV}$) is generally larger than $K_0\approx230\,\rm{MeV}$, $J_0\approx-300\,\rm{MeV}$, $S\approx30\,\rm{MeV}$, $L\approx 60\,\rm{MeV}$, $K_{\rm{sym}}\approx-100\,\rm{MeV}$ and $J_{\rm{sym}}\approx300\,\rm{MeV}$\,\cite{LCXZ21}, one can introduce small quantities $K_0/\overline{M}_{\rm{N}}$, $J_0/\overline{M}_{\rm{N}}$, $S/\overline{M}_{\rm{N}}$, $L/\overline{M}_{\rm{N}}$, $K_{\rm{sym}}/\overline{M}_{\rm{N}}$ and $J_{\rm{sym}}/\overline{M}_{\rm{N}}$ to expand $a_4$ to dig out the main features of its dependence on the empirical EOS characteristics:
\begin{widetext}
\begin{align}
a_4\approx&-\frac{\Lambda}{54(\Lambda\beta-1)^2\widehat{\rho}_0^3}
\left\{+
2
\left(\left(\Lambda\beta-1\right)\eta^2+\frac{2\beta\xi}{3}\right)
\left[\left(\frac{J_0}{\overline{M}_{\rm{N}}}\right)+
\left(\frac{J_{\rm{sym}}}{\overline{M}_{\rm{N}}}\right)\delta^2
\right]\right.\notag\\
&\hspace*{1.cm}-3\left(\left(\Lambda\beta-1\right)\eta^2+\beta\xi\right)
\left[\left(\frac{J_0}{\overline{M}_{\rm{N}}}\right)-3\left(\frac{K_0}{\overline{M}_{\rm{N}}}\right)
+\left(\left(\frac{J_{\rm{sym}}}{\overline{M}_{\rm{N}}}\right)-3\left(\frac{K_{\rm{sym}}}{\overline{M}_{\rm{N}}}\right)\right)\delta^2\right]\widehat{\rho}_0\notag\\
&\hspace*{1.cm}
+\left(\left(\Lambda\beta-1\right)\eta^2+2\beta\xi\right)
\left[\left(\frac{J_0}{\overline{M}_{\rm{N}}}\right)-6\left(\frac{K_0}{\overline{M}_{\rm{N}}}\right)
+\left(
\left(\frac{J_{\rm{sym}}}{\overline{M}_{\rm{N}}}\right)-6\left(\frac{K_{\rm{sym}}}{\overline{M}_{\rm{N}}}\right)
+18\left(\frac{L}{\overline{M}_{\rm{N}}}\right)
\right)\delta^2\right]\widehat{\rho}_0^2\notag\\
&\left.\hspace*{1.cm}
-54\beta\xi\left[\left(\Lambda\beta-1\right)+\frac{1}{162}\left(\frac{J_0}{\overline{M}_{\rm{N}}}\right)
-\frac{1}{18}\left(\frac{K_0}{\overline{M}_{\rm{N}}}\right)
+\left(
\frac{1}{162}\left(\frac{J_{\rm{sym}}}{\overline{M}_{\rm{N}}}\right)-\frac{1}{18}\left(\frac{K_{\rm{sym}}}{\overline{M}_{\rm{N}}}\right)+\frac{1}{3}\left(\frac{L}{\overline{M}_{\rm{N}}}\right)-\left(\frac{S}{\overline{M}_{\rm{N}}}\right)
\right)\delta^2\right]\widehat{\rho}_0^3\right\},\label{fff}
\end{align}
\end{widetext}
here $\Lambda=\overline{M}_{\rm{N}}\rho_{\rm{c}}/\varepsilon_{\rm{c}}$.
For a given set of nuclear EOS parameters $(K_0,J_0,S,L,K_{\rm{sym}},J_{\rm{sym}})$ and a central energy density $\varepsilon_{\rm{c}}$ (or a central density $\rho_{\rm{c}}$), 
the NS mass as well as its radius could be obtained.
The isospin asymmetry $\delta$ in charge neutral npe matter at $\beta$-equilibrium is determined by $4\delta E_{\rm{sym}}(\rho)\approx\mu_{\rm{e}}\approx[3\pi^2\rho(1-\delta)/2]^{1/3}$, here $\mu_{\rm{e}}$ is the electron chemical potential.

A main feature of Eq.\,(\ref{fff}) is that $J_0$ and $J_{\rm{sym}}$ are the leading-order contributions to $a_4$, since they appear at $\widehat{\rho}_0^0$ order in the curly brackets. Other characteristics such as $K_0$, $L$,  $K_{\rm{sym}}$, etc., appear at higher-orders of $\widehat{\rho}_0$.
We take $\rho_{\rm{c}}\approx5\rho_0\approx0.8\,\rm{fm}^{-3}$ (equivalently $\widehat{\rho}_0\approx1/5$) and $K_0\approx230\,\rm{MeV}$, $J_0\approx-300\,\rm{MeV}$, $S\approx30\,\rm{MeV}$, $L\approx 60\,\rm{MeV}$, $K_{\rm{sym}}\approx-100\,\rm{MeV}$ together with $J_{\rm{sym}}\approx300\,\rm{MeV}$\,\cite{LCXZ21}, then $\delta\approx0.6$ is obtained at $\rho_{\rm{c}}$, the central energy density $\varepsilon_{\rm{c}}$ and the central pressure $P_{\rm{c}}$ are obtained as $\varepsilon_{\rm{c}}\approx 863\,\rm{MeV}/\rm{fm}^3$ and $P_{\rm{c}}\approx 182\,\rm{MeV}/\rm{fm}^3$, respectively. Consequently, $\Lambda=\overline{M}_{\rm{N}}\rho_{\rm{c}}/\varepsilon_{\rm{c}}\approx0.86$,  $\widehat{P}_{\rm{c}}\approx0.21$ and $\eta\approx-0.74$, $\beta\approx0.86$ and $\xi\approx0.10$.
In FIG.\,\ref{fig_a4_ll},  we show the dependence of $a_4$ on the nuclear EOS parameters $K_0$, $J_0$, $S$, $L$, $K_{\rm{sym}}$ and $J_{\rm{sym}}$ according to Eq.\,(\ref{fff}), within their currently known uncertainty ranges.

We first notice that $a_4\sim\mathcal{O}(1)$ and it can take either positive or negative values, depending on the empirical parameters.
We can further find, for instance,  $a_4$ can be positive with the increasing of skewness parameter $J_0$ of SNM or the skewness parameter $J_{\rm{sym}}$ of the symmetry energy,  e.g., $J_{\rm{sym}}\gtrsim560\,\rm{MeV}$ is needed for $a_4\gtrsim0$ (while fixing other parameters at their default values).
A more positive $J_{\rm{sym}}$ or $J_0$ makes $a_4$ more positive, tending to generate the peak in $s^2$ at a smaller density\,\cite{ZLi23}.
In addition, the lower-order parameters such as $S$ or $L$ have little impact on the evolution of $a_4$.
It is necessary to point out that graphs of FIG.\,\ref{fig_a4_ll} are obtained by truncating the relevant expansions to linear order, which is expected to well behave for $S$ or $L$, e.g.,  $S/\overline{M}_{\rm{N}}\approx3.3\%$ or $L/\overline{M}_{\rm{N}}\approx6.5\%$ for $S\approx30\,\rm{MeV}$ or $L\approx60\,\rm{MeV}$, respectively.
However, for terms like $J_{\rm{sym}}/\overline{M}_{\rm{N}}\approx33\%$ (adopting $J_{\rm{sym}}\approx300\,\rm{MeV}$) and $J_0/\overline{M}_{\rm{N}}\approx-33\%$ (with $J_0\approx-300\,\rm{MeV}$), the corresponding higher-order corrections are necessarily needed. Based on these results, we conclude that the maximum absolute value of $a_4$ is about 1 within the current uncertain ranges of all EOS model parameters considered.


\begin{references}

\bibitem{Baym18}G. Baym {et al.}, \href{https://iopscience.iop.org/article/10.1088/1361-6633/aaae14}{Rep. Prog. Phys. \textbf{81}, 056902 (2018).}

\bibitem{Tews18} I. Tews {et al.}, \href{https://iopscience.iop.org/article/10.3847/1538-4357/aac267}{Astrophys. J. \textbf{860}, 149 (2018).}

\bibitem{McL19}L. McLerran and S. Reddy, \href{https://journals.aps.org/prl/abstract/10.1103/PhysRevLett.122.122701}{Phys. Rev. Lett. \textbf{122}, 122701 (2019).}

\bibitem{Baym19}G. Baym {et al.}, \href{https://iopscience.iop.org/article/10.3847/1538-4357/ab441e}{Astrophys. J. \textbf{885}, 42 (2019).}

\bibitem{Cas19}D. Alvarez-Castillo et al., \href{https://journals.aps.org/prd/abstract/10.1103/PhysRevD.99.063010}{Phys. Rev. D \textbf{99}, 063010 (2019).}

\bibitem{Fer20}M. Ferreira, R. Pereira, and C.  Provid$\hat{\rm{e}}$ncia,  \href{https://journals.aps.org/prd/abstract/10.1103/PhysRevD.102.083030}{Phys. Rev. D \textbf{102}, 083030 (2020).}

\bibitem{Dua20}D. Duarte, S. Hernandez-Ortiz, and K. S. Jeong, \href{https://journals.aps.org/prc/abstract/10.1103/PhysRevC.102.025203}{Phys. Rev. C \textbf{102}, 025203 (2020).}

\bibitem{Mal20}G. Malfatti et al., \href{https://journals.aps.org/prd/abstract/10.1103/PhysRevD.102.063008}{Phys. Rev. D \textbf{102}, 063008 (2020).}

\bibitem{Mot20}A. Motornenko et al., \href{https://journals.aps.org/prc/abstract/10.1103/PhysRevC.101.034904}{Phys. Rev. C \textbf{101}, 034904 (2020).}

\bibitem{Zhao20}T.Q. Zhao and J. Lattimer, \href{https://journals.aps.org/prd/abstract/10.1103/PhysRevD.102.023021}{Phys. Rev. D \textbf{102}, 023021 (2020).}

\bibitem{Jok21}P. Jokobus et al., \href{https://link.springer.com/article/10.1140/epjc/s10052-020-08779-x}{Eur. Phys. J. C \textbf{81}, 41 (2021).}

\bibitem{Min21}T. Minamikawa, T. Kojo, and M. Harada, \href{https://journals.aps.org/prc/abstract/10.1103/PhysRevC.103.045205}{Phys. Rev. C \textbf{103}, 045205 (2021).}

\bibitem{Kap21}J. Kapusta and T. Welle, \href{https://journals.aps.org/prc/abstract/10.1103/PhysRevC.104.L012801}{Phys. Rev. C \textbf{104}, L014801 (2021).}

\bibitem{Sen21}S. Sen and L. Sivertsen, \href{https://iopscience.iop.org/article/10.3847/1538-4357/abff4c}{Astrophys. J. \textbf{915}, 109 (2021).}

\bibitem{Stone21}J. Stone {et al.}, \href{https://academic.oup.com/mnras/article/502/3/3476/6061395}{Mon. Not. Roy. Astron. Soc. \textbf{502}, 3476 (2021).}

\bibitem{Tan22}H. Tan {et al.}, \href{https://journals.aps.org/prl/abstract/10.1103/PhysRevLett.128.161101}{Phys. Rev. Lett. \textbf{128}, 161101 (2022).}

\bibitem{Tan22-a}H. Tan {et al.}, \href{https://journals.aps.org/prd/abstract/10.1103/PhysRevD.105.023018}{Phys. Rev. D \textbf{105}, 023018 (2022).}

\bibitem{Alt22}S. Altiparmak, C. Ecker,  and L. Rezzolla, \href{https://iopscience.iop.org/article/10.3847/2041-8213/ac9b2a}{Astrophys. J. Lett. \textbf{939}, L34 (2022).}

\bibitem{Dri22}C. Drischler, S. Han, and S. Reddy, \href{https://journals.aps.org/prc/abstract/10.1103/PhysRevC.105.035808}{Phys. Rev. C \textbf{105}, 035808 (2022).}

\bibitem{Huang22}Y.J. Huang {et al.}, \href{https://journals.aps.org/prl/abstract/10.1103/PhysRevLett.129.181101}{Phys. Rev. Lett. \textbf{129}, 181101 (2022).}

\bibitem{Kojo22}T. Kojo, G. Baym, and T. Hatsuda, \href{https://iopscience.iop.org/article/10.3847/1538-4357/ac7876}{Astrophys. J. \textbf{934}, 46 (2022).}

\bibitem{Ecker22}C. Ecker and L. Rezzolla, \href{https://iopscience.iop.org/article/10.3847/2041-8213/ac8674/meta}{Astrophys. J. Lett. \textbf{939}, L35 (2022).}

\bibitem{Ecker23}C. Ecker and L. Rezzolla, \href{https://academic.oup.com/mnras/article/519/2/2615/6957261}{Mon. Not. Roy. Astron. Soc. \textbf{519}, 2615 (2023).}

\bibitem{Fuji22}Y. Fujimoto {et al.}, \href{https://journals.aps.org/prl/abstract/10.1103/PhysRevLett.129.252702}{Phys. Rev. Lett. \textbf{129}, 252702 (2022).}

\bibitem{Fuji23}Y. Fujimoto et al.,  \href{https://journals.aps.org/prl/abstract/10.1103/PhysRevLett.130.091404}{Phys. Rev. Lett. \textbf{130}, 091404 (2023).}

\bibitem{Han23}M.Z. Han et al., \href{https://www.sciencedirect.com/science/article/abs/pii/S2095927323002475}{Science Bulletin \textbf{68}, 913 (2023).}


\bibitem{Mar23}M. Marczenko et al., \href{https://journals.aps.org/prc/abstract/10.1103/PhysRevC.107.025802}{Phys. Rev. C \textbf{107}, 025802 (2023).}

\bibitem{Som23}R. Somasundaram, I. Tews,  and J. Margueron, \href{https://journals.aps.org/prc/abstract/10.1103/PhysRevC.107.025801}{Phys. Rev. C \textbf{107}, 025801 (2023).}



\bibitem{Pro23}C. Provid$\hat{\rm{e}}$ncia et al., \href{https://arxiv.org/pdf/2307.05086.pdf}{arXiv:2307.05086 (2023).}


\bibitem{Liu23} H. Liu et al., \href{https://journals.aps.org/prd/abstract/10.1103/PhysRevD.108.034004}{Phys. Rev. D \textbf{108}, 034004 (2023).}



\bibitem{Chen23}M.J. Chen et al., \href{https://arxiv.org/pdf/2309.11245.pdf}{arXiv:2309.11245 (2023).}


\bibitem{Ann18}E. Annala {et al.}, \href{https://journals.aps.org/prl/abstract/10.1103/PhysRevLett.120.172703}{Phys. Rev. Lett. \textbf{120}, 172703 (2018).}

\bibitem{Ann20N}E. Annala {et al.}, \href{https://www.nature.com/articles/s41567-020-0914-9}{Nat. Phys. \textbf{16}, 907 (2020).}

\bibitem{Ann23}E. Annala et al., \href{https://www.nature.com/articles/s41467-023-44051-y}{Nat. Comm. \textbf{14}, 8451 (2023).}

\bibitem{Gorda23}T. Gorda, O. Komoltsev, and A. Kurkela, \href{https://iopscience.iop.org/article/10.3847/1538-4357/acce3a}{Astrophys. J. \textbf{950}, 107 (2023).}

\bibitem{Gorda21}T. Gorda et al., \href{https://journals.aps.org/prl/abstract/10.1103/PhysRevLett.127.162003}{Phys. Rev. Lett. \textbf{127}, 162003 (2021).}


\bibitem{ZLi23}N.B. Zhang and B.A.  Li, \href{https://link.springer.com/article/10.1140/epja/s10050-023-01010-x}{Eur. Phys. J. A  \textbf{59}, 86 (2023).}

\bibitem{Cao23}Z. Cao and L.W. Chen, \href{https://arxiv.org/pdf/2308.16783.pdf}{arXiv:2308.16783 (2023).}

\bibitem{Mro23}D. Mroczek et al.,  \href{https://arxiv.org/pdf/2309.02345.pdf}{arXiv:2309.02345 (2023).}

\bibitem{Abbott2017}B. Abbott {et al.}, \href{https://journals.aps.org/prl/abstract/10.1103/PhysRevLett.119.161101}{Phys. Rev. Lett. \textbf{119}, 161101 (2017).}

\bibitem{Abbott2018}B. Abbott {et al.}, \href{https://journals.aps.org/prl/abstract/10.1103/PhysRevLett.121.161101}{Phys. Rev. Lett. \textbf{121}, 161101 (2018).}

\bibitem{Abbott2020}B. Abbott et al., \href{https://iopscience.iop.org/article/10.3847/2041-8213/ab75f5}{Astrophys. J. L. \textbf{892}, L3 (2020).}

\bibitem{Brandes23}L. Brandes, W. Weise, and N. Kaiser, \href{https://journals.aps.org/prd/abstract/10.1103/PhysRevD.107.014011}{Phys. Rev. D \textbf{107}, 014011 (2023).}

\bibitem{Brandes23-a}L. Brandes, W. Weise, and N. Kaiser, \href{https://arxiv.org/abs/2306.06218}{arXiv:2306.06218 (2023).}

\bibitem{Tak23}J. Takatsy et al., \href{https://arxiv.org/pdf/2303.00013.pdf}{arXiv:2303.00013 (2023).}

\bibitem{Pang23}P. Pang et al., \href{https://arxiv.org/pdf/2308.15067.pdf}{arXiv:2308.15067 (2023).}

\bibitem{Fan23}Y.Z. Fan et al., \href{https://journals.aps.org/prd/abstract/10.1103/PhysRevD.109.043052}{Phys. Rev. D \textbf{109}, 043052 (2024).}

\bibitem{Shapiro1983}S. Shapiro and S. Teukolsky, \textit{Black Holes, White Dwarfs and Neutron Stars}, Wiley-VCH, 1983, Chap. 2.

\bibitem{Ess21}R.  Essick et al., \href{https://journals.aps.org/prl/abstract/10.1103/PhysRevLett.127.192701}{Phys. Rev. Lett. \textbf{127}, 192701 (2021).}


\bibitem{Fon21} E. Fonseca {et al.},  \href{https://iopscience.iop.org/article/10.3847/2041-8213/ac03b8}{Astrophys. J. Lett. \textbf{915}, L12 (2021).}

\bibitem{Riley21}T. Riley {et al.}, \href{https://iopscience.iop.org/article/10.3847/2041-8213/ac0a81}{Astrophys. J. Lett. \textbf{918}, L27 (2021).}

\bibitem{Miller21}M. Miller {et al.}, \href{https://iopscience.iop.org/article/10.3847/2041-8213/ac089b}{Astrophys. J. Lett. \textbf{918}, L28 (2021).}

\bibitem{Salmi22}T. Salmi et al., \href{https://iopscience.iop.org/article/10.3847/1538-4357/ac983d/pdf}{Astrophys. J. \textbf{941}, 150 (2022).}


\bibitem{Dem10}P. Demorest et al., \href{https://www.nature.com/articles/nature09466}{Nature \textbf{467}, 1081 (2010).}

\bibitem{Ant13} J. Antoniadis {et al.}, \href{https://www.science.org/doi/10.1126/science.1233232}{Science \textbf{340}, 1233232 (2013).}


\bibitem{TOV39-1}R. Tolman, \href{https://journals.aps.org/pr/abstract/10.1103/PhysRev.55.364}{Phys. Rev. \textbf{55}, 364 (1939).}
\bibitem{TOV39-2}
J. Oppenheimer and G. Volkoff, \href{https://journals.aps.org/pr/abstract/10.1103/PhysRev.55.374}{Phys. Rev. \textbf{55}, 374 (1939).}

\bibitem{Misner1973}C. Misner, K. Thorne, and J. Wheeler, \textit{Gravitation},  Princeton University Press, 2017, Section 23.5.


\bibitem{Walecka1974}J. Walecka, \href{https://www.sciencedirect.com/science/article/pii/0003491674902085}{Ann. Phys. \textbf{83}, 491 (1974).}

\bibitem{Chin1976}S. Chin, \href{https://www.sciencedirect.com/science/article/pii/0003491677900161}{Ann. Phys. \textbf{108}, 301 (1976).}

\bibitem{Baym1976}G. Baym and S. Chin, \href{https://www.sciencedirect.com/science/article/pii/0370269376905177}{Phys. Lett. \textbf{B62}, 241 (1976).}


\bibitem{Freedman1977}B. Freedman and L. McLerran, \href{https://journals.aps.org/prd/abstract/10.1103/PhysRevD.16.1130}{Phys. Rev. D \textbf{16}, 1130 (1977);} \href{https://journals.aps.org/prd/abstract/10.1103/PhysRevD.16.1147}{1147 (1977);} \href{https://journals.aps.org/prd/abstract/10.1103/PhysRevD.16.1169}{1169 (1977).}


\bibitem{Akmal1998}A. Akmal, V. Pandharipande, and D. Ravenhall, \href{https://journals.aps.org/prc/abstract/10.1103/PhysRevC.58.1804}{Phys. Rev. C \textbf{58}, 1804 (1998).}


\bibitem{LP01}J. Lattimer and M. Prakash, \href{https://iopscience.iop.org/article/10.1086/319702}{Astrophys. J. \textbf{550}, 426 (2001).}


\bibitem{Alford2008}M. Alford {et al.}, \href{https://journals.aps.org/rmp/abstract/10.1103/RevModPhys.80.1455}{Rev. Mod. Phys. \textbf{80}, 1455 (2008).}

\bibitem{LCK08} B.A. Li, L.W. Chen, and C.M. Ko, \href{https://www.sciencedirect.com/science/article/pii/S0370157308001269}{Phys. Rep. \textbf{464}, 113 (2008).}

\bibitem{Wat16}A. Watts et al., \href{https://journals.aps.org/rmp/abstract/10.1103/RevModPhys.88.021001}{Rev. Mod. Phys. \textbf{88}, 021001 (2016).}

\bibitem{Oertel2017} M. Oertel {et al.}, \href{https://journals.aps.org/rmp/abstract/10.1103/RevModPhys.89.015007}{Rev. Mod. Phys.  \textbf{89}, 015007 (2017).}

\bibitem{Isa18}I. Vida$\widetilde{\rm{n}}$a, \href{https://royalsocietypublishing.org/doi/10.1098/rspa.2018.0145}{Proc. Roy. Soc. Lond. A \textbf{474}, 0145 (2018).}


\bibitem{Dri21}C. Drischler, J. Holt, and C. Wellenhofer, \href{https://www.annualreviews.org/doi/10.1146/annurev-nucl-102419-041903}{Annu. Rev. Nucl. Part. Sci. \textbf{71}, 403 (2021).}

\bibitem{Lovato22}A. Lovato {et al.},  \href{https://arxiv.org/abs/2211.02224}{arXiv:2211.02224 (2022).}

\bibitem{Soren2023}A. Sorensen {et al.}, \href{https://doi.org/10.1016/j.ppnp.2023.104080}{Prog. Part. Nucl. Phys.  \textbf{134}, 104080 (2024).}

\bibitem{CLZ23-a}B.J. Cai, B.A. Li, and Z. Zhang, \href{https://iopscience.iop.org/article/10.3847/1538-4357/acdef0}{Astrophys. J. \textbf{952}, 147 (2023).}
\bibitem{CLZ23-b}B.J. Cai, B.A. Li, and Z. Zhang, \href{https://doi.org/10.1103/PhysRevD.108.103041}{Phys. Rev. D \textbf{108}, 103041 (2023).}

\bibitem{Lat05} J. Lattimer and M. Prakash, \href{https://journals.aps.org/prl/abstract/10.1103/PhysRevLett.94.111101}{Phys. Rev. Lett. \textbf{94}, 111101 (2005).}

\bibitem{Of1} D. Ofengeim, \href{https://journals.aps.org/prd/abstract/10.1103/PhysRevD.101.103029}{Phys. Rev. D \textbf{101}, 103029 (2020).}

\bibitem{Of2} D. Ofengeim,  P. Shternin,  and T. Piran,  \href{https://link.springer.com/article/10.1134/S1063773723100055}{Astron. Lett. \textbf{49}, 567 (2023).}


\bibitem{Jim}J. Lattimer, private communications, Jan. 2024.

\bibitem{Lat24-talk}J. Lattimer,  slide 15, talk given at the \href{https://indico.mitp.uni-mainz.de/event/380/contributions/4746/attachments/3474/4433/Bormio24Lattimer.pdf}{60th International Winter Meeting in Nuclear Physics, Jan. 22-26, 2024, Bormio, Italy.}


\bibitem{Sil04}R. Silbar and S. Reddy,  \href{https://pubs.aip.org/aapt/ajp/article/72/7/892/1056102/Neutron-stars-for-undergraduates}{Am. J. Phys. \textbf{72}, 892 (2004).}

\bibitem{Smi12}A. Smith, \href{http://www.as.utexas.edu/astronomy/education/spring13/bromm/secure/TOC_Supplement.pdf}{Online Lecture Notes, Univ. of Texas at Austin}


\bibitem{Chan10}S. Chandrasekhar, \textit{An Introduction to the Study of Stellar Structure}, Dover Publications, 2010, Chap. 3.


\bibitem{MPA1}H.  M$\ddot{\rm{u}}$ther, M. Prakash, and T. Ainsworth, \href{https://www.sciencedirect.com/science/article/pii/037026938791611X}{Phys. Lett. \textbf{B199}, 469 (1987).}

\bibitem{ENG}L. Engvik et al., \href{https://ui.adsabs.harvard.edu/abs/1996ApJ...469..794E/abstract}{Astrophys. J.  \textbf{469}, 794 (1996).}

\bibitem{Sky}J. Stone and P. Reinhard, \href{https://www.sciencedirect.com/science/article/pii/S0146641006000627}{Prog. Nucl. Part. Phys. \textbf{58}, 587 (2007).}


\bibitem{SLy}F.  Douchin and P.  Haensel, \href{https://arxiv.org/abs/astro-ph/0111092}{Astron. Astrophys. \textbf{380}, 151 (2001).}

\bibitem{Serot1986}B. Serot and J. Walecka, \href{https://inspirehep.net/literature/207866}{Adv. Nucl. Phys. \textbf{16}, 1 (1986).}


\bibitem{Glen85}N. Glendenning, \href{https://ui.adsabs.harvard.edu/abs/1985ApJ...293..470G/abstract}{Astrophys. J. \textbf{293}, 470 (1985).}

\bibitem{H4}B.  Lackey, M. Nayyar, and B. Owen, \href{https://journals.aps.org/prd/abstract/10.1103/PhysRevD.73.024021}{Phys. Rev. D \textbf{73}, 024021 (2006).}

\bibitem{NL3}G. Lalazissis, J.  K$\ddot{\rm{o}}$nig,  and P.  Ring,  \href{https://journals.aps.org/prc/abstract/10.1103/PhysRevC.55.540}{Phys. Rev. C \textbf{55}, 540 (1997).}

\bibitem{FSU}B. Todd-Rutel and J. Piekarewicz, \href{https://journals.aps.org/prl/abstract/10.1103/PhysRevLett.95.122501}{Phys. Rev. Lett. \textbf{95}, 122501 (2005).}




\bibitem{ALi20}A. Li et al.,  \href{https://www.sciencedirect.com/science/article/pii/S2214404820300355}{J. High Energy Astrophys. \textbf{28}, 19 (2020).}


\bibitem{ALF2}M. Alford et al., \href{https://iopscience.iop.org/article/10.1086/430902}{Astrophys. J. \textbf{629}, 969 (2005).}

\bibitem{NASA} \href{https://nssdc.gsfc.nasa.gov/planetary/factsheet/sunfact.html}{Data from NASA.}


\bibitem{Lightman1975}A. Lightman et al., \textit{Problem Book in Relativity and Gravitation},  Princeton University Press, 1975, Problem 16.13.



\bibitem{Bjorken83}J. Bjorken, \href{https://journals.aps.org/prd/abstract/10.1103/PhysRevD.27.140}{Phys. Rev. D \textbf{27}, 140 (1983).}

\bibitem{Kur10}A. Kurkela, P. Romatschke, and A. Vuorinen, \href{https://journals.aps.org/prd/abstract/10.1103/PhysRevD.81.105021}{Phys. Rev. D \textbf{81}, 105021 (2010).}

\bibitem{Gorda23-a}T. Gorda et al., \href{https://journals.aps.org/prl/abstract/10.1103/PhysRevLett.131.181902}{Phys. Rev. Lett. \textbf{131}, 181902 (2023).}

\bibitem{Bed15}P. Bedaque and A. Steiner, \href{https://journals.aps.org/prl/abstract/10.1103/PhysRevLett.114.031103}{Phys. Rev. Lett. \textbf{114}, 031103 (2015).}


\bibitem{Olii23}D. Oliinychenko et al., \href{https://journals.aps.org/prc/pdf/10.1103/PhysRevC.108.034908}{Phys. Rev. C \textbf{108}, 034908 (2023).}

\bibitem{Kut22}M. Kuttan et al.,  \href{https://journals.aps.org/prl/abstract/10.1103/PhysRevLett.131.202303}{Phys. Rev. Lett. \textbf{131}, 202303 (2023).}

\bibitem{Bra23}B. Brandt et al., \href{https://link.springer.com/article/10.1007/JHEP07(2023)055}{JHEP \textbf{7}, 055 (2023).}
\bibitem{Aya23} A. Ayala et al., \href{https://arxiv.org/pdf/2310.13130.pdf}{arXiv:2310.13130 (2023).}

\bibitem{NAP2011} The National Academies Press, {\it New Worlds, New Horizons in Astronomy and Astrophysics}, 2011.

\bibitem{LCXZ21}B.A. Li et al., \href{https://www.mdpi.com/2218-1997/7/6/182}{Universe \textbf{7}, 182 (2021).}

\bibitem{He14}X.T. He et al., 
\href{https://journals.aps.org/prc/abstract/10.1103/PhysRevC.91.015810}{Phys. Rev. C \textbf{91}, 015810 (2015).}

\bibitem{Li19}B.A.Li et al., \href{https://www.mdpi.com/2218-1997/7/6/182}{Eur. Phys. J. A \textbf{55}, no.7, 117 (2019).}
\end{references}
\end{document}